\magnification=\magstephalf
\input amstex
\loadbold
\documentstyle{amsppt}
\refstyle{A}
\NoBlackBoxes

\vsize=7.5in

\def\pf{\hfill $\square$}
\def\c{\cite}
\def\fb{\frak{b}}
\def\fg{\frak{g}}

\def\fk{\frak{k}}

\def\end{\text{End}}

\def\IR{\Bbb R}
\def\IC{\Bbb C}
\def\ID{\Bbb D}
\def\IP{\Bbb P}
\def\bk{\boldkey k}

\def\bb{\boldkey b}

\def\gtw{{{\tilde{\fg}}^{\Bbb{R}}_{w}}}
\def\Gtw{{{\tilde{G}}^{\Bbb{R}}_{w}}}
\def\ktw{{{\tilde{\frak{k}}}_{w}}}
\def\Ktw{{{\tilde{K}}_{w}}}
\def\btw{{{\tilde{\frak{b}}}_{w}}}
\def\Btw{{{\tilde{B}}_{w}}}
\def\piktw{{\Pi_{\ktw}}}
\def\pibtw{{\Pi_{\btw}}}
\def\psij{\psi^{[j]}}
\def\Ej{{\Cal E}^{[j]}}
\def\E{\Cal{E}}
\def\bds{\partial{\Bbb D}}

\topmatter
\title The periodic defocusing Ablowitz-Ladik equation and the geometry of 
Floquet CMV matrices\endtitle
\leftheadtext{L.-C. Li, I. Nenciu}
\rightheadtext{The periodic AL equation and Floquet CMV matrices}

\author Luen-Chau Li and Irina Nenciu\endauthor
\address{L.-C. Li, Department of Mathematics,Pennsylvania State University
University Park, PA  16802, USA}\endaddress
\email luenli\@math.psu.edu\endemail
\address{I. Nenciu, Department of Mathematics, University of Illinois at
Chicago, Chicago, IL, USA and Institute of Mathematics ``Simon Stoilow'' of 
the Romanian Academy, Bucharest, Romania}\endaddress
\email nenciu\@uic.edu\endemail

\abstract  In this work, we show that the periodic defocusing Ablowitz-Ladik equation
can be expressed as an isospectral deformation of Floquet CMV matrices.  We then
introduce a Poisson Lie group whose underlying group is a loop group and show
that the set of Floquet CMV matrices is a Coxeter dressing orbit of this Poisson Lie
group.   By using the group-theoretic framework, we establish the Liouville integrability 
of the equation by constructing action-angle variables, we also solve the Hamiltonian 
equations generated by the commuting flows via Riemann-Hilbert factorization problems.
\endabstract
\endtopmatter

\document
\subhead
1. \ Introduction
\endsubhead

\bigskip

The defocusing Ablowitz-Ladik (AL) equation (a.k.a. defocusing discrete nonlinear Schr\"odinger equation)
is the system of differential-difference equations given by
$$ -i\dot\alpha_n = \alpha_{n+1} -2\alpha_{n} + \alpha_{n-1} -|\alpha_n|^2 (\alpha_{n-1} +\alpha_{n+1})
\eqno(1.1)$$
where $\alpha_n$ is a sequence of numbers inside the unit disk $\Bbb D.$   It was introduced
by Ablowitz and Ladik in \c{AL} as a spatial discretization of the defocusing nonlinear Schr\"odinger
equation and since then has been the subject of numerous studies.   In particular, much attention
has been focused on the inverse scattering method for solving the equation in the two-sided case
where the index $n$ ranges over the set of integers. (See \c{APT} and the references therein.)
By contrast, the literature on the periodic problem is relatively sparse. (See, for example, \c{K},
\c{MEKL}, \c{N}, \c{GHMT}.)

In recent years, one of the interesting developments in the arena of the defocusing AL equation
has been the connection with the theory of orthogonal polynomials on the circle (OPUC) and
the so-called CMV matrices  \c{S2}, \c{N}, and our work here is a continuation of this development.   Since we are
dealing with the periodic defocusing AL equation here, let us begin with a set of
Verblunsky coefficients $\{\alpha_j\}_{j=0}^{\infty}$ satisfying the periodicity condition $\alpha_{j+p}=\alpha_j,
j=0, 1, 2, \cdots.$   Without loss of generality, we may assume $p$ is even.   In \c{S2},  Simon
introduced the discriminant $\Delta(z)$ associated with $\{\alpha_j\}_{j=0}^{\infty}$ and together with
the second author, they obtained the following involution theorem \c{S2}, \c{N}
$$\{\Delta(z), \Delta(w)\}_{AL} =0,\quad \{P, \Delta(z) \}_{AL} =0\eqno(1.2)$$
by calculating with Wall polynomials.    Here $\{\cdot, \cdot\}_{AL}$ is the Ablowitz-Ladik Poisson bracket
\c{KM}, \c{S2} and $P = \prod_{j=0}^{p-1} \rho_j,$ where $\rho_j = \sqrt{1-|\alpha_j|^{2}}.$  This
prompted the search for an integrable system which is related to OPUC in the same way
the Toda lattice is related to orthogonal polynomials on the line.   As it turned out, the sought-after
integrable system is the periodic defocusing AL equation \c{N}.   In \c{N}, the Lax equations for
the commuting flows were expressed in terms of the extended CMV matrix $\Cal E$ with periodic Verblunsky coefficients. 
However, as $\Delta(z)$ is related to the Floquet CMV matrix $\Cal E(h)$  (which is a unitary
loop with spectral parameter $h$) through the characteristic polynomial $\hbox{det}(zI -\Cal E(h)),$
it is natural to ask if the same equations can be rewritten as isospectral deformations of 
$\Cal E(h).$    As the reader will see in Section 2, this is indeed the case and the result is
the point of departure in this work.   More precisely, the result not only suggests that the set of 
$p\times p$ Floquet CMV matrices should have some Poisson geometric meaning,
but also points to the linearization of such flows on geometric objects related to the 
Jacobi varieties of the underlying spectral curves.    Thus our goal in this work is two-fold.
First of all, we will link the Floquet CMV matrices to Poisson Lie groups, analogous to
what we did in our earlier work on finite CMV matrices. (See \c{L1} and \c{KN}.)  Secondly, by using the
group-theoretic framework, we will study the defocusing AL equation with regard to
action-angle variables.   We will also solve the commuting Hamiltonian
flows via Riemann-Hilbert factorization problems.   At this juncture, let us mention some
earlier works related to the integration of the periodic defocusing AL equation which is part of our second
goal here.  To start with,
it has been known for quite some time that the defocusing AL equation (1.1) can be
represented as a Lax system on a lattice (or discrete zero curvature representation),
where the Lax operator $L_j(z)$ associated to site $j$ of the lattice is given by (see, for example, \c{AL1}, \c{AL2}
and \c{FT})
$$L_j(z) = \pmatrix  z  &  \bar \alpha_j\\
                              \alpha_j  &  z^{-1}\endpmatrix  .\eqno(1.3)$$
Therefore, in the periodic case with period $p,$ the monodromy matrix
$M(z) = L_{p-1}(z)\cdots L_{0}(z)$
undergoes an isospectral deformation which means that an equation
in Lax pair form (and different from the one we are using here) is known
for the periodic defocusing AL equation.  In \c{MEKL},
a transformation of a natural generalization of (1.3) was discovered and
the result was applied in the construction of finite genus solutions of 
a more general version of the AL equation.  In particular, the authors in
\c{MEKL} were able to write down the solution of the initial value
problem for the periodic defocusing AL equation itself.   On the other hand,
from a different direction, the authors in \c{GHMT} considered a more
general version of the AL hierarchy, and discussed the problem of solving 
the $r$-th AL flow when the initial data is the  
stationary solution of the $p$-th equation of the 
hierarchy.   As the reader will see in Section 6 below, our approach
in solving the commuting Hamiltonian flows associated with the
periodic defocusing AL equation is quite different from those
in these earlier works.

The paper is organized as follows.   In Section 2, we begin by recalling the notion of CMV matrices,
extended CMV matrices and Floquet CMV matrices.   Then we show how to rewrite the Lax equations
for $\Cal E$ of the commuting flows associated with the periodic defocusing AL equation as 
isospectral deformations of $\Cal E(h).$   
To prepare for what we need in subsequent sections, we 
also discuss the structure of the powers of $\Cal E(h).$    In Section 3, we have two main goals.  
The first goal is to show that
the set of $p\times p$ Floquet CMV matrices is a symplectic leaf of a Poisson Lie group whose
underlying group is a loop group $\widetilde G^{\Bbb R}_{w}.$    Indeed, as one would expect
from results in \c{L1} and \c{KN} concerning the finite dimensional case, the Poisson structure
here is also a Sklyanin structure.   In fact, it is the Sklyanin structure associated with the
Iwasawa decomposition of the loop group  $\widetilde G^{\Bbb R}_{w}$: $\widetilde G^{\Bbb R}_{w}=
\widetilde K_{w} \widetilde B_{w}.$  (The decomposition was established in \c{GW}.)   However,
in order to write down this Sklyanin structure $\{\cdot, \cdot \}_{J^{\sharp}},$  we find it necessary
to restrict ourselves to a subclass of functions $\Cal F(\widetilde G^{\Bbb R}_w)$ of
$C^{\infty}(\widetilde G^{\Bbb R}_{w}).$  Fortunately, $\Cal F(\widetilde G^{\Bbb R}_{w})$  forms
an algebra of functions under ordinary multiplication and is closed under $\{\cdot,\cdot\}_{J^{\sharp}}.$
Hence $\{\cdot,\cdot\}_{J^{\sharp}}$ defines a Poisson bracket on $\Cal F(\widetilde G^{\Bbb R}_{w}).$
Now note that although we are dealing with a restricted class of functions here, it can be checked
that the notion of Poisson Lie groups can be extended to this infinite dimensional context in a
rigorous way.   Moreover,  we can check by hand that the symplectic leaves of 
$(\widetilde G^{\Bbb R}_{w}, \{\cdot,\cdot\}_{J^{\sharp}})$ are still given by the orbits of
the dressing action.   With this preparation, the technique in \c{L1} can be naturally extended
to show that the set of $p\times p$  Floquet CMV matrices is a dressing orbit through a Coxeter
element $x_f$ of the affine Weyl group $W_{aff}.$   Indeed, the induced Poisson structure on
$\widetilde K_{w}$ is a loop group analog of the Bruhat Poisson structure in \c{LW} and \c{Soi}
and we can show that the set of $p\times p$ Floquet CMV matrices is a product of two dimensional
orbits.   In the rest of the section, our goal is to clarify the relation between the AL bracket and the
Sklyanin bracket $\{\cdot, \cdot\}_{J^{\sharp}}$, and to describe the Hamiltonian equations generated
by the central functions on $\widetilde G_{w},$ thus connecting the group-theoretic framework with the
equations in Section 2.     

In Section 4,  we study the analytical  properties of the Bloch solution of $\Cal E u = zu,$ which play an important role
in subsequent sections. Since $\Cal E$ defines a (pentadiagonal) periodic difference operator, the 
seminal work of van Moerbeke and Mumford \c{MM} comes to mind.  However, we note 
that neither $\Cal E$ nor its factors $\Cal L$ and $\Cal M$ in the theta-factorization
of $\Cal E$ satisfy the genericity assumption in \c{MM}.  So the analysis in this section
is more delicate than the standard case \c{MM}, \c{AM}.   In Section 5, we start with a simple proof of the
involution theorem in (1.2), which is possible because of the group-theoretic
setup in Section 3.  Then we proceed to construct the angle variables.   To compute the Poisson
brackets between the conserved quantities and the various quantities related to
the putative angles, we make use of a device introduced in \c{DLT}.  We would
like to point out that in general, such computations could be difficult
because they may require detailed information on the asymptotics of the 
normalized eigenvectors in neighborhoods of the points at infinity of the Riemann surface.  In our
case, asymptotics beyond the leading order are difficult to get because 
we are in a non-generic situation, but fortunately we are saved by some
special structure.
Finally, in Section 6 we solve the commuting Hamiltonian flows via
Riemann-Hilbert factorization problems, which again are suggested by
the group-theoretic framework.  We remark that it is in this very
last section that we find it advantageous to think of our flows
on $\Cal E(h)$ as flows on the factors $g^e$ and $g^o(h)$ in the
theta-factorization of $\Cal E(h).$  This is precisely the reason why
we introduce Lax systems on a period $2$ lattice in Section 3.

\bigskip

\subhead
2. \ Preliminaries
\endsubhead
\bigskip

In this section, for the convenience of the reader, we begin with some background material on
CMV matrices and the involution theorem of Nenciu-Simon. (Good references are \c{S2} and
\c{S3}.)  Then we will show how to rewrite the Lax equation in \c{N} for the periodic defocusing AL equation 
(in terms of the extended CMV matrix $\Cal E$) as an isospectral deformation of the Floquet CMV matrix $\Cal E(h)$.
We will also present a result on the structure of the powers of $\Cal E(h)$ which we will use 
in Sections 5 and 6.

The CMV matrices are the unitary analogs of Jacobi matrices \c{S3} and made their debut
 in the numerical linear algebra literature. (See \c{B-GE} and in particular \c{W}.)  
 Subsequently, they were rediscovered by Cantero, Moral and Val\'azquez \c{CMV} in the
 context of the theory of orthogonal polynomials on the circle (OPUC).   To introduce these objects, let
 $\Bbb D =\{ z\in \Bbb C\mid |z| < 1\},$ and let $d\mu$ be a nontrivial probability measure on $\bds,$
 then one can produce an orthonormal basis of $L^{2}(\bds, d\mu)$ by applying the Gram-Schmidt process
 to $1,z,z^{-1},z^2,z^{-2},\cdots.$   As it turns out \c{CMV}, the matrix representation of the operator 
 $f(z)\mapsto zf(z)$ in $L^2(\bds, d\mu)$ with respect to this orthonormal basis  is the infinite CMV matrix
 $${\Cal C} = \pmatrix \bar\alpha_0 & \rho_0\bar\alpha_1 & \rho_0\rho_1 & 0 & 0 & \cdots\\
                      \rho_0& -\alpha_0\bar\alpha_1 & -\alpha_0\rho_1 & 0 & 0 & \cdots\\
                       0  & \rho_1\bar\alpha_2 &-\alpha_1\bar\alpha_2  & \rho_2\bar\alpha_3 &\rho_2\rho_3 & \cdots\\
                       0 & \rho_1\rho_2& -\alpha_1\rho_2 & -\alpha_2\bar\alpha_3 & -\alpha_2\rho_3 &\cdots\\
                        \cdots & \cdots & \cdots & \cdots & \cdots & \cdots\\
                      \endpmatrix = \widetilde{\Cal L}\widetilde{\Cal M}\eqno(2.1)$$
where $\alpha_j\in \Bbb D$ are the so-called Verblunsky coefficients, $\rho_j =\sqrt{1-|\alpha_j|^2},$
$j=0,1,\cdots,$  and where $\widetilde{\Cal L} = \hbox{diag}(\theta_0,\theta_2,\cdots),$
$\widetilde{\Cal M} = \hbox{diag}(1, \theta_1,\theta_3,\cdots),$ with
$$\theta_j = \pmatrix   \bar \alpha_j& \rho_j\cr
  \rho_j& -\alpha_j\cr
  \endpmatrix , \quad j=0,1,\cdots .\eqno(2.2)$$
The factorization on the right hand side of (2.1) is called the $theta$-factorization and lends itself
to generalization.   Indeed, if we now have a two-sided sequence $\{\alpha_j\}_{j= -\infty}^{\infty}$
with $\alpha_j\in \Bbb D$ for all $j,$ then we can define the extended (two-sided) CMV matrix
$\Cal E$ by extending $\widetilde{\Cal L}$ and $\widetilde{\Cal M}$ to doubly-infinite matrices
in the obvious way. (Of course, the $1\times 1$ block will not appear in this extension.)
In this work, we are mainly interested in the case where the sequence $\{\alpha_j\}_{j=0}^{\infty}$
of Verblunsky coefficients is periodic of period $p$ and in this context, it is convenient to
extend the one-sided sequence to a two-sided seqence $\{\alpha_j\}_{j=-\infty}^{\infty}$ 
satisfying the periodicity condition $\alpha_{j+p} =\alpha_j$ for all $j\in \Bbb Z.$
Thus correspondingly, we have an extended CMV matrix with periodic Verblunksky coefficients
and such matrices have been used to formulate the Lax equation for the periodic defocusing
AL equation in \c{N}.    Now suppose $\Cal E$ is an extended CMV matrix with periodic
Verblunsky coefficients with period $p.$   Without loss of generality, we will assume from
now onwards that $p$ is even (otherwise, we just replace $p$ by $2p$).   Note that if $T$ is
the operator on $l^{\infty}(\Bbb Z)$ defined by $(Tu)_j = u_{j+p},$ then $\Cal E T = T \Cal E.$
Therefore, if for $h\in \bds,$ we define
$$X_{(h)} = \{u\in l^{\infty}(\Bbb Z)\mid Tu = h^{-1} u\},\eqno(2.3)$$
then the finite dimensional space $X_{(h)}$ is invariant under $\Cal E.$   A basis of $X_{(h)}$ is
given by the vectors
$$\delta_k = \sum_{j= -\infty}^{\infty} h^{-j} e_{jp + k},\,\,\,k=0,\cdots, p-1\eqno(2.4)$$
where $e_j$ is the vector in $l^{\infty}(\Bbb Z)$ with $j$-th component equal to $1$ and zeros
elsewhere.  By definition, the Floquet CMV matrix $\Cal E(h)$ is the matrix of $\Cal E\mid X_{(h)}$
with respect to the ordered basis $(\delta_0,\cdots,\delta_{p-1}),$ i,e.
$$\Cal E \delta_k = \sum_{j=0}^{p-1} (\Cal E(h))_{jk} \delta_j,\,\,\,k=0,\cdots, p-1.\eqno(2.5)$$
Fom this, it is clear that the matrix  of $\Cal E^n\mid X_{(h)}$ with respect
to the same ordered basis is $\Cal E(h)^n.$  Thus the entries of $\Cal E(h)^n$ are related to those
of $\Cal E^n$ by the formula
$$(\Cal E(h)^n)_{jk} = \sum_{\l\in \Bbb Z} h^{-l} (\Cal E^n)_{j, k + lp}\eqno(2.6)$$
for $0\leq j,k \leq p-1.$
Finally, the Floquet CMV matrix $\Cal E(h)$ also has a theta-factorization $\Cal E(h) = g^e g^o(h)$, where
$$g^e = \hbox{diag}(\theta_0,\theta_2,\cdots,\theta_{p-2}),\eqno(2.7)$$
and
$$g^o(h) = \pmatrix
    -\alpha_{p-1} & 0 & \cdots & 0 & \rho_{p-1} h \\
    0 & \theta_1  &   &   & 0 \\
    \vdots &  & \ddots &  & \vdots \\
    0 &   &   &  \theta_{p-3} & 0 \\
    \rho_{p-1} h^{-1} & 0 & \cdots & 0 & \bar\alpha_{p-1} \\
\endpmatrix.\eqno(2.8)$$
This is of course a consequence of the factorization for the corresponding extended CMV matrix
$\Cal E.$  In case we want to emphasize the dependence of $g^e$ and $g^o$ on 
$\underline{\alpha} = (\alpha_o,\cdots, \alpha_{p-1})\in \Bbb D^p,$ we also write
$g^e = g^e(\underline{\alpha})$ and $g^o =g^o(\underline{\alpha}).$

Another very important notion associated with periodic Verblunsky coefficients is that of the
discriminant introduced in \c{S2}:
$$\Delta(z) = z^{-p/2} \hbox{tr}\,(T_p(z)),\,\,\,z\in \Bbb C\setminus\{0\}\eqno(2.9)$$
where
$$T_p(z) = \frac{1}{\prod_{j=0}^{p-1} \rho_j} \pmatrix  z & -\bar\alpha_{p-1}\\
                                                                            -\alpha_{p-1} z & 1\endpmatrix
    \cdots \pmatrix   z  &   -\bar\alpha_0\\
                    -\alpha_0 z &  1\endpmatrix \eqno(2.10)$$
is the transfer matrix.    In \c{S2},  by seeking a Poisson bracket on $\Bbb D^p$ so that the
modulus $P = \prod_{j=0}^{p-1} \rho_j$ generates the Aleksandrov flow, the author arrives
at the Ablowitz-Ladik bracket (see \c{KM} for more general versions of this structure)
$$\{f_1, f_2\}_{AL} = 2 i\sum_{j=0}^{p-1} \rho_j^2\left({\frac{\partial f_1}{\partial \alpha_j}}
{\frac{\partial f_2}{\partial \bar \alpha_j}} -{\frac{\partial f_1}{\partial \bar \alpha_j}}
{\frac{\partial f_2}{\partial \alpha_j}}\right).\eqno(2.11)$$
We recall the involution theorem of Nenciu-Simon which was obtained by calculating with
Wall polynomials.

\proclaim
{Theorem 2.1 \c{N}, \c{S2}}  For all $z, w\in \Bbb C\setminus \{0\},$
$$\{\,\Delta(z), \Delta(w)\,\}_{AL} =0,\,\,\, \{\,P, \Delta(z)\,\}_{AL} =0.\eqno(2.12)$$
Hence if $P\cdot \Delta(z) = \sum_{j= -p/2}^{p/2} I_j z^j,$  the functions
$P, I_0, \hbox{Re}\, I_j, \hbox{Im}\,I_j, j=1,\cdots, p/2-1$ Poisson commute with
each other.
\endproclaim

This result, when combined with the proof that the functions $P, I_0, \hbox{Re}\, I_j, \hbox{Im}\,I_j,$ $ j=1,\cdots, p/2-1$
are functionally independent on an open dense subset of $\Bbb D^p$ \c{S2}, shows that any of the functions in
the above list generates a completely integrable Hamiltonian system.   To write down the Lax pairs, the
author in \c{N} actually considered a different, but equivalent set of commuting Hamiltonians, which
are constructed from the real and imaginary parts of 
$$K_n = \frac{1}{n} \sum_{k=0}^{p-1} (\Cal E^n)_{kk},\quad 1\leq n\leq p/2 -1\,,$$
together with $K_{p/2}$ and $P.$   Indeed, an easy computation shows that
$$\{\hbox{Re}(K_1), \alpha_j\}_{AL} = i\rho_j^2(\alpha_{j-1}+\alpha_{j+1}),\,\,\,\,\,\{\log(P), \alpha_j\}_{AL}
 = i\alpha_j\eqno(2.13)$$
for all $0\leq j\leq p-1.$   Hence the periodic defocusing AL equation is generated by the Hamiltonian
$\hbox{Re}(K_1) - 2\log(P).$

To relate the $K_n$'s to the Floquet CMV matrix, and to the coefficients of $P\cdot \Delta(z),$ first recall
that \c{S2}
$$\hbox{det}(zI -\E(h)) = \left(\prod_{j=0}^{p-1} \rho_j\right) z^{p\over 2}[\Delta(z)
   -(h + h^{-1})].\eqno(2.14)$$
In view of (2.13), it is clear we must consider the structure of the powers of $\Cal E.$   We will
skip the proof of the following result which can be established by induction on $n.$

\proclaim
{Lemma 2.2}  For $n\geq 1,$  $(\Cal E^n)_{j,k}$ is identically zero if one of the following holds:
\smallskip
\noindent (a) $|j-k| \geq 2n + 1$, or
\newline
(b) $j-k = 2n,$ where $j$ and $k$ are both even, or
\newline
(c) $j-k = -2n$ where $j$ and $k$ are both odd.
\endproclaim
As a consequence of this result, note that for $n\leq \frac{p}{2}-1$ and for $0\leq k\leq p-1,$ we have
$$(\Cal E(h)^n)_{kk} = (\Cal E^n\delta_k)_k = \sum_{q\in \Bbb Z} h^{-q} (\Cal E^n)_{k,qp+k} = (\Cal E^n)_{kk}
\eqno(2.15)$$
since $|k-(pq+k)|\geq p\geq 2n+2$ for all $q\neq 0.$   Hence it follows that (cf. (5.4))
$$K_n = \frac{1}{n} \oint_{|h|=1} \hbox{tr}(\Cal E(h)^n) \frac{dh}{2\pi i h},\,\,\, n=1,\cdots, p/2 -1.\eqno(2.16)$$
From the relation in (2.14), we see that the integral in (2.16) for $n =p/2$ are also relevant, as this is related to 
$I_0$ and $P.$    But by an induction argument similar to the proof of Lemma 2.2 above, we can
show that
$$(\Cal E^m)_{k,k+2m} = \prod_{j=k}^{k+2m-1} \rho_j\quad \hbox{if}\,\, k \,\,\hbox{is even}\eqno(2.17)$$
and 
$$(\Cal E^m)_{k,k-2m} = \prod_{j=k-2m}^{k-1} \rho_j \quad \hbox{if}\,\, k \,\, \hbox{is odd}.\eqno(2.18)$$
Since $P = \prod_{j=k}^{k+p-1} \rho_j$ for any $k\in \Bbb Z,$ it follows that
$$(\Cal E(h)^{p/2})_{kk} = \cases (\Cal E^{p/2})_{kk} + h^{-1}\cdot P & \quad \text{if}\,\, k \,\,\text{is even}\\
                                                      (\Cal E^{p/2})_{kk} + h\cdot P  & \quad \text{if} \,\, k \,\,\text{is odd}
                                                      \endcases \eqno(2.19)$$
 and therefore
 $$\hbox{tr}(\Cal E(h)^{p/2}) = \hbox{tr}(\Cal E^{p/2}) + \frac{p}{2} (h + h^{-1})P.\eqno(2.20)$$
 Thus it follows from (2.20) and (2.15) that
 $$K_{p/2} = \frac{2}{p} \oint_{|h|=1} \hbox{tr}(\Cal E(h)^{p/2}) \frac{dh}{2\pi i h}\eqno(2.21)$$
 and so the set of Hamiltonians in Theorem 2.1 is equivalent to
 $P$, $\hbox{Re}\, K_j,$ $\hbox{Im}\,K_j,$ $j=1,\cdots, p/2-1,$ $K_{p/2}.$
 
 The next result gives the Lax equations of the Hamiltonian systems generated by the above set of functions
 and is the central result of \c{N}.   We will use the following notation: for an infinite two-sided matrix $\Cal A,$
 $\Pi_{u} (\Cal A) = \frac{1}{2} \Cal A_{0} + \Cal A_{+},$ where $\Cal A_{+}$ is the upper triangular part of 
 $\Cal A$ and $\Cal A_0$ is the diagonal part.
 
\proclaim
{Theorem 2.3}   For $1\leq n \leq p/2,$ 
\newline
(a) the Hamiltonian equation  generated by $\hbox{Re}\, K_n$  can be expressed as
$$\dot \Cal E = [\,\Cal E, i \Pi_{u} (\Cal E^n) + i ((\Pi_{u}(\Cal E^n))^*\,],\eqno(2.22)$$
\newline
(b) the Hamiltonian equation generated by $\hbox{Im} \,K_n$  can be expressed as
$$\dot \Cal E = [\, \Cal E,  \Pi_{u} (\Cal E^n) - ((\Pi_{u}(\Cal E^n))^*\,].\eqno(2.23)$$
\endproclaim

Our next goal is to compute the evolution of $\Cal E(h)$ under (2.22) and (2.23).   In order to do this,
we have to establish the following result.

\proclaim
{Proposition 2.4 } Let $n\leq p/2-1,$ then the structure of
$\Cal E(h)^n$ as a Laurent polynomial in $h$ is given by
$$\Cal E(h)^n=\Cal{A}_0(n)+ h\Cal{A}_1(n)+ h^{-1}\Cal{A}_{-1}(n),\eqno(2.24 )$$
where $\Cal{A}_1(n)$ is strictly upper triangular and $\Cal{A}_{-1}(n)$ is
strictly lower triangular. 
\endproclaim

\demo
{Proof}  In order to prove (2.24), let $0\leq j,k\le p-1$ be two indices. Then as in (2.15), we have
$$\bigl(\Cal E(h)^n\bigr)_{jk}=\sum_{q\in\Bbb Z} h^{-q}\bigl(\E^n\bigr)_{j,k+pq}\,.
\eqno(2.25)$$
Now, note that for $|q|\ge2$ and any $j,k$,  we have the inequality
$$\bigl|j-(k+pq)\bigr|\geq 2p-|j-k|\geq 2p-(p-1)=p+1\geq 2n+1$$
and therefore $\bigl(\Cal E^n\bigr)_{j,k+pq}\equiv 0$. 
Hence formula (2.24) holds for some matrices $\Cal A_{\pm1}(n)$. To find the structure of the 
matrices $\Cal A_{\pm1}(n)$, note that
for $j\geq k,$  we have 
$\bigl|j-(k-p)\bigr|=p+(j-k)\ge p\ge 2n+1$
and so it follows from Lemma 2.2 that $\bigl(\Cal E^n\bigr)_{j,k-p}\equiv 0$.  Consequently, 
$$\bigl(\Cal E(h)^n\bigr)_{jk}=\bigl(\Cal E^n\bigr)_{jk}+h^{-1}\bigl(\Cal E^n\bigr)_{j,k+p}\quad \hbox{for}\,\, 
j\geq k. \eqno(2.26)$$
A similar argument shows that 
$$\bigl(\E(h)^n\bigr)_{jk}=\bigl(\E^n\bigr)_{jk}+h\bigl(\E^n\bigr)_{j,k-p}\quad \hbox{for}\,\,
j\leq k.\eqno(2.27)$$
These last two equations then establish our claim about the triangularity of $\Cal A_{\pm1}(n)$. In fact, we find that
$$\bigl(\Cal A_0(n)\bigr)_{jk}=\bigl(\E^n\bigr)_{jk}\qquad\text{for all}\,\,\, 0\le j,k\le p-1\,,\eqno(2.28)$$
$$\bigl(\Cal A_1(n)\bigr)_{jk}= \cases
\bigl(\Cal E^n\bigr)_{j,k-p} &\quad \text{if}\,\,\, j<k;\\
0 &\quad \text{if}\,\,\, j\geq k\,,
\endcases\eqno(2.29)$$
and
$$\bigl(\Cal A_{-1}(n)\bigr)_{jk}= \cases
\bigl(\E^n\bigr)_{j,k+p} &\quad \text{if}\,\,\, j>k;\\
0 &\quad \text{if}\,\,\, j\le k\,.
\endcases\eqno(2.30)$$
This completes the proof.
\pf
\enddemo 

We are now ready to give the result alluded to above which is the point of departure in this work.  We will
make use of the projections $\Pi_{\frak k}$ and  $\Pi_{\widetilde{\frak k}_{w}}$ introduced in Section 3
below. (See the second paragraph of Section 3 and (3.12).)

\proclaim
{Proposition 2.5}   
(a) If $\Cal E$ evolves according to (2.22), then 
$$\dot \Cal E(h) = [\,\Cal E(h), \Pi_{\widetilde{\frak k}_{w}} (i\Cal E(h)^n)\,].\eqno(2.31)$$
\newline
(b)  If $\Cal E$ evolves according to (2.23), then
$$\dot \Cal E(h) =  [\,\Cal E(h), \Pi_{\widetilde{\frak k}_{w}} (\Cal E(h)^n)\,].\eqno(2.32)$$
\endproclaim

\demo
{Proof}  Let $Q_n(h)$ be the matrix of $\Pi_{u}(\Cal E^n)\mid X_{(h)}$ with respect to the ordered
basis $(\delta_0,\cdots, \delta_{p-1}).$  To establish (2.31) and (2.32), it suffices to prove that
$$Q_n(h) = \frac{1}{2}(\Cal{A}_0(n))_{0} + (\Cal{A}_0(n))_{+} + h^{-1}\Cal{A}_{-1}(n),\eqno(2.33)$$
where $\Cal{A}_0(n)$ and $\Cal{A}_{-1}(n)$ are the matrices in (2.24), and where $(\Cal{A}_0(n))_{+}$
and $(\Cal{A}_0(n))_{0}$ are respectively the upper triangular part and diagonal part of $\Cal{A}_0(n).$
But by a direct calculation and making use of Lemma 2.2, we find
$$(\Cal E^n)_{+}\delta_k = h^{-1} \sum_{j=0}^{p-1} (\Cal E^n)_{j-p,k} \delta_j + \sum_{j=0}^{k-1} (\Cal E^n)_{jk}
\delta_j.\eqno(2.34)$$
Hence from (2.28) and (2.30), we conclude that the matrix of $(\Cal E^n)_{+}\mid X_{(h)}$  with respect
to the ordered basis $(\delta_0,\cdots, \delta_{p-1})$ is given by 
$h^{-1} \Cal{A}_{-1}(n)  + \Cal A_0(n)_{+}.$   On the other hand, it is easy to see that the matrix
of $(\Cal E^n)_{0}\mid X_{(h)}$ with respect to the same basis is $(\Cal{A}_0(n))_{0}.$   Hence
(2.33) follows.
\pf
\enddemo

We close this section with an important remark about the periodic defocusing AL equation itself.
Namely, if  $\dot \alpha_j  = i\rho_j^2(\alpha_{j-1} + \alpha_{j+1}),$  or equivalently,
$\dot \Cal E(h) =  [\,\Cal E(h), \Pi_{\widetilde{\frak k}_{w}} (i\Cal E(h))\,],$ then
$\beta_j(t) = e^{-2it} \alpha_j(t)$ satisfies the periodic defocusing AL equation and vice versa.
Thus in order to solve the periodic defocusing AL equation, it suffices to solve
$\dot \Cal E(h) =  [\,\Cal E(h), \Pi_{\widetilde{\frak k}_{w}} (i\Cal E(h))\,],$
which is generated by the Hamiltonian  $\,\hbox{Re}\, K_1 = -\,\hbox{Re}\, I_{p/2-1}(\Cal E(h)).$
(Compare (2.16) and (5.4).)   We will solve the Hamiltonian equations generated by the
conserved quantities in Theorem 2.1 in Section 6 below.

\bigskip

\subhead
3. \ Floquet CMV matrices, dressing orbits and Hamiltonian flows
\endsubhead

\bigskip

The first goal of this section is to show that the set of Floquet CMV 
matrices is a symplectic leaf of a Poisson Lie group whose underlying 
group is a loop group.  Indeed, by following the method of investigation
in \c{L1}, we will show that there exist symplectic leaves
of a more elementary nature in terms of which we can describe the 
collection of Floquet CMV matrices.  

As explained in Section 2 above, we can assume that $p$ is even.
Let $G^{\IR}$ be $GL(p,\IC)$ considered as a real Lie group, and let
$K$ and $B$ be respectively the unitary group $U(p)$ and the lower
triangular group of $p\times p$ matrices with positive diagonal entries.  
It is well-known that $G^{\IR}$  admits the 
Iwasawa decomposition $G^{\IR} = KB.$  On the Lie algebra level,
this corresponds to $\fg^{\IR} = \fk \oplus \fb$ (with associated
projections $\Pi_{\fk}$,  $\Pi_{\fb}$), where $\fg^{\IR}$, $\fk$ and
$\fb$ are, respectively, the Lie algebras of $G^{\IR}$, $K$ and $B$.
Later on in the section, we will also need the maximal torus $T$
of $K,$ consisting of unitary diagonal matrices.

Let $\tilde{G}^{\IR} = C^{\infty}(S^1, G^{\IR})$ be the smooth loop
group with the $C^{\infty}$ topology. $\tilde{G}^{\IR}$ is a Fr\"echet
Lie group with the Lie algebra $\tilde{\fg}^{\IR} =
C^{\infty}(S^1,\fg^{\IR}).$  We will use the following nondegenerate
ad-invariant pairing on $\tilde{\fg}^{\IR}$:
$$
(X,Y) = Im \oint_{|h|=1} \hbox{tr} (X(h)Y(h)) \frac{dh}{2\pi i h}.
\eqno(3.1)
$$
As the reader will see, this choice is critical for what we have
in mind.

Following \cite{GW}, choose a symmetric weight function
$w:\Bbb{Z}\longrightarrow \IR_{+}$, which is rapidly increasing in the
sense that
$$\lim_{n\to \infty} w(n) n^{-s} = \infty, \quad \forall s > 0.
\eqno(3.2)
$$
Also, assume that $w$ is of non-analytic type:
$$\lim_{n\to \infty} w(n)^{1/n} = 1.
\eqno(3.3)
$$
For $X\in {\tilde{\fg}}^{\IR}$ given by $X(h) =
\sum_{j=-\infty}^{\infty} X_{j} h^{j}$,    we define
$$
\|X\|_{w} = \sum_{j=-\infty}^{\infty} \|X_j\|w(j),
\eqno(3.4)
$$
where
$\|\cdot\|$ is a norm on $\fg^{\IR}$.  Also, set
$$
(P_{+}X)(h)=\sum_{j>0} X_{j}h^{j},\quad (P_{-}X)(h)= \sum_{j<0} X_{j}h^{j},
\quad (P_{0}X)(h) = X_{0}.
\eqno(3.5)
$$

Consider the Banach Lie group
$$\Gtw =\{\, g\in \tilde{G}^{\IR}\mid \|g\|_{w}<\infty \,\}
\eqno(3.6)
$$
with Lie algebra
$$\gtw =\{\, X\in {\tilde{\fg}}^{\IR} \mid \|X\|_{w} <\infty \,\}.
\eqno(3.7)
$$
From \cite{GW}, we have the Iwasawa decomposition for the loop group
$\Gtw$ and its Lie algebra
$$\Gtw = \Ktw \cdot \Btw, \quad \gtw = \ktw \oplus \btw
\eqno(3.8)
$$
where
$$\Ktw =\{\,g\in \Gtw\mid g^{*}g = I\,\},\quad
     \Btw =\{\,g\in \Gtw\mid P_{-} g = 0,\,P_{0} g \in B\,\}
\eqno(3.9)
$$
are Banach Lie subgroups of $\Gtw$ and $\ktw$, $\btw$ are their
respective Lie algebras.  Denote by $\bk:\Gtw\longrightarrow \Ktw$,
$\bb:\Gtw\longrightarrow \Btw$ the analytic maps defined by the
factorization $g = \bk(g)\bb(g)^{-1},\, g\in \Gtw.$  Also, denote by
$\piktw$ and $\pibtw$ the projection maps relative to the splitting
$\gtw = \ktw \oplus \btw$.  Then from standard classical r-matrix theory
\cite{STS1}, \cite{STS2},
$$J^{\sharp} = \piktw -\pibtw
\eqno(3.10)
$$
is a solution of the modified Yang-Baxter equation (mYBE).  Hence we
can equip $\gtw$ with the $J^{\sharp}$-bracket
$$[X,Y]_{J^{\sharp}} = \frac {1}{2}([J^{\sharp}X,Y] + [X,J^{\sharp}Y]).
\eqno(3.11)
$$
We will denote the vector space $\gtw$ equipped with the Lie bracket
$[\cdot, \cdot]_{J^{\sharp}}$
by $(\gtw)_{J^{\sharp}}.$  In fact, it is easy to check from (3.11)
that $(\gtw)_{J^{\sharp}} = \ktw \ominus \btw$ (Lie algebra antidirect sum).
Note that explicitly, the projection maps 
$\piktw$, $\pibtw$ are given by the
formulas
$$\aligned
      &\piktw X = P_{-}X + \Pi_{\fk} X_{0} -(P_{-}X)^{*}\\
      &\pibtw X = P_{+}X + \Pi_{\fb} X_{0} +(P_{-}X)^{*},
\endaligned
\eqno(3.12)
$$
where  $P_{\pm}$ are defined in (3.5).  

In order to introduce 
the Poisson structure on $\Gtw,$ it is necessary to restrict ourselves
to a subclass of functions in $C^{\infty}(\Gtw).$  We say that a
function $\varphi\in C^{\infty}(\Gtw)$ is {\sl smooth at $g\in \Gtw$} iff 
there exists $D\varphi(g)\in \gtw$ (called the right gradient
of $\varphi$ at $g$) such that 
$${d\over dt}{\Big|_{t=0}} \varphi(e^{tX}g) = 
(D\varphi(g), X),\quad X\in \gtw \eqno(3.13)$$
where $(\cdot, \cdot)$ is the pairing in (3.1).
If $\varphi\in C^{\infty}(\Gtw)$ is smooth at $g$ for all 
$g\in \Gtw,$  then we say it is {\sl smooth on} $\Gtw.$ 
Note that the nondegeneracy of $(\cdot, \cdot)$ implies that 
the map
$$i:\gtw\longrightarrow (\gtw)^{*},\quad X\mapsto (X,\cdot)\eqno(3.14)$$ 
is an isomorphism 
onto a subspace of $(\gtw)^{*}$ which we will call
the smooth part of $(\gtw)^{*}.$
Thus $\varphi\in C^{\infty}(\Gtw)$ is smooth at $g$
iff $T_{e}^{*}r_{g} d\varphi(g)$ is in
the smooth part of $(\gtw)^{*}$ and we can
define the left gradient of such a function at $g$ by 
$${d\over dt}{\Big|_{t=0}} \varphi(ge^{tX}) = 
(D' \varphi(g), X),\quad X\in \gtw .\eqno(3.15)$$
For each $g\in\Gtw,$ we will denote the collection of all 
smooth functions at $g$ by ${\Cal F}_{g}(\Gtw)$ and we set
${\Cal F}(\Gtw) =\cap_{g\in \Gtw} {\Cal F}_{g}(\Gtw).$  
With the above considerations, it is easy
to check that ${\Cal F}(\Gtw)$ is non-empty and forms an algebra under ordinary
multiplication of functions. 

\proclaim 
{Proposition 3.1} (a) For $\varphi, \psi\in {\Cal F}(\Gtw)$ and $g\in \Gtw,$
define
$$\{\varphi, \psi\}_{J^{\sharp}}(g) ={1\over 2} (J^{\sharp}(D'\varphi(g)), 
D'\psi(g)) - {1\over 2}(J^{\sharp}(D\varphi(g)), D\psi(g)).
\eqno(3.16)$$
Then $\{\varphi, \psi\}_{J^{\sharp}}\in {\Cal F}(\Gtw)$ and hence
$\{\cdot,\cdot\}_{J^{\sharp}}$ defines a Poisson bracket on
${\Cal F}(\Gtw).$
\newline
(b) The Hamiltonian equation of motion generated by 
$\varphi\in {\Cal F}(\Gtw)$ is given by
$$\aligned
\dot g &= g\, (\piktw(D'\varphi(g)))-(\piktw(D\varphi(g)))\,g\\
       &= (\pibtw(D\varphi(g)))\,g - g\,(\pibtw(D'\varphi(g))).\\
\endaligned                                        
\eqno(3.17)$$
\endproclaim

\demo
{Proof} (a) A straight forward  calculation shows
that for $\varphi,\psi\in {\Cal F}(\Gtw),$ $D\{\varphi,\psi\}_{J^{\sharp}}(g)$
exists for each $g\in \Gtw$ and is given by
$$\eqalign{
D\{\varphi,\psi\}_{J^{\sharp}}(g) = & Ad_{g}[D'\varphi(g), D'\psi(g)]_{J^{\sharp}} 
+ Ad_{g} \frac{d}{ds} \mid_{s=0} D'\varphi(e^{s\eta(g)D\varphi(g)} g)\cr
& - Ad_{g}\frac{d}{ds} \mid_{s=0}  D'\psi(e^{s\eta(g)D\varphi(g)} g),\cr}
\eqno(3.18)$$
where
$$\eta(g) = {1\over 2} Ad_{g}\circ J^{\sharp}\circ Ad_{g^{-1}}
            -{1\over 2} J^{\sharp}.\eqno(3.19)$$
This shows $\{\varphi, \psi\}_{J^{\sharp}}\in {\Cal F}(\Gtw).$
To prove the second half of (a), first note that
$\hbox{Tr}(X(h)Y(h))\in \IR$ for $X,Y\in \ktw.$
From this, it follows that $(X,Y)=0.$  Consequently, 
$\ktw$ is an isotropic subalgebra of $\gtw$ relative to
the pairing $(\cdot, \cdot).$    On the other hand, if
$X,Y\in \btw,$ we have
$$\oint_{|h|=1} \hbox{tr}(X(h)Y(h)){dh\over 2\pi ih}
=\hbox{tr}(X_0Y_0)\in \IR$$
because $X_0$, $Y_0$ are lower triangular with real diagonal
entries.  So $\btw$ is also
an isotropic subalgebra of $\gtw.$  Combining these two
facts, we can now conclude that $J^{\sharp}$ is skew-symmetric
relative to $(\cdot,\cdot).$  
Finally, the Jacobi identity and the derivation property 
now follow from
standard calculations in \c{STS2} which work without change in
our infinite dimensional context.
\newline
(b) This derivation of the Lax equation from the Poisson
structure is standard.

\pf
\enddemo

Note that although we are dealing with a restricted class of functions,
the notion of Poisson submanifolds can be defined analogously to the
standard case.
Now it is easy to check that if $\varphi\in {\Cal F}(\Gtw),$ then
so are $\varphi\circ r_{g}$ and $\varphi\circ l_{g}$ for all 
$g\in \Gtw,$ where $r_g$ and $l_g$ denote right 
and left translation by $g,$ respectively. Hence the
notion of Poisson Lie group can be extended to this infinite
dimensional context and $(\Gtw, \{\cdot,\cdot\}_{J^{\sharp}})$ is a 
coboundary Poisson Lie group.  On the infinitesimal level, we will call
$(\gtw, (\gtw)_{J^{\sharp}})$ the tangent Lie
bialgebra of  $(\Gtw, \{\cdot,\cdot\}_{J^{\sharp}})$ as 
the map $\rho:\gtw\longrightarrow L(\gtw,\gtw)$ ($L(\gtw,\gtw)$ is
the space of linear maps on $\gtw$) given
by $\rho(X)= {1\over 2}(ad_{X}\circ J^{\sharp} - J^{\sharp}\circ ad_{X})$
and satisfying the relation
$([Y,Z]_{J^{\sharp}}, X) = (Z, \rho(X)(Y))$ 
is a $1$-coboundary (and hence a $1$-cocycle)
with respect to the adjoint representation.  Thus
in speaking of a Lie bialgebra here, the underlying vector spaces 
of the pair of Lie algebras involved are only required
to be in duality with respect to $(\cdot,\cdot)$ and this
is what we will continue to do.
\proclaim
{Corollary 3.2} (a) $\Ktw$ is a Poisson Lie subgroup of 
$(\Gtw, \{\cdot, \cdot\}_{J^{\sharp}})$ in the sense that
$\Ktw$ is a Lie subgroup of $\Gtw$ which is also a Poisson
submanifold of $(\Gtw, \{\cdot, \cdot\}_{J^{\sharp}}).$
Moreover, the tangent Lie bialgebra $(\ktw, (\gtw)_{J^{\sharp}}/\ktw^{\perp})$
of $\Ktw$ (where $\ktw^{\perp}$ is defined relative to $(\cdot,\cdot)$)
is isomorphic to
$(\ktw, \btw^{-})$ where $\btw^{-}$ is $\btw$ but
equipped with the $-$ bracket. 
\newline
(b) The underlying group of the Poisson group $(\Gtw)_{J^{\sharp}}$ dual to
 $(\Gtw, \{\cdot,\cdot\}_{J^{\sharp}})$ consists of $\Gtw$ equipped
with the multiplication
$$g\ast h \equiv \bk(g)h\bb(g)^{-1}.\eqno(3.20)$$

\endproclaim
 
\demo
{Proof} (a) To show that $\Ktw$
is a Poisson Lie subgroup, it is enough to check that
$\Ktw$ is a Poisson submanifold of  $(\Gtw, \{\cdot,\cdot \}_{J^{\sharp}})$
and this can be done by using the expression for the Hamiltonian
vector field in Proposition 3.1(b).  To show that the tangent
Lie bialgebra $(\ktw, (\gtw)_{J^{\sharp}}/\ktw^{\perp})$ of $\Ktw$ is isomorphic
to $(\ktw, \btw^{-}),$ first note that $\ktw^{\perp} = \ktw.$
Hence we have $X + \ktw^{\perp} = \Pi_{\btw} X + \ktw$ for all
$X\in (\fg^{\Bbb R}_{w})_{J^{\sharp}}.$  Therefore, the induced
Lie bracket on $(\gtw)_{J^{\sharp}}/\ktw^{\perp}$ is given by
$$\align
[X + \ktw, Y + \ktw] & = [\Pi_{\btw} X, \Pi_{\btw} Y]_{J^{\sharp}} + \ktw\\
                     & = - [\Pi_{\btw} X, \Pi_{\btw} Y] + \ktw.\\
\endalign
$$
Conseqently, the map $X + \ktw \mapsto \Pi_{\btw} X$ is an isomorphism,
when $\btw$ is equipped with the $-$ bracket.
\newline
(b) The formula is a consequence of the fact that $\gtw = \ktw\ominus \btw$
and can be verified easily.
\pf
\enddemo

\remark
{Remark \rom{3.3}}  In view of the second half of Corollary 3.2 (a), the 
induced structure on $\Ktw$ is the loop group analog of the Bruhat
Poisson structure in \c{LW} and \c{Soi}.
\endremark
\smallskip

We next turn to the description of the symplectic leaves of the
Sklyanin structure in (3.16).  Unfortunately, we cannot assume
the general results in \c{STS2} and \c{LW} apply to our case without some
verification, because the analysis in these works is for finite 
dimensional Poisson Lie groups.
In this regard, let us also remark that as far as we know,
the integrability of the characteristic distribution of
a Poisson structure on an infinite dimensional manifold
is by no means automatic because an analog of the 
Stefan-Sussmann result \c{St, Su} is not available.   In the
following, we will check things by hand.  So let us define
$$S_{x} = \{\,X_{\varphi}(x)\mid \varphi\in {\Cal F}_{x}(\Gtw)\,\}\eqno(3.21)$$
for each $x\in \Gtw.$  Then the characteristic distribution of the
Poisson bracket $\{\cdot,\cdot\}_{J^{\sharp}}$ is given by
$S = \cup_{x\in\Gtw} S_{x}.$   On the other hand, if 
$(\Gtw)^{o}_{J^{\sharp}}$  denotes the identity component
of $(\Gtw)_{J^{\sharp}}$, 
the right dressing action \c{STS2} of $(\Gtw)^{o}_{J^{\sharp}}$ on
$\Gtw$ is defined by the formula
$$\eqalign{
\Phi_{g}(x) &=\bk(g)^{-1}x\bk(x^{-1}gx)\,\cr
           &=\bb(g)^{-1}x\bb(x^{-1}gx)\,\cr}
\eqno(3.22)$$
and the infinitesimal generator of this action corresponding to 
$\xi\in \gtw$ is the vector field $\xi_{\Gtw}$ on $\Gtw,$ where
$$\xi_{\Gtw}(x) = {1\over 2} xJ^{\sharp}(x^{-1}\xi x) -{1\over 2}
J^{\sharp}(\xi)x.\eqno(3.23)$$  
For each $x\in \Gtw,$ let $F_{x}$ be the subspace of $T_{x}\Gtw$
spanned by the vectors $\xi_{\Gtw}(x).$   Then 
$F = \cup_{x\in\Gtw} F_{x}$ is an integrable generalized
distribution whose leaves are the orbits of the
dressing action $\Phi.$  

\proclaim
{Proposition 3.4}  For each $x\in \Gtw,$
$T_{x} {\Cal O} = S_{x},$ where ${\Cal O}$ is the orbit of the dressing
action containing $x.$   Hence the characteristic distribution
$S$ is integrable and the leaves of this distribution are given
by the orbits of the dressing action $\Phi.$

\endproclaim

\demo
{Proof} From (3.7),
$$X_{\varphi}(x) = {1\over 2} x J^{\sharp}(x^{-1} D\varphi(x) x)-{1\over 2} 
      J^{\sharp}(D\varphi(x))x.\eqno(3.24)$$
To show that $S_{x} =  F_{x},$ it suffices to show that for each $\xi\in \gtw,$
there exists $\varphi\in {\Cal F}_{x}(\gtw)$ such that $D\varphi(x) =\xi.$
To do so, we use the fact that the exponential map 
$\exp: \gtw \longrightarrow \Gtw$ is a diffeomorphism of a
neighborhood $U$ of $0$ onto a neighborhood $V$ of the identity
element  of $\Gtw$ \c{GW}. Clearly, $Vx$ is a neighborhood
of $x$ and in this neighborhood, define 
$\bar\varphi_{\xi}(g)= (\xi, \log (g x^{-1}))$
and extend this to a function $\varphi_{\xi}\in C^{\infty}(\Gtw).$
Then $\varphi_{\xi}\in {\Cal F}_{x}(\Gtw)$ with
$D\varphi_{\xi}(x) = \xi.$  This completes the proof.
\pf
\enddemo

Let ${\Cal O}$ be a dressing orbit,
as in the proposition above. We next show that ${\Cal O}$ is 
a symplectic leaf of $(\Gtw, \{\cdot,\cdot\}_{J^{\sharp}}),$ i.e.,
there exists a weak (resp.~ strong) symplectic form (see, for example, \c{OR}
for such matters)
$\omega_{\Cal O}$ on $\Cal O$ consistent with the Poisson
bracket $\{\cdot,\cdot\}_{J^{\sharp}}$ in the case when ${\Cal O}$
is infinite (resp.~ finite) dimensional.  For this purpose,
let $(\gtw)^{*}_{S}$ denote the smooth part of $(\gtw)^{*}$
and let $(T^{*}\Gtw)_{S}= \cup_{g\in \Gtw} T^{*}_{g} r_{g^{-1}} (\gtw)^{*}_{S}.$
Also, denote by $\pi^{\sharp}: (T^{*}\Gtw)_{S}\longrightarrow T\Gtw$
the bundle map corresponding to $\{\cdot,\cdot\}_{J^{\sharp}}$ and
let $j$ be the left inverse of $i:\gtw\longrightarrow (\gtw)^{*}.$
For each $x\in {\Cal O}$, we define a skew-symmetric bilinear form
$\omega_{x}$ on $S_{x}$ by the formula
$$\eqalign{
& \omega_{x}(\pi^{\sharp}(x)(\alpha),\pi^{\sharp}(x)(\beta))\cr
= & \,(T_{x}r_{x^{-1}} \pi^{\sharp}(x)(\alpha), j(T^{*}_{e}r_{x}\beta))\cr
= & - (j(T^{*}_{e}r_{x}\alpha),T_{x}r_{x^{-1}} \pi^{\sharp}(x)(\beta)).\cr}
\eqno(3.25)
$$
Clearly, the value of the above expression depends only on the
values of 
$\pi^{\sharp}(x)(\alpha)$ and
$\pi^{\sharp}(x)(\beta).$
Thus $\omega_{x}$ is a well-defined skew-symmetric bilinear form
on $S_{x}$.
Now suppose $ \omega_{x}(\pi^{\sharp}(x)(\alpha),\pi^{\sharp}(x)(\beta))=0$
for all $\pi^{\sharp}(x)(\beta).$  Then from (3.25) and the nondegeneracy
of $(\cdot,\cdot)$, it follows that $j(T^{*}_{e}r_{x}\alpha)=0$
which in term implies $\pi^{\sharp}(x)(\alpha)=0.$  So this 
establishes the nondegeneracy of $\omega_{x}.$  Thus there
exists a $2$-form $\omega_{{\Cal O}}$ on ${\Cal O}$ such that
$\omega_{{\Cal O}}(x) = \omega_{x}$ for each $x\in {\Cal O}.$
Now the argument that $\omega_{{\Cal O}}$ is differentiable
and closed follows as in the finite dimensional case in
\c{Ko}.  Conseqently, $\omega_{{\Cal O}}$ defines a Poisson
structure $\{\cdot,\cdot\}_{\Cal O}$ on ${\Cal O}$ and we have
$$\{\varphi, \psi\}_{J^{\sharp}} \mid {\Cal O} = 
\{\varphi\mid {\Cal O}, \psi\mid {\Cal O}\}_{\Cal O}\eqno(3.26)
$$
for all $\varphi,\psi\in {\Cal F}(\Gtw).$
Hence we have established the following result.

\proclaim
{Proposition 3.5} The symplectic leaves of $(\Gtw, \{\cdot,\cdot\}_{J^{\sharp}})$
are given by the orbits of the dressing action $\Phi.$
\endproclaim

In our next step, we will make the connection between certain symplectic 
leaves of 
$(\Gtw, \{\cdot,\cdot \}_{J^{\sharp}})$ (these are also
those of the Poisson Lie subgroup $\Ktw$) and the Floquet CMV matrices.  
We begin
with some notations. As in \c{L1}, we will denote by
$g^e$ any $p\times p$ block diagonal matrices with
 $2\times2$ diagonal blocks of the form
$$\pmatrix
\bar\alpha & \rho\\ \rho & -\alpha
\endpmatrix,
\qquad \hbox{with}\quad \alpha\in\ID\quad\hbox{and}\quad\rho=
\sqrt{1-|\alpha|^2}.
\eqno(3.27)
$$
We will denote the collection of such matrices by $\Cal{T}^e$.
On the other hand, we will denote by $g^{o}(h)$ any loops
in $\Ktw$ of the form
$$
\pmatrix
    \bar\alpha & 0 & \cdots & 0 & \rho h \\
    0 &   &   &   & 0 \\
    \vdots &  & \tilde b &  & \vdots \\
    0 &   &   &   & 0 \\
    \rho h^{-1} & 0 & \cdots & 0 & -\alpha \\
\endpmatrix,
\eqno(3.28)
$$
where $\alpha\in \ID,$ $\rho=\sqrt{1-|\alpha|^2}$ and $\tilde b$ is
a $(p-2)\times (p-2)$ block diagonal matrix with $2\times 2$ blocks 
of the same kind as in $g^e.$   We will denote the collection 
of such unitary loops by $\Cal{T}^0.$
Clearly, for given $g^e\in \Cal{T}^e$ and $g^o\in \Cal{T}^o$, the
product $g^eg^o(h)$ is a Floquet CMV matrix.  Indeed,
the map
$$\eqalign{
   m\mid {\Cal T}^{e}\times {\Cal T}^{o}:&{\Cal T}^{e}\times {\Cal T}^{o}
   \longrightarrow \{p\times p\,\,\hbox{Floquet CMV}\,\,\hbox{matrices}\}\cr
    &(g^{e},g^{o})\mapsto g^{e}g^{o}\cr}\eqno(3.29)$$
is a diffeomorphism, where $m:\Ktw\times \Ktw\longrightarrow \Ktw$
is the multiplication map of the Poisson Lie subgroup $\Ktw$.
Finally, we will denote the dressing orbit through $x\in \Gtw$
by ${\Cal O}_{x}.$

In analogy to formula (2.21) in \c{L1}, we
introduce the following special Floquet CMV matrix
$$x_{f}(h) = x^{e}_{f} x^{o}_{f}(h)
\eqno(3.30)
$$
corresponding to
$${\underline\alpha} = (0,0,\cdots,0).
\eqno(3.31)
$$
In other words,
$$x^{e}_{f} = \text{diag}(w^*,w^*,\cdots)
\eqno(3.32)
$$
and
$$x^{o}_{f}(h)= \pmatrix
                0& 0 & \dots& 0 & h\\
                0& w^* &\dots &0 & 0\\
                \vdots&\vdots & \ddots &\vdots& \vdots\\
                0& 0 & \dots& w^* & 0\\
                h^{-1}& 0&\dots&0&0\\
                \endpmatrix
\eqno(3.33)
$$
where
$$w^* = \pmatrix
           0 & 1\\
           1 & 0\\
     \endpmatrix.
\eqno(3.34)
$$

These matrices are elements of the affine Weyl group 
$W_{aff} = W \ltimes \Check{T}$ \c{PS}, where $W = N(T)/T$
is the Weyl group of $K,$ and $\Check{T}$ is the lattice
of homomorphisms $S^{1}\longrightarrow T.$  Indeed, if
$\frak h$ is the Cartan subalgebra of $\frak g$ consisting
of diagonal matrices, $\lambda_i-\lambda_j,$ $i\neq j$
are the roots corresponding to the pair $(\frak g, \frak h)$,
then in terms of the simple roots 
$\alpha_i = \lambda_{i}-\lambda_{i+1}$, $i=1,\cdots, p-1$
and the highest root $\theta =\lambda_1 -\lambda_p,$ we have
$$x^{e}_{f} = w_{\alpha_1} w_{\alpha_3}\cdots w_{\alpha_{p-1}},
\eqno(3.35)$$
and 
$$x^{o}_{f}(h) = w_{\alpha_2}w_{\alpha_4}\cdots w_{\alpha_{p-2}}
\exp(h(E_{00}-E_{p-1,p-1}))w_{\theta}.\eqno(3.36)
$$
Here for each $j=1,\cdots, p-1,$
$w_{\alpha_j} = \hbox{diag} (I_{j-1}, w^*,I_{p-j-1})$
is the element in $W$ which corresponds to 
the simple reflection $s_{\alpha_j},$ while 
$$w_{\theta} = \pmatrix  0 & 0 & \cdots & 0 & 1 \\
    0 &   &   &   & 0 \\
    \vdots &  & I_{p-2} &  & \vdots \\
    0 &   &   &   & 0 \\
    1  & 0 & \cdots & 0 & 0 \endpmatrix
\eqno(3.37)
$$
is the element in $W$ which corresponds to the
reflection $s_{\theta}.$  Finally, the element $\exp(h(E_{00}-E_{p-1,p-1}))$
is in $\Check{T},$ where $E_{jj}$ denote the
diagonal matrix with a $1$ in the $(j,j)$ position
and zeros elsewhere.  But now recall that there is an additonal
element  $\alpha_0 =\delta -\theta$ in the simple system
of affine roots in addition to those given by the
extensions of the $\alpha_j$'s, $j=1,\cdots, p-1.$
(see, for example, \c{Mac} for details on affine Lie algebras).  
With $\alpha_0,$ we can 
interpret the product $w_{\alpha_0} =\exp(h(E_{00}-E_{p-1,p-1}))w_{\theta}$ as
corresponding 
to $s_{\alpha_0}.$  Therefore, by (3.35) and (3.36), we conclude
that the CMV matrix
$x_{f}(h)$ introduced above is a Coxeter element of
the affine Weyl group.  With this background, we are
now ready to accomplish our first goal of this section.

\proclaim
{Theorem 3.6} (a)  ${\Cal{O}}_{x^{e}_{f}} = \Ktw \cap \Btw\, x^{e}_{f}\, \Btw
 = {\Cal{T}}^{e}.$
\newline
(b) ${\Cal{O}} _{x^{o}_{f}} = \Ktw \cap \Btw\, x^{o}_{f}\, \Btw
= {\Cal{T}}^{o}.$
\newline
(c) Equip $\Ktw\times\Ktw$ with the product structure, then
${\Cal{O}}_{x^{e}_{f}}\times {\Cal{O}} _{x^{o}_{f}}$ is a
symplectic leaf of $\Ktw\times \Ktw.$   Moreover,
the collection of $p\times p$ Floquet CMV matrices is the image of
${\Cal{O}}_{x^{e}_{f}}\times {\Cal{O}} _{x^{o}_{f}}$ under the Poisson
automorphism  
$m\mid {\Cal{O}}_{x^{e}_{f}}\times {\Cal{O}} _{x^{o}_{f}}: 
{\Cal{O}}_{x^{e}_{f}}\times {\Cal{O}} _{x^{o}_{f}}\longrightarrow
\{p\times p\,\, \hbox{Floquet CMV matrices}\,\,\},$
where $m$ is the multiplication map of $\Ktw.$
Hence $\{p\times p\,\, \hbox{Floquet CMV matrices}\,\,\} ={\Cal O}_{x_{f}}
= \Ktw \cap \Btw\,x_{f}\,\Btw,$
a Coxeter dressing orbit.

\endproclaim

\demo
{Proof}
(a)  Take an arbitrary element
$$\aligned
          a  &=\bk(g)^{-1}x^{e}_{f}\bk((x^{e}_{f})^{-1}gx^{e}_{f})\\
             &=\bb(g)^{-1}x^{e}_{f}\bb((x^{e}_{f})^{-1}gx^{e}_{f})
\endaligned
\eqno(3.38)
$$
in the dressing orbit through $x^{e}_{f}$.  From the first line of
the above expression, it is clear that $a$ is unitary. On the other
hand, it follows from the second line of the above expression that
$P_{-} a = 0$ and that $P_{0} a$ is block lower triangular with
$2\times 2$ blocks on the diagonal.  Moreover, from the fact that
the elements in $B$ have positive diagonal entries, it follows that
each of the $2\times 2$ blocks on the main diagonal of $P_{0} a$ has
the following properties: (i) the entry in the upper right hand
corner is positive, (ii) the determinant is negative (since
$det(w^*) = -1$). But $P_{-} a = 0$ implies $P_{-} a^{-1} =0.$  As
$P_{-} a^{-1} = P_{-} a^* = (P_{+} a)^{*}$, we conclude that $P_{+}
a = 0$  and hence $a = P_{0} a.$ But then $(a^{*})^{-1} = ((P_{0}
a)^{*})^{-1}$ is upper block triangular with diagonal blocks having
the same properties.  Since $a = (a^{*})^{-1}$, it follows that $a$
must be block diagonal, i.e.,
$$a = \text{diag} (\phi_{0},\phi_{2},\cdots, \phi_{p-2})
\eqno(3.39)
$$
where for each $j$, $\phi_{2j}$ is a unitary matrix with a positive
entry in the upper right hand corner and whose determinant is $-1.$
Consequently, $\phi_{2j}$ must be of the form
$$\phi_{2j} = \pmatrix
       \bar \alpha_{2j}& \rho_{2j}\\
       \rho_{2j}& -\alpha_{2j}
       \endpmatrix
\eqno(3.40)
$$
for some $\alpha_{2j}\in \ID$, where $\rho_{2j}
=(1-|\alpha_{2j}|^{2})^{\frac{1}{2}}$. Hence we have shown that
${\Cal O}_{x^{e}_{f}} \subset {\Cal T}^{e}.$ The reverse
inclusion ${\Cal T}^{e}\subset {\Cal O}_{x^{e}_{f}}$ follows
exactly as in the proof of the corresponding assertion in Theorem
2.4 (a) of \cite{L1}.
\newline
\noindent (b)  Take an arbitrary element
$$\aligned
b  &=\bk(g)^{-1}x^{o}_{f}\bk((x^{o}_{f})^{-1}gx^{o}_{f})\\
   &=\bb(g)^{-1}x^{o}_{f}\bb((x^{o}_{f})^{-1}gx^{o}_{f})
\endaligned
\eqno(3.41)
$$
in the dressing orbit through $x^{o}_{f}$. From the first line of
the above expression, $b$ is unitary.  On the other hand, it follows
from the second line of the same expression that
$$b(h) = \sum_{j= -1}^{\infty} b_{j} h^{j} .
\eqno(3.42)
$$
Now,
$$b^{-1} = (\bb((x^{o}_{f})^{-1}gx^{o}_{p}))^{-1}(x^{o}_{f})^{*}\,\bb(g)
\eqno(3.43)
$$
since $x^{o}_{f}$ is unitary.  From this, it follows that
$$(P_{+} (h^{-1}b))^{*}= P_{-}((\bar{h}b)^{*}) = P_{-}(h b^{-1}) =0.
\eqno(3.44)
$$
Therefore, when we combine this with (3.42), we conclude
that
$$b(h) = b_{-1} h^{-1} + b_{0} + b_{1} h.
\eqno(3.45)
$$

Considering the coefficient of $h^{-1}$ in the second line of (3.41), we 
see that
$$
b_{-1}=\left(\bb(g)^{-1}\right)_0 (x^{o}_{f})_{-1} \left(\bb((x^{o}_{f})^{-1}gx^{o}_{p})
\right)_0,
$$
as the first and last factors in the second line of (3.39) only 
contain nonnegative powers of $h$. Since
$(x^{o}_{f})_{-1}$ has only one nonzero entry in its bottom left corner, and it is
multiplied by lower triangular matrices with positive diagonal entries, we 
see that all the entries of
$b_{-1}$ are zero, except $(b_{-1})_{p-1,0}=\gamma>0$.

Further note that, if $x(h)$ is unitary for all $h\in S^1$, then an element
$y(h)\in\Cal{O}_x$ if and only if
$y(h)^*\in\Cal O_{x^*}$. Indeed,
$$
y(h)=\bk(g)^{-1}x\bk(x^{-1}gx) \quad \Longleftrightarrow \quad
y(h)^*=\bk(g^{\prime})^{-1}x^*\bk(x{g^{\prime}} x^*),
$$
where $g^{\prime}=x^{-1}gx$. In our case, $(x_f^o)^*=x_f^o$, and hence
$$
b(h)^*=b_1^* h^{-1}+b_0^*+b_{-1}^* h\,\in\Cal{O}_{x_f^o}.
$$
So we see, by the above argument applied to $b_1^*$, that $b_1$ must have only one
nonzero entry, in its upper right-hand corner: $(b_1)_{0,p-1}=\delta>0$.

It remains to understand $b_0$. Recall that $b(h)$ is unitary and expand the right-hand side of
$1=b(h)^*\cdot b(h)=b(h)\cdot b(h)^*$ in powers of $h$. Given the
structure of the matrices $b_{\pm1}$, it is immediate that the coefficients of
$h^{\pm2}$ are both zero. The coefficients of $h^{\pm1}$ must also be zero. This translates into
$$
b_1^* b_0+b_0^* b_{-1}=0
$$
and
$$
b_{-1} b_0^*+b_0 b_1^*=0.
$$
Taking into account the shape of $b_{\pm1}$, and writing down explicitly these
relations leads to the fact that
$$
0-\hbox{th}\,\, \hbox{row of }\,\,b_0= (\eta,0\ldots,0),\eqno(3.46)
$$
$$
(p-1)-\hbox{th}\,\, \hbox{row of}\,\,b_0= (0,\ldots,0,\nu),\eqno(3.47)
$$
where
$$
\delta \eta+\gamma\bar\nu=0.\eqno(3.48)
$$

Therefore, when we combine our analysis above on $b_{\pm 1},$  together with
(3.46) and (3.47), our conclusion is that $b(h)$ looks like
$$
b(h)=
\pmatrix
    \eta & 0 & \cdots & 0 & \delta h \\
    0 &   &   &   & 0 \\
    \vdots &  & \tilde b &  & \vdots \\
    0 &   &   &   & 0 \\
    \gamma h^{-1} & 0 & \cdots & 0 & \nu \\
\endpmatrix,
\eqno(3.49)
$$
where $\tilde b$ is a $(p-2)\times(p-2)$ matrix which is independent of $h$. 
But a unitary matrix can only be a direct sum of unitary matrices, and 
$b(h)$ is unitary. Unitarity of the
$2\times2$ piece at $h=1$, together with the fact that $\gamma,\delta>0$, 
means that
$$
\pmatrix \eta & \delta h\\ \gamma h^{-1} & \nu
\endpmatrix
=\pmatrix
-\alpha_{p-1} & \rho_{p-1} h \\ \rho_{p-1}h^{-1} & \bar\alpha_{p-1}
\endpmatrix
\eqno(3.50)
$$
for some complex number $\alpha_{p-1}\in\ID$ and $\rho_{p-1}=\sqrt{1-|
\alpha_{p-1}|^2}$.

The last step requires that we understand the shape of $\tilde b$. This is 
achieved by
writing the coefficient of $h^0$ in $b(h)$ and the second line of (3.41). 
Again using the fact that the coefficients of $h^0$ in the external factors 
on the right-hand side of (3.41) are lower triangular with positive diagonal
entries, 
and keeping track of all the (potentially) nonzero entries, we obtain that 
$\tilde b$ is a unitary matrix
which is block-lower triangular, each $2\times2$ diagonal block having a 
positive entry in its upper right-hand corner. Hence the same argument as in 
part (a) of this theorem shows that
$$
\tilde b=\hbox{diag} (\phi_1,\phi_3,\ldots),\eqno(3.51)
$$
where
$$
\phi_k=
\pmatrix
\bar\alpha_k & \rho_k \\ \rho_k & -\alpha_k
\endpmatrix.
\eqno(3.52)$$
Inserting this  into (3.49) and using (3.50) shows that $b\in \Cal{T}^o$.

The proof is finished once we prove that $\Cal{T}^o\subset\Cal{O}_{x_f^o}$. 
This is done
as in the previous cases. More precisely, if
$b=\theta_1 \oplus \theta_3 \oplus \cdots \oplus \theta_{p-1}(h)$ is an element
of $\Cal{T}^o$,
choose a matrix $g=l_1\oplus\cdots\oplus l_{p-3} \oplus l_{p-1}(h)$,
where, for $0\leq j\leq (p-4)/2$,
$$
l_{2j+1}=
\pmatrix
\rho_{2j+1} & 0\\-\alpha_{2j+1} & 1
\endpmatrix
\eqno(3.53)$$
is a $2\times2$ block situated between the rows and columns $2j+1$ and 
$2j+2$, and
$$
l_{p-1}(h)=
\pmatrix
1 & \bar\alpha_{p-1} h\\ 0 & \rho_{p-1}
\endpmatrix.
\eqno(3.54)$$
Note that $l_{2j+1}$ are all independent of $h$ and lower triangular with 
positive diagonal entries. Further
note that
$$
l_{p-1}(h)=
\pmatrix
1 & 0\\ 0 & \rho_{p-1}
\endpmatrix+ h
\pmatrix
0 & \bar\alpha_{p-1}\\ 0 & 0
\endpmatrix.
\eqno(3.55)$$
In particular, this implies that $g\in\tilde B_w$ and hence
$$
\bk(g)^{-1}x^{o}_{f}\bk((x^{o}_{f})^{-1}gx^{o}_{f})=\bk(g x^o_f).
\eqno(3.56)$$

Furthermore, we have the factorizations
$$
l_{2j+1}w^*=\theta_{2j+1}
\pmatrix
\rho_{2j+1} & 0 \\ -\bar\alpha_{2j+1} & 1
\endpmatrix
\eqno(3.57)$$
for $0\leq j\leq (p-4)/2$, and
$$
l_{p-1}(h)
\cdot
\pmatrix
0 & h\\ h^{-1} & 0
\endpmatrix
=
\theta_{p-1}(h)
\cdot
\pmatrix
1 & \alpha_{p-1} h\\ 0 & \rho_{p-1}
\endpmatrix.
\eqno(3.58)$$
In other words $g\cdot x_f^o=b\cdot \tilde g$, where $\tilde g$ is in 
$\Btw$ for the same reason as $g$ is.
We therefore conclude that
$$
\bk(g)^{-1}x^{o}_{f}\bk((x^{o}_{p})^{-1}gx^{o}_{f})=\bk(g x^o_f)=b
\eqno(3.59)$$
is indeed an element of $\Cal{O}_{x^o_f}$.
\newline
\noindent (c) It is easy to see that 
$\Cal{O}_{x_{p}} = \Cal{O}_{x^e_f}\cdot \Cal{O}_{x^o_f}.$
The rest of the assertion is clear from what we have 
already done.
\pf
\enddemo

\remark
{Remark \rom{3.7}}  By essentially following the same argument, we 
can also establish the following fact:
$$\{p\times p\,\, \hbox{Floquet CMV matrices}\,\,\} =\Cal{O}_{w_{\alpha_1}}
\cdots\Cal{O}_{w_{\alpha_{p-1}}}\cdot\Cal{O}_{w_{\alpha_2}}\cdots
\Cal{O}_{w_{\alpha_{p-2}}}\cdot\Cal{O}_{w_{\alpha_{0}}},\eqno(3.60)$$
where each orbit on the right hand side is two-dimensional.
\endremark
\smallskip

In the next result, we clarify the relation between the Ablowitz-Ladik bracket 
in (2.11) and the Sklyanin bracket $\{\cdot,\cdot\}_{J^{\sharp}}.$

\proclaim
{Theorem 3.8} The map 
$$\eqalign{&
\Bbb D^p\longrightarrow \widetilde{G}^{\Bbb R}_{w}, \cr
& \underline{\alpha}=(\alpha_0,\cdots,\alpha_{p-1})\mapsto \Cal{E}(h) =g^e(\underline{\alpha})g^o(\underline{\alpha})(h)\cr}
\eqno(3.61)$$
is a Poisson embedding, when $\Bbb D^p$ is equipped with the Ablowitz-Ladik bracket,
and $\widetilde{G}^{\Bbb R}_{w}$ is equipped with Sklyanin structure $\{\cdot,\cdot\}_{J^{\sharp}}.$

\endproclaim

\demo
{Proof}  As the multiplication map of $\widetilde G^{\Bbb R}_{w}$ is a Poisson map, it is enough
to show that the map $\underline{\alpha}\mapsto (g^e(\underline\alpha), g^o(\underline\alpha))$ is Poisson,
when $\widetilde G^{\Bbb R}_{w}\times \widetilde G^{\Bbb R}_{w}$ is equipped with the product
structure.  For this purpose, denote by $E_{ab}$ the $p\times p$ matrix whose $(a,b)$ entry
is equal to $1$ and whose other entries are zero. For $j\,\, \hbox{even}, j\in \{0,\cdots, p-1\}$
and for  $l\,\, \hbox{odd}, l\in \{0,\cdots, p-1\},$  introduce the following functions on 
$\widetilde{G}^{\Bbb R}_{w} \times \widetilde{G}^{\Bbb R}_{w}$:
$$\eqalign{
& F_j(A,B) =\hbox{Im} \,\oint_{|h|=1}\!\!\!\! \hbox{tr}\,\big(A(h)E_{jj}\big) \,\frac{dh}{2\pi i h},
\,\, G_j(A,B) =\hbox{Re} \,\oint_{|h|=1}\!\!\!\! \hbox{tr}\,\big(A(h)E_{jj}\big) \,\frac{dh}{2\pi i h},\cr
& F_l(A,B) =\hbox{Im} \,\oint_{|h|=1}\!\!\!\! \hbox{tr}\,\big(B(h)E_{ll}\big) \,\frac{dh}{2\pi i h},
\,\, G_l(A,B) =\hbox{Re} \,\oint_{|h|=1}\!\!\!\! \hbox{tr}\, \big(B(h)E_{jj}\big) \,\frac{dh}{2\pi i h}.\cr}\eqno(3.62)$$
Then we have 
$$\eqalign{
& F_j(g^e, g^o) = -\hbox{Im}\, \alpha_j, \,\, G_j(g^e,g^o) = \hbox{Re}\, \alpha_j\cr
& F_l(g^e, g^o) = -\hbox{Im}\, \alpha_l, \,\, G_l(g^e,g^o) = \hbox{Re}\, \alpha_l.\cr}\eqno(3.63)$$
In view of this, it suffices to compute the Poisson brackets of these functions at
$(g^e, g^o)\in \widetilde{G}^{\Bbb R}_{w} \times \widetilde{G}^{\Bbb R}_{w}.$
For a function $\varphi$ on $\widetilde{G}^{\Bbb R}_{w} \times \widetilde{G}^{\Bbb R}_{w}$
which is smooth in both variables, we
denote by $D_i \varphi (A,B)$ (resp. $D_i^{\prime} \varphi (A,B)$) its right gradient
(resp. left gradient) with respect to the $i$-th variable, $i=1,2.$
Then we have
$$\eqalign{
& D_1 F_j(g^e,g^o) = g^e E_{jj} = \bar\alpha_j E_{jj} + \rho_j E_{j+1,j},\cr
& D_1^{\prime} F_j(g^e,g^o) = E_{jj} g^e = \bar\alpha_j E_{jj} + \rho_jE_{j,j+1},\cr
& D_2 F_j(g^e,g^o) =0,\,\, D_2^{\prime} F_j(g^e,g^o) =0,\cr}\eqno(3.64)$$
and similarly,
$$\eqalign{
& D_2 F_l(g^e,g^o) = g^o(h) E_{ll} = \bar\alpha_l E_{ll} + \rho_l E_{l+1,l},\cr
& D_2^{\prime} F_l(g^e,g^o) = E_{ll} g^o(h) = \bar\alpha_l E_{ll} + \rho_lE_{l,l+1},\,\,l \,\,\hbox{odd}, 1\leq l\leq p-3,\cr
& D_2 F_{p-1}(g^e,g^o) = g^o(h) E_{p-1,p-1} = \rho_{p-1}h E_{0,p-1} + \bar\alpha_{p-1}E_{p-1,p-1},\cr
& D_2^{\prime} F_{p-1}(g^e,g^o) = E_{p-1,p-1} g^o(h) = \rho_{p-1} h^{-1} E_{p-1,0} +\bar\alpha_{p-1}E_{p-1,p-1},\cr
& D_1 F_l(g^e,g^o) =0,\,\, D_1^{\prime} F_l(g^e,g^o) =0,\,\, l\,\,\hbox{odd}, 1\leq l\leq p-1.\cr}
\eqno(3.65)$$
On the other hand, it is clear that
$$\eqalign{
& D_1 G_j(g^e,g^o) = i D_1 F_j(g^e,g^o), D_1^{\prime} G_j( g^e,g^o) = i D_1^{\prime} F_j(g^e,g^o),\cr
& D_2 G_j(g^e,g^o) =0,\,\, D_2^{\prime} G_j(g^e,g^o) =0,\cr}\eqno(3.66)$$
and similarly,
$$\eqalign{
& D_2 G_l(g^e,g^o) = i D_2 F_l(g^e,g^o), D_2^{\prime} G_l( g^e,g^o) = i D_2^{\prime} F_l(g^e,g^o)\cr
& D_1 G_l(g^e,g^o) =0,\,\, D_1^{\prime} G_l(g^e,g^o) =0.\cr}\eqno(3.67)$$

In what follows, the indices $j$ and $k$ are even, while the indices
$l$ and $m$ are odd, and all indices are from $\{0,\cdots, p-1\}.$
Let $\{\cdot,\cdot\}_{*}$ denote the product structure on $\Gtw\times\Gtw,$
then it is immediate from the definition of $\{\cdot,\cdot\}_{*}$ that
$$\{F_j, F_l\}_{*} =0,\, \{G_j,G_l\}_{*}=0,\, \{F_j, G_l\}_{*}=0,\, 
\{G_j, F_l\}_{*} =0.\eqno(3.68)$$
Now, from (3.64), $D_1 F_j(g^e,g^o)$ and $D_1^{\prime} F_j(g^e,g^o)$
are constant loops, so it follows from (3.12) and equation (2.6) of 
\c{L1} that
$$\eqalign{
& J^{\sharp}(D_1 F_j(g^e,g^o)) = -\alpha_j E_{jj} -\rho_j E_{j+1,j},\cr
& J^{\sharp}(D_1^{\prime} F_j(g^e,g^o)) = \rho_jE_{j,j+1} -2\rho_j E_{j+1,j}
-\alpha_jE_{jj}.\cr}\eqno(3.69)$$
Therefore, on using (3.69) and (3.64), we find for $j\neq k$ that
$$\eqalign{&
\hbox{tr}\,\,(J^{\sharp}(D_1^{\prime} F_j(g^e,g^o))D_1^{\prime}
  F_k(g^e,g^o))\cr
 = & (D_1^{\prime} F_k(g^e,g^o))_{kk}
  (J^{\sharp}(D_1^{\prime} F_j(g^e,g^o))_{kk} 
 + (D_1^{\prime} F_k (g^e,g^o))_{k,k+1}(J^{\sharp}(D_1^{\prime}F_j(g^e,g^o))_{k+1,k}\cr
  =\, & 0,\cr}\eqno(3.70)$$
and similarly,
$$\eqalign{&
\hbox{tr}\,\,(J^{\sharp}(D_1 F_j(g^e,g^o))D_1F_k(g^e,g^o))\cr
 = & (D_1 F_k(g^e,g^o))_{kk}(J^{\sharp}(D_1 F_j(g^e,g^o))_{kk} 
 + (D_1 F_k (g^e,g^o))_{k+1,k}(J^{\sharp}(D_1 F_j(g^e,g^o))_{k,k+1}\cr
  =\, & 0.\cr}\eqno(3.71)$$
Thus it follows from (3.70),(3.71) and (3.66) that
$$\{F_j, F_k\}_{*} (g^e, g^o) =0, \,\,\{G_j,G_k\}_{*}(g^e,g^o) =0,\,\, 
  \{F_j, G_k\}_{*} (g^e,g^o) =0,\,\,\hbox{for}\,\, j\neq k.\eqno(3.72)$$
In a similar fashion, by using (3.64), (3.66). and (3.69), we
find
$$\eqalign{
& \{F_j, G_j\}_{*}(g^e,g^o)\cr
= & \frac{1}{2} \hbox{tr}\,\,(i\,J^{\sharp}(D_1^{\prime} 
F_j(g^e,g^o))D_1^{\prime}F_j(g^e,g^o))- \frac{1}{2} 
\hbox{tr}\,\,(i\,J^{\sharp}(D_1 F_j(g^e,g^o))D_1F_j(g^e,g^o))\cr
= & -\rho_j^2.\cr}\eqno(3.73)$$
Analogously, for the odd indices $l,m$, with $l,m\in \{0,\cdots, p-3\},$
we have
$$\{F_l, F_m\}_{*} (g^e, g^o) =0, \,\,\{G_l,G_m\}_{*}(g^e,g^o) =0,\,\, 
  \{F_l, G_m\}_{*} (g^e,g^o) =0,\,\,\hbox{for}\,\, l\neq m.\eqno(3.74)$$
Also,
$$\{F_l, G_l\}_{*}(g^e,g^o) = -\rho_l^2.\eqno(3.75)$$
Next, we consider brackets with the quantities $F_{p-1}$ and $G_{p-1}.$
To this end, we note the formulas
$$\eqalign{
& J^{\sharp}(D_2 F_{p-1}(g^e,g^o)) = -\alpha_{p-1} E_{p-1,p-1} -\rho_{p-1}h E_{0,p-1},\cr
& J^{\sharp}(D_2^{\prime} F_{p-1}(g^e,g^o)) = \rho_{p-1} h^{-1}E_{p-1,0} -
\alpha_{p-1}E_{p-1,p-1} -2\rho_{p-1}h E_{0,p-1}.\cr}\eqno(3.76)$$
Therefore, if $l$ is odd, $l\leq p-3,$ a calculation similar to (3.70) and 
(3.71) above shows that
$$\{F_{p-1}, F_l\}_{*}(g^e,g^o)=0,\,\, \{G_{p-1},G_l\}_{*}(g^e,g^o)=0,\,\,
  \{F_{p-1}, G_l\}_{*}(g^e, g^o)=0.\eqno(3.77)$$
Now, by (3.76), (3.67) and (3.65), we have
$$\hbox{tr}\,\,(J^{\sharp}(D_2 F_{p-1}(g^e,g^o))D_2G_{p-1}(g^e,g^o))
 =\,  -i |\alpha_{p-1}|^2,\eqno(3.78)$$
while
$$\hbox{tr}\,\,(J^{\sharp}(D_2^{\prime} F_{p-1}(g^e,g^o))D_2^{\prime}G_{p-1}(g^e,g^o))
 =\, -2i \rho_{p-1}^2 -i |\alpha_{p-1}|^2.\eqno(3.79)$$
Consequently,
$$\{F_{p-1}, G_{p-1}\}_{*}(g^e,g^o) = -\rho_{p-1}^2.\eqno(3.80)$$
Assembling the calculations, we conclude that for all
$a,b\in \{0,\cdots, p-1\},$ we have
$$\eqalign{&
\{F_a, F_b\}_{*} (g^e,g^o)=0,\,\,\{G_a, G_b\}_{*}(g^e,g^o)=0,\,\, \{F_a, G_b\}_{*}(g^e,g^o) =0, \,\,\hbox{if}\,\,a\neq b,\cr
& \{F_a, G_a\}_{*}(g^e,g^o) = -\rho_a^2.\cr}\eqno(3.81)$$
So finally, we obtain the following Poisson bracket relations
$$\eqalign{&
\{G_a -iG_a, G_b-iF_b\}_{*}(g^e,g^o) =0, \,\{G_a + iF_a, G_b+iF_b\}_{*}(g^e,g^o) =0,
\,\, \hbox{if}\,\, a\neq b,\cr
& \{G_a - iF_a, G_b + iF_b\}_{*}(g^e,g^o) = 2i \delta_{ab}\rho_a^2,\cr}\eqno(3.82)$$
as desired.

\pf
\enddemo

Finally we describe the Hamiltonian equations generated by central
functions on $\Gtw$ in the above framework.   We also introduce
the kind of equations which we will need to use in Section 6 below.

\proclaim 
{Proposition 3.9} (a) The Hamiltonian equation of motion generated
by a central function $\varphi$ on $\Gtw$ is given by the Lax equation
$$\aligned
\dot g & = g\,(\piktw(D\varphi(g)))-(\piktw(D\varphi(g)))\,g\\
       & = (\pibtw(D\varphi(g)))\,g -g\,(\pibtw(D\varphi(g)))\\
\endaligned
\eqno(3.83)$$
\newline
(b) Consider $(\Gtw, \{\cdot,\cdot\}_{J^{\sharp}})$
and equip the group $\Gtw\times \Gtw$ with the product Poisson structure.  If
$\varphi$ is a central function on $\Gtw,$ then the Lax system 
$$\eqalign{
\dot g_1 & = g_1\left(\piktw (D\varphi (g_2 g_1))\right) -
\left(\piktw (D\varphi(g_1g_2))\right)g_1\cr
         & = \left(\pibtw (D\varphi(g_1g_2))\right)g_1 -
 g_1\left(\pibtw (D\varphi (g_2 g_1))\right),\cr
\dot g_2 &= g_2 \left(\piktw (D\varphi (g_1 g_2))\right)
 -\left(\piktw (D\varphi(g_2g_1))\right)g_2\cr
         & = \left(\pibtw (D\varphi(g_2g_1))\right)g_2 -
g_2 \left(\pibtw (D\varphi (g_1 g_2))\right).\cr}
\eqno(3.84)$$
is the Hamiltonian equation on $\Gtw\times\Gtw$ generated by
$H_{\varphi}(g_1,g_2) = \varphi(g_1g_2).$
Moreover, under the Hamiltonian flow defined by (3.84), 
$g= g_1 g_2$ evolves according to (3.83).
\newline
(c) If $k_i(t),$ $b_i(t),$ $i=1,2$ are the solutions of the
factorization problems
$$e^{tD\varphi(g_1(0) g_2(0))} = k_1(t)b_{1}^{-1}(t),\quad
e^{t D\varphi(g_2(0) g_1(0))} = k_2(t) b_{2}^{-1}(t),\eqno(3.85)$$
where $k_i(t)\in \Ktw,$ $b_i(t)\in \Btw,$  then the flow defined
by (3.84) is given by
$$\aligned
g_1(t) & = k_1(t)^{-1} g_1(0) k_2(t) = b_1(t)^{-1} g_1(0) b_2(t),\\
g_2(t) & = k_2(t)^{-1} g_2(0) k_1(t) = b_2(t)^{-1} g_2(0) b_1(t).\\
\endaligned
\eqno(3.86)$$
\endproclaim
To conclude this section, we remark that equations of the 
type in (3.84) are a special case of
so-called Lax systems on a periodic lattice or difference Lax equations 
and we refer the reader to \c{STS2} and \c{LP} for the general
theory. In Section 6 below, we will show how to solve
the factorization problems for the flows generated by the
commuting integrals of the periodic defocusing Ablowitz-Ladik equation
by means of Riemann theta functions associated with a 
hyperelliptic curve.

\bigskip

\subhead
4. \ Analytical properties of the Bloch solution
\endsubhead

\bigskip

For any $z\in \Bbb C,$ consider the equation
$${\Cal E} u = z u, \eqno(4.1)$$
where ${\Cal E}$ is the extended CMV matrix with periodic
Verblunsky coefficients with period $p$, as in Section 2.   
Since ${\Cal E}$ admits a $\theta$-factorization 
${\Cal E} = {\Cal L}{\Cal M}$ \c{S2}, as in the one-sided case,
it follows that (4.1) is equivalent to 
$${\Cal M} u = {\Cal L}^{*}u.\eqno( 4.2)$$
In terms of the components of $u$ and the entries of
${\Cal L}$ and ${\Cal M}$, (4.2) gives the three-term recurrence
relations
$$\eqalign{
& \rho_{2j-1} u_{2j-1} -\alpha_{2j-1} u_{2j} = z(\alpha_{2j}u_{2j} + \rho_{2j}
  u_{2j+1})\cr
& \bar{\alpha}_{2j+1} u_{2j+1} + \rho_{2j+1} u_{2j+2}= z(\rho_{2j}u_{2j} -
  \bar{\alpha}_{2j} u_{2j+1})\cr}\eqno(4.3)$$
for all $j\in \Bbb Z.$  Due to the equivalent form in (4.2), the
space of solutions of (4.1) is two dimensional.   Indeed, it is clear
from (4.3) that for given values of $u_{-1}$ and $u_{0}$,  we can
determine all other values of $u_{n}$ by recursion.   For our analysis,
we will fix a basis with the following initial conditions:
$$\eqalign{
& \phi_{-1}(z) =1,\quad \phi_{0}(z) =0,\cr
& \psi_{-1}(z) =0,\quad \psi_{0}(z) =1.\cr}\eqno(4.4)$$
By using the first relation in (4.3) corresponding to $j=0,$ we have
$$\phi_{1}(z) =\frac{\rho_{p-1}}{z\rho_{0}}.\eqno(4.5)$$   
In general, an easy induction using (4.3) gives the following result.

\proclaim
{Proposition 4.1} For all $j\geq 1,$
$$\eqalign{
& \phi_{2j}(z) = -\frac{\bar{\alpha}_{0}\rho_{p-1}}{\rho_{0}\cdots \rho_{2j-1}}
   z^{j-1} -\cdots -\frac{\bar{\alpha}_{2j-1}\rho_{p-1}}
   {\rho_{0}\cdots \rho_{2j-1}}\frac{1}{z^{j}},\cr
&  \phi_{2j+1}(z) = \frac{\bar{\alpha}_{0}\alpha_{2j}\rho_{p-1}}
   {\rho_{0}\cdots \rho_{2j}}z^{j-1} + \cdots +\frac{\rho_{p-1}}
   {\rho_{0}\cdots \rho_{2j}}\frac{1}{z^{j+1}}.\cr}\eqno(4.6)$$
In particular,
$$\phi_{p-1}(z) = \frac{\bar{\alpha}_{0}\alpha_{p-2}\rho_{p-1}}
   {\rho_{0}\cdots \rho_{p-2}}z^{p/2-2} + \cdots +\frac{\rho_{p-1}}
   {\rho_{0}\cdots \rho_{p-2}}\frac{1}{z^{p/2}}.\eqno(4.7)$$
\endproclaim

Similarly, we have
$$\psi_{1}(z) = -\frac{\alpha_0}{\rho_0} -\frac{\alpha_{p-1}}{\rho_0}
\frac{1}{z},\eqno(4.8)$$
and by induction, we obtain the following analog of Proposition 4.1.

\proclaim 
{Proposition 4.2} For all $j\geq 1,$
$$\eqalign{
& \psi_{2j}(z) = \frac{z^j}{\rho_{0}\cdots\rho_{2j-1}} + \cdots +
  \frac{\bar{\alpha}_{2j-1}\alpha_{p-1}}{\rho_{0}\cdots\rho_{2j-1}}\frac{1}
  {z^j},\cr
& \psi_{2j+1}(z) = -\frac{\alpha_{2j}}{\rho_{0}\cdots\rho_{2j}}z^j - \cdots -
  \frac{\alpha_{p-1}}{\rho_{0}\cdots\rho_{2j}}\frac{1}{z^{j+1}}.\cr}\eqno(4.9)
$$
In particular,
$$\psi_{p-1}(z) = -\frac{\alpha_{p-2}}{\rho_{0}\cdots \rho_{p-2}}z^{p/2-1}
  -\cdots -\frac{\alpha_{p-1}}{\rho_{0}\cdots \rho_{p-2}}\frac{1}{z^{p/2}},
  \eqno(4.10)$$
and
$$\psi_{p}(z) = \frac{z^{p/2}}{\rho_{0}\cdots \rho_{p-1}} + \cdots +
  \frac{|\alpha_{p-1}|^2}{\rho_{0}\cdots \rho_{p-1}}\frac{1}{z^{p/2}}.
\eqno(4.11)
$$
\endproclaim
Now by the periodicity of the Verblunsky coefficients, we have
$$\pmatrix \phi_{j+p}(z) & \psi_{j+p}(z) \endpmatrix = \pmatrix \phi_{j}(z) &
 \psi_{j}(z)\endpmatrix M(z)\eqno(4.12)$$
for all $j,$ where 
$$M(z) = \pmatrix \phi_{p-1}(z) & \psi_{p-1}(z) \\
                  \phi_{p}(z)   & \psi_{p}(z)\endpmatrix\eqno(4.13)$$
is the monodromy matrix.

\proclaim
{Proposition 4.3} For all $z,$ $\det M(z) =1.$ 
\endproclaim

\demo
{Proof} Let $W_{j}(z) = 
\rho_{j}(\phi_{j}(z)\psi_{j+1}(z) - \phi_{j+1}(z)\psi_{j}(z)).$  
Then from the first relation in (4.3), we have
$$W_{2j-1}(z) = -z W_{2j}(z)\eqno(4.14)$$
for all $j.$   Similarly, from the second relation in (4.3), we find that
$$W_{2j+1}(z) = - z W_{2j}(z)\eqno(4.15)$$
for all $j.$
As the right hand sides of (4.14) and (4.15) are equal, it follows 
that $W_{2j-1}(z)$ is independent of $j$ and consequently
$W_{p-1}(z) = W_{-1}(z)$ from which the assertion follows.
\pf
\enddemo

From Proposition 4.3, the eigenvalues of the monodromy matrix (i.e.,
the Floquet multipliers)
are the roots of the characteristic polynomial
$$h^2 -\hbox{tr}\, M(z) h + 1 =0.\eqno(4.16)$$
If $T:l^{\infty}(\Bbb Z)\longrightarrow l^{\infty}(\Bbb Z)$ denote the 
shift operator defined by $(Tu)_{j} = u_{j+p},$
then the unique solution of the problem
$${\Cal E}f = z f, \quad Tf = h^{-1} f, \,\,\, f_{p-1} =1\eqno(4.17)$$
is called the Bloch solution and finding this solution
is equivalent to considering the spectrum of the corresponding
Floquet CMV matrix ${\Cal E}(h)$,
$${\Cal E}(h) \widehat v = z \widehat v\eqno(4.18)$$
where
$$\widehat v = \pmatrix f_0 \\
                        \vdots \\ 
                        f_{p-2} \\
                        1\endpmatrix . \eqno(4.19)$$
Hence the ordered pair $(z,h)$  in (4.17) must obey the equation
$$\hbox{det}(zI -\E(h)) = \left(\prod_{j=0}^{p-1} \rho_j\right) z^{p\over 2}
[\Delta(z)-(h + h^{-1})] =0,\eqno(4.20)$$
where the discriminant $\Delta(z)$ is related to the transfer matrix
$$T_{p}(z) = {1\over {\prod_{j=0}^{p-1} \rho_j}}
\pmatrix z & -\bar{\alpha}_{p-1}\\
       -\alpha_{p-1} z & 1\endpmatrix  \cdots
\pmatrix z & -\bar{\alpha}_{0}\\
       -\alpha_{0} z & 1\endpmatrix\eqno(4.21)$$
by the formula \c{S2}
$$\Delta(z) = z^{-{p/2}}\,\hbox{tr}\,\, T_{p}(z).\eqno(4.22)$$
By comparing (4.16) and (4.20), we therefore conclude that
$\hbox{tr}\, M(z) = \Delta (z)$ and this relates the
multiplier curve and the spectral curve.  We will make the
genericity assumption
\smallskip
\noindent $(GA)_1$ the roots of $P(z) = \left(\prod_{i=0}^{p-1} \rho_i\right)^{2}
(\Delta(z)^{2} z^{p} - 4 z^{p}) =\prod_{i=1}^{2p} (z -\lambda_{i})$ 
are distinct.
\smallskip
Then it is straightforward to check that the affine curve as defined by
the equation 
$$I(h,z):= h\cdot \hbox{det}(zI -\E(h)) =0 \eqno(4.23)$$ 
is smooth with branch
points located at $\lambda_1,\cdots, \lambda_{2p}.$   We
will denote by $C$ the hyperelliptic Riemann surface of genus $g = p-1$ corresponding to this
affine curve.  In order to find the divisor structure of 
$f_j, j=0,\cdots, p-2$ on $C,$  let
$$(z) = Q_{+} + Q_{-}-P_{+} - P_{-}\eqno(4.24)$$
where $P_{+}, Q_{+}$ are on the $+$ sheet and $P_{-}, Q_{-}$ are on the $-$ sheet of
$C.$  (The $\pm$-sheets correspond to the choice of sign in front of the radical in the
first line of (4.25).)  Solving for $h$ in terms of of $z$ from (4.20), we find
$$\eqalign{
h(z) & = \frac{z^{p/2} \Delta(z) \pm \sqrt{\Delta(z)^{2} z^{p} -4z^{p}}}
         {2z^{p/2}}\cr
     & = \frac{z^{p/2} \Delta(z) \pm z^{p/2}\Delta(z)\left[ 1 -\frac{2z^p}
          {z^p\Delta(z)^2} + \cdots \right]}{2 z^{p/2}}\quad \hbox{for}\,z\,\,
          \hbox{near}\,\infty,\cr
     & = \frac{z^{p/2}}{\prod_{j=0}^{p-1} \rho_j} + \cdots \quad \hbox{for}\,
          z\,\, \hbox{near}\, \infty\, \hbox{on the} + \,\hbox{sheet},\cr
     & = \frac{\prod_{j=0}^{p-1} \rho_j}{z^{p/2}} + \cdots \quad \hbox{for}\,
          z\,\, \hbox{near}\, \infty\, \hbox{on the} - \,\hbox{sheet}.\cr}
\eqno(4.25)
$$
On the other hand, it is clear from (5.1) that  $P\,z^{p/2}\Delta(z)\sim 1$ as $z\to 0.$
Therefore, it follows from the first line of (4.25) that $h(z) \sim \frac{1}{P\,z^{p/2}}$
as $z\to 0$ on the $+$ sheet, while $h(z)\sim P\,z^{p/2} $ as $z\to 0$ on the $-$ sheet.
Thus we have
$$(h) = -\frac{p}{2} P_{+} + \frac{p}{2} P_{-}-\frac{p}{2}Q_{+} + \frac{p}{2}Q_{-}.\eqno(4.26)$$

\proclaim
{Proposition 4.4} For each $0\leq j\leq p-2,$
$$f_j(P) = h(P)\phi_j(z(P))  + \frac{1 - h(P)\phi_{p-1}(z(P))}{\psi_{p-1}
(z(P))}\psi_j(z(P)), \,\, P\in C. \eqno(4.27)$$
\endproclaim

\demo
{Proof} Since $\phi$ and $\psi$ form a basis of solutions of 
the equation ${\Cal E}u = zu,$ we must have
$f_j = c_1 \phi_j  + c_2 \psi_j$ for some constants $c_1$ 
and $c_2.$   Putting $j= -1$ and $0$ in the above expression
and using the initial conditions in (4.4), we find
$f_j = h \phi_j + f_0 \psi_j$.  As the vector with
components $h$ and $f_0$ is an eigenvector of the
monodromy matrix $M(z)$ with eigenvalue $h^{-1},$ 
we find that $\phi_{p-1} h + \psi_{p-1} f_{0} =1.$
Solving for $f_{0}$ from this expression, we obtain the
desired expression for $f_j.$
\pf
\enddemo

We next make the following assumption.
\smallskip
\noindent $(GA)_2$\,\,\,\,  $\alpha_{j} \neq 0$ for $j=0,\cdots, p-1.$
\smallskip
Note that in particular, $\alpha_{p-2}\neq 0$ and so the degree of the
polynomial $z^{p/2} \psi_{p-1}(z)$ is  $p-1.$  The roots of this polynomial
will be denoted by $z_k,$ $k=1,\cdots, p.$ (In general, the $z_k$'s 
are not necessarily all distinct.)

\proclaim
{Proposition 4.5} For each $0\leq j\leq p-2,$ $f_j(P)$ is
a single-valued meromorphic function on the Riemann
surface $C$.   On the finite part of $C$ away from $Q_+,$ $f_j(P)$ has at worst
poles at the points $P_k = (\phi_{p-1}(z_k), z_k),$
where $\psi_{p-1}(z_k) =0$ for $k=1,\cdots, p-1.$  Moreover,
the $z_k$'s coincide with the  eigenvalues of the
Dirichlet problem
$${\Cal E} u = zu, \quad u_{-1} =0, u_{p-1} =0.\eqno(4.28)$$
Equivalently, the $z_k$'s are the zeros of the equation
$$\hbox{det}\,\,(z \widehat{(g^e)^*} - \widehat{ g^o(h)}) =0,\eqno(4.29)$$
where $\widehat{(g^e)^*}$ and $\widehat{ g^o(h)}$ are $(p-1)\times (p-1)$
matrices obtained from $(g^e)^*$ and $g^o(h)$ by removing their last
row and last column.
\endproclaim

\demo
{Proof} For each $j,$ it follows from (4.27) that $f_j(P)$ is meromorphic
on $C.$    Let us consider a point $z_k$ where $\psi_{p-1}(z_k) =0.$  From 
(4.13), we see that  at such a point, the Floquet multipliers are given by
$\phi_{p-1}(z_k)$ and $\phi_{p-1}(z_k)^{-1}$ and hence
$(\phi_{p-1}(z_k), z_k)$ and $(\phi_{p-1}(z_k)^{-1},z_k)$ are 
points on $C.$    Clearly,
$1 - h\phi_{p-1}(z)$ vanishes at $(\phi_{p-1}(z_k)^{-1}, z_k)$.
Therefore, provided that $\phi_{p-1}(z_k)\neq \pm 1$ or
equivalently, $\Delta(z_k) \neq \pm 2,$  $f_j(P)$  has a pole at 
$P_k = (\phi_{p-1}(z_k), z_k).$    Since  $\psi_{-1}=0,$ we see 
that the solutions of $\psi_{p-1}(z)=0$ coincide with the
eigenvalues of the Dirichlet problem in (4.28).  On the other hand, 
observe that if $\psi_{p-1}(z_k) =0,$ then
$$\hbox{diag}(-\alpha_{p-1}, \theta_1,\cdots, \theta_{p-3})
  \pmatrix \psi_{0}(z_k) \\
           \vdots \\
            \psi_{p-2}(z_k) \endpmatrix
  = z_k\,\hbox{diag}(\theta_0^*,\cdots, \theta_{p-4}^*, \alpha_{p-2})
   \pmatrix \psi_{0}(z_k) \\
            \vdots \\
            \psi_{p-2}(z_k) \endpmatrix$$
from the connection with (4.28) where $\psi_{0} =1.$
As the matrix on the left hand side of the above formula is $\widehat{g^o(h)},$
while the one on the right hand side is $\widehat{(g^e)^*},$  the last
assertion in the proposition follows.
\pf
\enddemo

\remark
{Remark 4.6}   The Dirichlet eigenvalues above should not be confused with the
Dirichlet data in Chapter 11 of Simon \c{S2}.     While the latter quantities always
reside on the unit circle and the number of such quantities is equal to $p$, it is not the 
case for the $z_k$'s, as is evident from the relation $\prod_{k=1}^{p-1} z_k =  -{\alpha_{p-1}\over \alpha_{p-2}}.$
\endremark

In what follows, we will denote by $h^{\pm}(z)$ (resp. $f_j^{\pm}(z)$)
the values of the function $h(P)$ (resp. $f_j(P)$)
on the $\pm$ sheets of the Riemann surface. 
\proclaim
{Proposition 4.7} For $j=0,\cdots, p/2-1,$ we have
$$\eqalign{
& f^{-}_{2j}(z) \sim -{\frac{\left(\prod_{i=2j}^{p-2} \rho_{i}\right)}
  {\alpha_{p-2}}} z^{-(p/2 -j-1)},\cr
& f^{-}_{2j+1}(z)\sim {\frac{\alpha_{2j}}{\alpha_{p-2}}} \left(\prod_{i=2j+1}^{p-2} 
  \rho_{i}\right) z^{-(p/2-j-1)}\cr}\eqno(4.30)$$
as $z\to\infty.$  Hence $f_{2j}(P)$ and $f_{2j+1}(P)$ have
zeros of order $p/2-j-1$ at $P_{-}.$
\endproclaim

\demo
{Proof} Consider first the even case.  By using (4.25) and Propositions
4.1 and 4.2, we have
$$\frac{1 -h^{-}(z) \phi_{p-1}(z)}{\psi_{p-1}(z)}\psi_{2j}(z)\sim -
\frac{\left(\prod_{i=2j}^{p-2} \rho_i\right)}{\alpha_{p-2}}.\frac{1}
{z^{p/2-j-1}}\eqno(4.31)$$
as $z\to \infty.$  Similary, 
$$h^{-}(z)\phi_{2j}(z) \sim -\frac{\bar{\alpha}_{0}\rho_{p-1}
\left(\prod_{i=2j}^{p-1} \rho_i\right)}{z^{p/2-j+1}}\eqno(4.32)
$$
as $z \to \infty.$  Therefore, on comparing (4.31) and (4.32), the assertion
for the even case follows. We will skip the details for the odd case
as it proceeds in the same way.

\pf
\enddemo
To investigate the behaviour of $f^{+}_{j}(z)$ as $z\to \infty,$
we will first establish an identity for the product
$f^{+}_{j}(z)f^{-}_{j}(z).$   To this end, observe that
$$\eqalign{
& \frac{1-h^{\pm}\phi_{p-1}(z)}{\psi_{p-1}(z)}\cr
= & \frac{1}{2\psi_{p-1}(z)}\left[(2-\phi_{p-1}(z)(\phi_{p-1}(z)+\psi_{p}(z)))
    + \phi_{p-1}(z)\sqrt{\Delta(z)^{2} -4}\,\right].\cr}\eqno(4.33)$$
Therefore, by a direct multiplication and using (4.31), we find
$$\eqalign{
f^{+}_{j}(P)f^{-}_{j}(P) =& \frac{-\psi_{j}(\phi_j\phi_{p-1} + \psi_j\phi_p) +
    \phi_j(\psi_j\psi_p+\phi_j\psi_{p-1})}
    {\psi_{p-1}}\cr
=& \frac{-\psi_j \phi_{j+p} + \phi_j\psi_{j+p}}{\psi_{p-1}}\cr}\eqno(4.34)$$
where on the right hand side, we have omitted the variable $z$
throughout.  Note that in going from the first line of (4.34) to the
second line, we have used (4.12).  Our next task is to interpret
the numerator of the right hand side of (4.34), which is necessary
because performing a direct asymptotic analysis of this quantity
by using Propositions 4.1 and 4.2 proves to be difficult.  That 
this is so is due to the degeneracy of the tridiagonal matrices
${\Cal L}$ and ${\Cal M}.$ (Note that neither ${\Cal E}$ nor its
factors $\Cal L$ and $\Cal M$ satisfy the genericity assumption in \c{MM}.) 
For this purpose, we introduce for each $1\leq j \leq p-2$ the shifted matrix
$\Ej$ whose $(k,l)$ entry is given by
$$(\Ej)_{kl} = {\Cal E}_{k+j, l+j}.\eqno(4.35)$$
On the other hand, let $\psij$ denote the solution of 
$$\Ej u = zu, \quad u_{-1} = 0, u_{0} =1.\eqno(4.36)$$
\proclaim
{Proposition 4.8} For each $0\leq j\leq p-2,$
$$f^{+}_{j}(z)f^{-}_j(z) = \frac{B_{j+1}(z)\psi^{[j+1]}_{p-1}(z)}{\psi_{p-1}(z)},
\eqno(4.37)$$
where
$$\eqalign{
B_{j+1}(z) =& \det \pmatrix \phi_j(z) & \psi_j(z)\\
                            \phi_{j+1}(z) & \psi_{j+1}(z)\endpmatrix \cr
          =& \cases \frac{\rho_{p-1}}{\rho_j}, &\text{for $j$ odd}\\
                    -\frac{\rho_{p-1}}{\rho_j z}, &\text{for $j$ even}.
           \endcases\cr}\eqno(4.38)
$$
\endproclaim

\demo
{Proof} From the definition of $\Ej$ and $\psi^{[j]},$ it is clear that
$$\psij_{k}(z) = c_1(z) \phi_{k + j}(z) + c_2(z)\psi_{k+j}(z)\eqno(4.39)$$
for some $c_1(z)$ and $c_2(z).$  By imposing the initial 
conditions in (4.36), we find that
$$c_1(z) = -\frac{\psi_{j-1}(z)}{B_j(z)},\,\,
  c_2(z) = \frac{\phi_{j-1}(z)}{B_j(z)}.\eqno(4.40)$$
Therefore, on substituting into (4.39), we obtain
$$B_{j}(z) \psij_{k}(z) = -\psi_{j-1}(z)\phi_{k+j}(z) + \phi_{j-1}(z)\psi_{k+j}(z).
\eqno(4.41)$$
Hence (4.37) follows from (4.34) if we replace $j$ by $j+1$ and let
$k=p-1$ in (4.41).  To complete the proof, it remains to calculate
$B_{j+1}(z).$  Here we make use of the quantity $W_{j}(z)$ introduced
in the proof of Proposition 4.3 which is related to $B_{j+1}(z)$ by
the relation $W_{j}(z) = \rho_j B_{j+1}(z).$  From the proof of 
Proposition 4.3 (see relations (4.14) and (4.15)), we learn 
that $W_{2j-1}(z)$ is independent of the value of $j,$ and the
same holds true for $W_{2j}(z).$  Consequently, we have
$$\rho_j B_{j+1}(z) = \rho_{j-2}B_{j-1}(z).\eqno(4.42)$$
From this, we find
$$\eqalign{
\rho_j B_{j+1}(z) = & \cases \rho_{-1} B_{0}(z), & \text{for $j$ odd},\\
                            \rho_{0} B_{1}(z), &\text{for $j$ even},
                     \endcases \cr
                 = & \cases \rho_{p-1}, & \text{for $j$ odd},\\
                            -\frac{\rho_{p-1}}{z}, &\text{for $j$ even}.
                      \endcases \cr}\eqno(4.43)$$
This completes the proof.
\pf
\enddemo
In our next result, we will analyze $\psij(z).$
 
\proclaim
{Proposition 4.9} (a) For $0\leq j\leq p/2 -1$ and  $0\leq k\leq p/2-1,$
$$\eqalign{
& \psi^{[2j]}_{2k}(z) = \frac{z^k}{\rho_{2j}\cdots\rho_{2j+2k-1}} + \cdots +
  \frac{\bar{\alpha}_{2j+2k-1}\alpha_{2j-1}}{\rho_{2j}\cdots\rho_{2j+2k-1}}\frac{1}
  {z^k},\cr
& \psi^{[2j]}_{2k+1}(z) = -\frac{\alpha_{2j+2k}}{\rho_{2j}\cdots\rho_{2j+2k}}z^k
 - \cdots -\frac{\alpha_{2j-1}}{\rho_{2j}\cdots\rho_{2j+2k}}\frac{1}{z^{k+1}}.\cr}
\eqno(4.44)$$
(b) For $0\leq j\leq p/2 -1$ and  $0\leq k\leq p/2-1,$
$$\eqalign{
&\psi^{[2j+1]}_{2k}(z) = \frac{\bar{\alpha}_{2j}{\alpha}_{2j+2k}}
{\rho_{2j+1}\cdots\rho_{2j+2k}} z^{k} + \cdots + \frac{1}{\rho_{2j+1}\cdots\rho_{2j+2k}}
\frac{1}{z^k},\cr
&\psi^{[2j+1]}_{2k+1}(z) = -\frac{\bar{\alpha}_{2j}}{\rho_{2j+1}\cdots\rho_{2j+2k+1}}
 z^{k+1} -\cdots - \frac{\bar{\alpha}_{2j+2k+1}}{\rho_{2j+1}\cdots\rho_{2j+2k+1}}
\frac{1}{z^k}.\cr}\eqno(4.45)$$
In particular, 
$$\psi^{[2j]}_{p-1}(z) = -\frac{\alpha_{2j-2}\rho_{2j-1}}{\prod_{i=0}^{p-1} \rho_i}
z^{p/2-1} -\cdots  -\frac{\alpha_{2j-1}\rho_{2j-1}}{\prod_{i=0}^{p-1} \rho_i}
\frac{1}{z^{p/2}}\eqno(4.46)$$
and
$$\psi^{[2j+1]}_{p-1}(z) = -\frac{\bar{\alpha}_{2j}\rho_{2j}}{\prod_{i=0}^{p-1}
\rho_i} z^{p/2} -\cdots - \frac{\bar{\alpha}_{2j-1}\rho_{2j}}{\prod_{i=0}^{p-1}
\rho_i}\frac{1}{z^{p/2-1}}.\eqno(4.47)$$
\endproclaim

\demo
{Proof} For the equation ${\Cal E}^{[i]} u =zu$ associated with the
shifted matrix ${\Cal E}^{[i]}$, the recurrence relations in
(4.3) have to be replaced by
$$\eqalign{
& \rho_{2k-1} u_{2k-1-i} -\alpha_{2k-1} u_{2k-i} = z(\alpha_{2k}u_{2k-i} + \rho_{2k}
  u_{2k+1-i})\cr
& \bar{\alpha}_{2k+1} u_{2k+1-i} + \rho_{2k+1} u_{2k+2-i}= z(\rho_{2k}u_{2k-i} -
  \bar{\alpha}_{2k} u_{2k+1-i}).\cr}\eqno(4.48)$$
For $i=2j,$ we can therefore obtain (4.44) from (4.9) by shifting
the indices.  For $i=2j +1,$ we have to solve (4.48) and an
inductive argument leads to (4.45).  
\pf
\enddemo

\proclaim
{Proposition 4.10} For $j=0,\cdots, p/2-1,$ we have
$$\eqalign{
& f^{+}_{2j}(z) \sim -\frac{\bar{\alpha}_{2j}} 
  {\left(\prod_{i=2j}^{p-2} \rho_{i}\right)}
   z^{p/2 -j-1},\cr
& f^{+}_{2j+1}(z)\sim \frac{1} {\left(\prod_{i=2j+1}^{p-2} 
  \rho_{i}\right)} z^{p/2-j-1}\cr}\eqno(4.49)$$
as $z\to\infty.$  Hence $f_{2j}(P)$ and $f_{2j+1}(P)$ have
poles of order $p/2-j-1$ at $P_{+}.$
\endproclaim

\demo
{Proof} From (4.46), (4,47), and Proposition 4.7, we have
$$f^{+}_{2j}(z)f^{-}_{2j}(z) \sim -\frac{\bar{\alpha}_{2j}}{\alpha_{p-2}},\quad
  f^{+}_{2j+1}(z)f^{-}_{2j+1}(z) \sim \frac{\alpha_{2j}}{\alpha_{p-2}},\eqno(4.50)$$
as $z\to \infty.$  Therefore the assertion follows from (4.50) and
(4.30).
\pf
\enddemo

We next investigate the behaviour of $f^{\pm}_j(z)$ as $z\to 0.$

\proclaim
{Proposition 4.11} For $j=0,\cdots, p/2-1,$ we have
$$\eqalign{& 
f^{-}_{2j}(z) \sim -\bar\alpha_{2j-1}\left(\prod_{i=2j}^{p-2} \rho_{i}\right)
     z^{p/2 -j},\cr
& f^{-}_{2j+1}(z)\sim  \left(\prod_{i=2j+1}^{p-2} \rho_{i}\right) z^{p/2-j-1}\cr}\eqno(4.51)$$
as $z\to 0.$  Hence at $Q_{-},$ $f_{2j}(P)$ has a zero of order $p/2-j,$ while
$f_{2j+1}(P)$ has a zero of order $p/2-j-1.$ 
\endproclaim

\demo
{Proof}  By (4.25), Proposition 4.1 and 4.2, we have
$$\frac{1 -h^{-}(z) \phi_{p-1}(z)}{\psi_{p-1}(z)}\psi_{2j}(z)\sim -
|\alpha_{p-1}|^2\bar\alpha_{2j-1} \left(\prod_{i=2j}^{p-2} \rho_i\right) z^{p/2-j}\eqno(4.52)$$
as $z\to 0.$  On the other hand,
$$h^{-}(z)\phi_{2j}(z) \sim -\bar \alpha_{2j-1} \rho_{p-1}^2\left(\prod_{i=2j}^{p-2} \rho_i\right) z^{p/2-j}.\eqno(4.53)$$
On combining (4.31) and (4.32) and simplify, we obtain the first relation in (4.50).   The other relation
in (4.51) follows in the same way.
\pf
\enddemo

\proclaim
{Proposition 4.12}  For $j=0,\cdots, p/2-1,$ we have
$$\eqalign{
& f^{+}_{2j}(z) \sim \frac{1} {\alpha_{p-1}\left(\prod_{i=2j}^{p-2}\rho_i\right)} z^{-(p/2-j)},\cr
& f^{+}_{2j+1}(z)\sim \frac{\alpha_{2j+1}} {\alpha_{p-1}\left(\prod_{i= 2j+1}^{p-2} \rho_i\right)}z^{-(p/2-j-1)}\cr}
\eqno(4.54)$$
as $z\to 0.$  Hence at $Q_{+},$ $f_{2j}(P)$  has a pole of order $p/2-j$ while $f_{2j+1}(P)$ has a
pole of order $p/2-j-1.$ 
\endproclaim

\demo
{Proof}  From (4.46), (4.47) and Proposition 4.7, we find
$$f^{+}_{2j}(z) f^{-}_{2j}(z) \sim -\frac{\bar\alpha_{2j-1}}{\alpha_{p-1}},\,\, f^{+}_{2j+1}(z) f^{-}_{2j+1}(z)\sim 
\frac{\alpha_{2j+1}}{\alpha_{p-1}}\eqno(4.55)$$
as $z\to 0.$  Therefore the assertion follows from (4.55) and (4.51).
\pf
\enddemo

Combining Propositons 4.5, 4.7, 4.10, 4.11 and 4.12, we obtain the main result
of the section.

\proclaim
{Theorem 4.13}  For $0\leq j\leq p/2 -1,$
$$\eqalign{
 (f_{2j}) \geq  - D  &-  \left({p\over 2} -j-1\right)P_{+} +
  \left({p\over 2} -j-1\right)P_{-}\cr
  & -\left({p\over 2}-j\right)Q_{+} + \left({p\over 2}-j\right)Q_{-},\cr
(f_{2j+1}) \geq  - D & - \left({p\over 2} -j-1\right)P_{+} +
  \left({p\over 2} -j-1\right)P_{-}\cr
   & -\left({p\over 2}-j-1\right)Q_{+} + \left({p\over 2}-j-1\right) Q_{-},\cr}\eqno(4.56)$$
where $D = \sum_{k=1}^{p-1} P_k.$
\endproclaim

\proclaim
{Corollary 4.14}  For $0\leq j \leq p/2 -1,$
$$\eqalign{
&(f_{2j+ p}) \geq - D + (j+1)P_{+} -(j+1)P_{-} + jQ_{+} - jQ_{-},\cr
&(f_{2j+1+p}) \geq - D + (j+1)P_{+} -(j+1)P_{-} + (j+1)Q_{+}  - (j+1)Q_{-}.\cr}\eqno(4.57)$$
\endproclaim

\demo
{Proof}  This follows from (4.51), the relation $f_{k + p} = h^{-1} f_k$ for all $k$
and (4.26).
\pf
\enddemo

To close, we present the following result which is essential in Section 6 below.

\proclaim
{Proposition 4.15 }   For each $0\leq j\leq {p/2} -1,$ the divisors
$$\Cal U_1^j =D + \left( \frac{p}{2} -j-1\right) P_{+}  + \left(\frac{p}{2}-j-1\right)Q_{+} -\left(\frac{p}{2} -j-1\right) P_{-}
-\left(\frac{p}{2} -j-1\right) Q_{-}$$ and
$$\Cal U_2^j=D + \left( \frac{p}{2} -j-1\right) P_{+}  + \left(\frac{p}{2}-j\right)Q_{+} -\left(\frac{p}{2} -j-1\right) P_{-}
-\left(\frac{p}{2} -j\right) Q_{-}$$
are general,  i.e.,
$$\hbox{dim} \,\, L(\Cal U_1^j) = \hbox{dim}\,\, L(\Cal U_2^j) =1\eqno(4.58)
$$ 
where for a divisor ${\Cal U}$  on $C,$
$$L({\Cal U}) = \{\hbox{meromorphic function}\,\,\phi\mid (\phi) \geq -{\Cal U}\,\}.$$
\endproclaim

\demo
{Proof}  We will adapt an argument of \c{MM} to our situation.  For a divisor 
${\Cal U}$ on $C,$ denote by $\Omega({\Cal U})$ the set of meromorphic
$1$-forms $\omega$ on $C$ such that $(\omega)\geq {\Cal U}.$   Take $k$ and $k^{\prime}$ such
that $k + k^{\prime}> g -2 = p-3,$ then  $D + k P_{-}+ k^{\prime} Q_{-}$ has degree $g +k+ k^{\prime} > 2g-2.$   
If $(\omega)\geq D + kP_{-} + k^{\prime} Q_{-},$
then $\omega$ must be $0$ since a holomorphic $1$-form can have at most
$2g-2$ zeros.  Thus $\hbox{dim}\, \Omega(D + kP_{-} + k^{\prime} Q_{-}) =0.$  By Riemann-Roch,
it follows that  $\hbox{dim}\, L(D + kP_{-} + k^{\prime} Q_{-}) = k + k^{\prime} + 1 > g -1.$   For concreteness,
take $k= k^{\prime} = {p\over 2},$  and we claim that
$$ \eqalign{
L(D + (j-1) P_{-} + (j-1) Q_{-}) &\subsetneq L(D + j P_{-} +(j-1)Q_{-})\cr
& \subsetneq L(D + jP_{-} + jQ_{-}), \quad 1\leq j\leq {p\over 2}.\cr}\eqno(4.59)$$
To establish this claim, we just have to observe that  by Corollary 4.14 above,
we have $f_{2(j -2)+1+ p} \in L(D + (j-1)P_{-}+(j-1)Q_{-}),$ but
$f_{2(j-2)+1+ p}\notin L(D + jP_{-} + (j-1) Q_{-}).$  Similarly,
$f_{2(j-1) + p}\in L(D + jP_{-} + (j-1)Q_{-}),$ but
$f_{2(j-1) + p} \notin L(D + jP_{-} + jQ_{-}).$
Thus it follows from $\hbox{dim}\, L(D + {p\over 2} P_{-} + {p\over 2}Q_{-}) = p+1$ and the
claim in (4.58) that $\hbox{dim}\, L(D) =1.$

We next show that $\hbox{dim}\, L(D + Q_{+} - Q_{-}) =1.$  Here we use the fact
that allowing an extra pole increases $\hbox{dim}\, L({\Cal U})$ by at most one.
Thus we have
$$1\leq \hbox{dim}\, L(D + Q_{+}- Q_{-}) \leq \hbox{dim}\, L(D -Q_{-}) +1.\eqno(4.60)$$
But $L(D-Q_{-}) \subsetneq L(D),$ as $f_{p-1} = 1\in L(D)$ but $f_{p-1}\notin L(D-Q_{-}).$
Hence we conclude from $\hbox{dim}\,L(D)=1$ that $\hbox{dim}\, L(D-Q_{-})=0.$
Consequently,  it follows from (4.54) that $\hbox{dim}\, L(D + Q_{+}- Q_{-}) =1.$
Based on this, we can establish $\hbox{dim}\, L(D + Q_{+} -Q_{-} + P_{+}- P_{-}) =1$
from the inequality
$$1\leq \hbox{dim}\, L(D + Q_{+}-Q_{-} +P_{+} -P_{-}) \leq \hbox{dim}\, L(D + Q_{+}-Q_{-}-P_{-}) +1\eqno(4.61)$$
and the observation that $L(D + Q_{+}-Q_{-} -P_{-})\subsetneq L(D + Q_{+} -Q_{-})$.
(This follows because $f_{p-2}\in L(D + Q_{+} -Q_{-})$ but $f_{p-2}\notin L(D+Q_{+} -Q_{-} -P_{-}).$)
Proceed inductively, we have the assertion.

\pf
\enddemo

\bigskip

\subhead
5. \ Action-angle variables
\endsubhead

\bigskip

As we saw in Section 2 above, the periodic Ablowitz-Ladik equation
can be expressed in Lax pair form with Lax operator given by
$\E(h).$   Therefore, the characteristic polynomial 
$\hbox{det}(zI - \E(h))$ is invariant under the Hamiltonian flow
and provides us with a collection of conserved quantities.  From \c{S2},
we have
$$\aligned
\hbox{det}(zI -\E(h)) &= \left(\prod_{j=0}^{p-1} \rho_j\right) z^{p\over 2}[\Delta(z)
   -(h + h^{-1})]\\
  &=\sum_{-{p\over 2}}^{p\over 2} I_{j} z^{j+{p\over 2}} - (h+h^{-1})z^{p\over 2}\left(\prod_{j=0}^{p-1} \rho_j\right)\\
\endaligned
\eqno(5.1)
$$
where the functions $I_j$ as defined in the second line of
(5.1) are such that 
$$I_{p/2} = I_{-{p/2}} =1,\quad \bar I_{j} = I_{-j},\quad j=0,\cdots, 
{p/2} -1.\eqno(5.2)$$
Moreover, they are all polynomials in the  $\alpha_j$'s, their
conjugates, and $P =\prod_{j=0}^{p-1} \rho_j,$ $j=0,\cdots, p-1.$

By using the fact that the collection of $p\times p$ Floquet CMV matrices
is a symplectic leaf of the Sklyanin bracket $\{\cdot, \cdot\}_{J^{\sharp}}$,
we begin by reproving the involution theorem in \c{N} and \c{S2}.

\proclaim
{Theorem 5.1} The functions $P, I_{0}, \hbox{Re}\, I_{j}, \hbox{Im}\,I_{j},
j=1,\cdots, p/2-1$ provide a collection of $p$ conserved quantities in 
involution for the Ablowitz-Ladik equation.
\endproclaim

\demo
{Proof} Write
$$\hbox{det}(zI -\E(h)) = \sum_{r=0}^{p} E_{r}(\E(h))z^{p-r}.\eqno(5.3)$$
Then up to signs, the $E_r$'s are the elementary symmetric functions.
From (5.1) and (5.3), we find that
$$\eqalign{
& I_j(\Cal E) = \oint_{|h|=1} E_{p/2 -j}(\E(h)) {dh\over 2\pi ih},\quad j=0,
\cdots, p/2 -1,\cr
& P(\Cal E) = - \oint_{|h|=1} E_{p/2}(\E(h)) {dh\over 2\pi i}.\cr}\eqno(5.4)$$
Hence the functions $P, I_{0}, \hbox{Re}\, I_{j}, \hbox{Im}\,I_{j},
j=1,\cdots, p/2-1$ are the pullbacks of central functions on $\Gtw$ to 
the $2p$ dimensional dressing orbit consisting of $p\times p$ Floquet CMV matrices.
Consequently, the assertion follows from the abstract involution theorem
in \c{STS2} and Theorem 3.6.  
\pf
\enddemo

\remark 
{Remark 5.2} (a) For readers who are not familiar with the abstract involution
theorem in \c{STS2}, let us remark that its demonstration is a one line
proof making use of the fact that for a central function, its left
gradient is the same as its right gradient.  (See the expression for
the Poisson bracket $\{\cdot,\cdot\}_{J^{\sharp}}$ in (3.16) above for our
case.)
\newline
(b)  Note that we are using the symbol $\Cal E$ as a shorthand for the unitary loop $\Cal E(\cdot)$
in (5.4) above and we will henceforth continue to use the symbol with this meaning.   Since we will not
be using the extended CMV matrices in what follows, this should not cause any confusion.

\endremark
\smallskip

Thus the number of commuting integrals as provided by the quantities
in the above theorem is exactly equal to one half the dimension of
the phase space. In the rest of the section, we will construct the variables
(essentially) conjugate to these actions.   As in Section 4, we denote by
$C$ the hyperelliptic Riemann surface of genus $g=p-1$ corresponding
to the affine curve $I(h,z) = h\,\hbox{det}(zI -\Cal E(h))=0.$    On $C,$
we introduce the holomorphic $1$-forms
$$\xi_{k} = \frac{z^{k-1}}{\frac{\partial I}{\partial h}} \, dz,\,\, k=1,\cdots, g=p-1.\eqno(5.5)$$
We also introduce the meromorphic $1$-form
$$\xi^{m} = (h + h^{-1})\frac{z^{p/2-1}}{\frac{\partial I}{\partial h}}\,dz \eqno(5.6)$$
with poles at $P_{\pm}$ and $Q_{\pm}.$   Pick a fix point $P_0$ on the finite part of $C$ and put $D_0 = g P_0.$
Then for $\Cal E(h)$ satisfying the genericity assumptions $(GA)_1$ and $(GA)_2,$
we define
$$\eqalign{
& \phi_k (\Cal E) = \int_{D_0}^{D} \xi_k,\,\,\, k=1, \cdots, g,\cr
& \nu (\Cal E) = \int_{D_0}^{D}  \xi^{m},\cr}
\eqno(5.7)$$
where $D = \sum_{k=1}^{g} P_k$ is the divisor of poles of $f_j, 0\leq j\leq g-1$ in the finite part
of $C.$ 
Note that in the definition of $\nu (\Cal E),$ the paths of integration going from the points
of $D_0$ to the points of $D$ must avoid the points $P_{\pm}.$   These multi-valued variables
are well defined because the points in $D_0$ and $D$ are in the finite part of $C.$
On the other hand, the multi-valuedness can be resolved in the standard way and we
will not try to get into the details here. (See, for example, \c{DLT} and \c{L2}.)
To compute the Poisson brackets between the conserved quantities in Theorem 5.1
and the variables in (5.7), we will make use of a device in \c{DLT} (which has also proved
to be successful in \c{L2}) which will allow us to simplify the calculation.   In the following,
we will deal with $\Cal E(h)$ for $h$ not necessarily on the unit circle.   Note that in
this general case, we have the relation
$$\Cal E(h) \Cal E(\bar h^{-1})^{*} = \Cal E (\bar h^{-1})^{*} \Cal E(h) = I,\,\, h\in \Bbb C\setminus \{0\}
\eqno(5.8)$$   
which can be checked by using the fact that $\Cal E(h)$ is unitary for $h\in \partial\Bbb D.$
For our purpose, we pick a fixed $h_0\in (-1, 1)\setminus\{0\},$ $z_0\in \partial \Bbb D$  such that
$(h_0, z_0)$ is not on $C$ and define
$$\eqalign{&
H_{h_0,z_0} (\Cal E) = \hbox{Re}\, \hbox{log} \,\hbox{det}\, (z_0 I - \Cal E(h_0)),\cr
& J_{h_0, z_0} (\Cal E) = \hbox{Im}\, \hbox{log} \, \hbox{det} \, (z_0 I - \Cal E(h_0)).\cr}\eqno(5.9)$$
As the reader will see in the calculation which follow, this choice of $h_0$ and $z_0$ is critical.

\proclaim
{Lemma 5.3}  (a)  The Hamiltonian equation generated by $H_{h_0, z_0}(\Cal E)$ is given by the
equation
$$ \dot \Cal E(h) = [\,\Cal E(h), B(h)\,] \eqno(5.10)$$
where
$$\eqalign{
B(h) = & \left( i(z_0 I -\Cal E(h_0))^{-1} \Cal E(h_0) \right)_{-} + i\,\hbox{Im}\, \left(i(z_0 I - \Cal E(h_0))^{-1}\Cal E(h_0)\right)_{0}\cr
& -\left[\left(i(z_0 - \Cal E(h_0))^{-1}\Cal E(h_0)\right)_{-}\right]^{*} + iz_0(z_0 I -\Cal E(h_0^{-1}))^{-1} \cdot 
\frac{1}{1-hh_0}.\cr}\eqno(5.11)$$
\newline
(b) The Hamiltonian equation generated by $J_{h_0,z_0}(\Cal E)$ is given by the equation
$$ \dot \Cal E(h) = [\, \Cal E(h), C(h)\,]\eqno(5.12)$$
where 
$$\eqalign{
C(h) = & \left((z_0 I - \Cal E(h_0))^{-1}\Cal E(h_0)\right)_{-} + i\,\hbox{Im}\, \left((z_0 I - \Cal E(h_0))^{-1}\Cal
E(h_0)\right)_{0}\cr
& -\left[\left((z_0 I -\Cal(h_0))^{-1}\Cal E(h_0)\right)_{-}\right]^{*} -z_0 (z_0 I - \Cal E(h_0^{-1}))^{-1} \cdot
\frac{1}{1-hh_0}.\cr}\eqno(5.13)$$
\endproclaim

\demo
{Proof} (a) If we let $\phi(g) = \hbox{Re}\,\hbox{log}\,\hbox{det} (z_0 I - g(h_0))$ for $g\in \widetilde G^{\Bbb R}_{w},$
then from the second equation in (3.60) and a direct calculation, we find that the Hamiltonian equation
generated by $H_{h_0,z_0}$ is given by (5.10), where
$$\eqalign{
B(h) =  & \Pi_{\widetilde{\frak b}_{w}} {\frac{ i(z_0 I - \Cal E(h_0))^{-1} \Cal E(h_0) h} {h-h_0}} \cr
        =   & (i(z_0 I -\Cal E(h_0))^{-1}\Cal E(h_0))_{0} + i \hbox{Im} \left(i(z_0 I -\Cal E(h_0))^{-1}\Cal E(h_0)\right)_{0}\cr
            & - \left[\left( i(z_0 I - \Cal E(h_0))^{-1}\Cal E(h_0)\right)_{-}\right]^{*}
                +\left( i(z_0 I -\Cal E(h_0))^{-1}\Cal E(h_0)\cdot \frac{h}{h-h_0}\right)^{*}.\cr}\eqno(5.14)$$
Note that in going from the first line of (5.14) to the second line, we have used (3.12), together with
the formula for $\Pi_{\frak b}$ from \c{L1}.   In the next step of the calculation, we will try to rewrite the
last term in the above expression in the desired form, and it is here that the choice of $h_0$ and
$z_0$ is important. To wit, by using (5.8), we have
$$\eqalign{
& \left( i(z_0 I -\Cal E(h_0))^{-1}\Cal E(h_0)\cdot \frac{h}{h-h_0}\right)^{*} \cr
= & \left( i(z_0\Cal E(h_0^{-1})^{*} - I)^{-1} \cdot \frac{h}{h-h_0}\right)^{*}\cr
= & -i(\bar z_0\Cal E(h_0^{-1}) - I)^{-1}\cdot \frac{\bar h}{\bar h -h_0}\cr
= & \, iz_0(z_0 I - \Cal E(h_0^{-1}))^{-1}\cdot \frac{1}{1-hh_0}.\cr}\eqno(5.15)$$
Hence (5.11) follows from (5.14) and (5.15).
\newline
(b) The proof is similar to (a) and so we will skip the details.

\pf
\enddemo

In our next two results, we will denote the Poisson bracket on the set of $p\times p$ Floquet
CMV matrices induced from $\{\cdot, \cdot\}_{J^{\sharp}}$ simply by $\{\cdot, \cdot\}.$

\proclaim
{Proposition 5.4}  For $j,k =1,\cdots, p/2-1,$ we have the following Poisson bracket relations
\newline
(a) $\left\{I_0, -\hbox{Im}\left(\frac{\phi_{p/2}}{2}\right)\right\} (\Cal E) =1, \left\{I_0, -\hbox{Im}\left(\frac{\phi_k+\phi_{p-k}}{2}\right)\right\} (\Cal E) =0,$
\newline 
$\phantom{abc}$ $\left \{I_0, -\hbox{Re}\left(\frac{\phi_k-\phi_{p-k}}{2}\right)\right\} (\Cal E)=0;$
\newline
(b) $\left\{\hbox{Re}\,I_j, -\hbox{Im}\left(\frac{\phi_{p/2}}{2}\right)\right\} (\Cal E)=0,\,\,\, \left\{\hbox{Re}\,I_j, -\hbox{Im}\left(\frac{\phi_k+\phi_{p-k}}{2}\right)\right\} (\Cal E)  = \delta_{j,p/2-k}, $
\newline
$\phantom{abc}$ $\left\{\hbox{Re} \,I_j, -\hbox{Re}\,\left(\frac{\phi_k-\phi_{p-k}}{2}\right)\right\}(\Cal E) =0;$
\newline
(c) $\left\{\hbox{Im}\,I_j, -\hbox{Im}\left(\frac{\phi_{p/2}}{2}\right)\right\} (\Cal E)=0, \left\{\hbox{Im} \,I_j, -\hbox{Im}\left(\frac{\phi_k+\phi_{p-k}}{2}\right)\right\} (\Cal E)= 0,$
\newline
$\phantom{abc}$ $\left\{\hbox{Im}\, I_j, -\hbox{Re}\left(\frac{\phi_k-\phi_{p-k}}{2}\right)\right\} (\Cal E)=\delta_{j,p/2-k};$
 \newline
(d) $\left\{P, -\hbox{Im}\left(\frac{\phi_{p/2}}{2}\right)\right\}(\Cal E) =0,$ $\left\{P, -\hbox{Im}\left(\frac{\phi_k+\phi_{p-k}}{2}\right)\right\}(\Cal E) =0,$
\newline
$\phantom{abc}$    $\left\{P, -\hbox{Re}\left(\frac{\phi_k-\phi_{p-k}}{2}\right)\right\} (\Cal E)=0.$

\endproclaim

\demo
{Proof}   In order to compute $\{\, H_{h_0,z_0}, \phi_k\,\}(\Cal E),$ it suffices to evaluate it on an
open dense set consisting of Floquet CMV matrices $\Cal E(h)$ for which 
\smallskip
\noindent (a) the points $P_j$ of the divisor $D$ are distinct,
\newline
(b) $\hbox{supp}\, D \cap \{ \frac{\partial I}{\partial h} =0\,\} =\emptyset,$
\newline
(c)  $\{Q_j\}_{j=1}^{p}\cap \,\,\hbox{supp}\, D = \emptyset,$  where $Q_j = (h_0^{-1}, z_j(h_0))\in C,$
$j=1,\cdots, p,$
\newline
(d) $\{Q_j\}_{j=1}^{p} \cap \{\frac{\partial I}{\partial h} =0\,\} =\emptyset.$
\smallskip
So suppose $\Cal E(h)$ satisfies (a)-(d) above.
Then in  the neighborhood of each $P_j,$ we can
take $z$ to be the local coordinate and express $h$ in terms of $z.$  Thus in particular, 
$h_j = h(z_j).$  Let $\Cal E(t)$ be the Hamiltonian flow
generated by $H_{h_0,z_0}$ and let $D(t) =\sum_{j=1}^{p-1} P_j(t)$ (where $t$
is small)  be the divisor of poles in the finite part of $C$ of the corresponding eigenvector 
with last component normalized to $1,$    $P_j(t) = (h_j(t), z_j(t)).$   Then
$$\eqalign{
\{\, H_{h_0,z_0}, \phi_k\,\}(\Cal E) = & {d\over dt}{\Big|_{\mid t =0}} \phi_k(\Cal E(t))\cr
= & \sum_{j=1}^{p-1} \frac{z_j^{k-1}}{\frac{\partial I}{\partial h}(h_j, z_j)}\cdot \frac{dz_j(t)}{dt}\Big|_{\mid t=0}.\cr}
\eqno(5.16)$$
To compute the rate of change of $z_j(t)$ at $t=0,$  consider an eigenvector $f(z,t)$  for $z$ in a neighborhood
of $z_j$ such that $(e_{p-1}, f(z_j(t), t)) =0$ for small values of $t.$  Differentiate this relation with respect
to $t$ at $t=0,$ we obtain
$$\frac{dz_j(t)}{dt}\Big|_{t=0} = \frac{(e_{p-1}, B(h_j) f(z_j,0))}{(e_{p-1}, \frac{\partial f}{\partial z}(z_j,0))}.$$
Therefore, on substituting this expression into (5.16), we find
$$\{\, H_{h_0,z_0}, \phi_k\,\}(\Cal E) = \sum_{j=1}^{p-1} \frac{z_j^{k-1}}{\frac{\partial I}{\partial h}(h_j,z_j)}\cdot
\frac{(e_{p-1}, B(h_j) f(z_j,0))}{(e_{p-1}, \frac{\partial f}{\partial z}(z_j,0))}.
\eqno(5.17)$$
Now, on using the fact that the last column of a strictly lower triangular matrix is the zero vector, and
$(e_{p-1}, \hbox{Re} \big(i(z_0 I -\Cal E(h_0))^{-1}\Cal E(h_0))_{0} f(z_j,0)\big) =0,$ it follows from (5.11) that
$$\eqalign{&
(e_{p-1}, B(h_j) f(z_j,0)) \cr
= & \left(e_{p-1}, \left(i\big(z_0 I -\Cal E(h_0)\big)^{-1}\Cal E(h_0) + iz_0\big(z_0 I -\Cal E(h_0^{-1})\big)^{-1}
\cdot \frac{1}{1-h_jh}\right) f(z_j,0)\right).\cr}\eqno(5.18)$$
Consequently, when we substitute (5.18) into (5.17), the result is
$$\{\, H_{h_0,z_0}, \phi_k\,\}(\Cal E)= i \sum_{j=1}^{p-1} \hbox{Res}_{P_j} F_k + 
i \sum_{j=1}^{p-1} \hbox{Res}_{P_j}  G_k,\eqno(5.19)$$
where 
$$\eqalign{&
F_k = \big(e_{p-1}, (z_0 I -\Cal E(h_0))^{-1}\Cal E(h_0)\widehat v(z,0)\big) \xi_k,\cr
& G_k =\big(e_{p-1}, z_0(z_0 I -\Cal E(h_0^{-1}))^{-1} \widehat v(z,0)\big)\frac{\xi_k}{1-h_0h}\cr}\eqno(5.20)$$
are meromorphic $1$-forms on $C.$  By a similar calculation, we also have
$$\{\,J_{h_0,z_0}, \phi_k\,\}(\Cal E) = \sum_{j=1}^{p-1} \hbox{Res}_{P_j} F_k 
- \sum_{j=1}^{p-1} \hbox{Res}_{P_j} G_k.\eqno(5.21)$$
Now from \c{S2}, we know that for $|z_0| =1, \Delta(z_0)\in \Bbb R,$ and so this implies $J_{h_0,z_0}$ is a constant
independent of $\Cal E(h).$   Consequently, $\{\,J_{h_0,z_0}, \phi_k\,\}(\Cal E) =0$ and hence
it follows from (5.21) that $\sum_{j=1}^{p-1} \hbox{Res}_{P_j} F_k = \sum_{j=1}^{p-1} \hbox{Res}_{P_j} G_k.$
Consequently, 
$$\{\, H_{h_0,z_0}, \phi_k\,\}(\Cal E)= 2 i \sum_{j=1}^{p-1} \hbox{Res}_{P_j}  G_k.\eqno(5.22)$$
Consider the meromorphic $1$-form $G_k.$   Since $(\frac{\partial I}{\partial h})^{\pm} \sim \mp 1$
as $z\to 0,$ it follows that $\xi^{\pm}_k\sim \mp z^{k-1}\,dz$ as $z\to 0.$  On the other hand, we
have 
$$\frac{1}{1- h_0h^{+}}\sim -\frac{P}{h_0} z^{p/2},\,\,\, \frac{1}{1- h_0h^{-}}\sim 1\quad \hbox{as}\,\, z\to 0.
\eqno(5.23)$$
Since the $f_j$'s are analytic at $Q_{-},$ it follows that $G_k$ has no poles at $Q_{-}.$   At $Q_{+},$
the most singular component of $\widehat v(z,0)$ is 
$f^{+}_0(z) \sim \frac{1} {\alpha_{p-1}\left(\prod_{i=0}^{p-2}\rho_i\right)} z^{-p/2}.$
Hence it follows from the asymptotics of $\xi^{+}_k,$ $f^{+}_0(z)$ and (5.23) that
$$G^{+}_k\sim \hbox{const.}\, z^{k-1}\quad \hbox{as}\,\, z\to 0.\eqno(5.24)$$
So again $G_k$ has no poles at $Q_{+}.$     Thus  $G_{k}$ has poles only at $D$ and
at $Q_j,$ $j=1,\cdots, p$  as we can similarly check that there are no poles at $P_{\pm}.$ 
 Hence by the residue theorem and (5.22),
$$\{\, H_{h_0,z_0}, \phi_k\,\}(\Cal E)=  - 2i \sum_{j=1}^{p} \hbox{Res}_{Q_j}  G_k .\eqno(5.25)$$
To simplify notation,  let
$b_j^{(k)} = \hbox{Res}_{Q_j}  G_k$ and put $b^{(k)} = \sum_{j=1}^{p} b_j^{(k)}.$
We first calculate $b_j^{(k)}.$    We have
$$\eqalign{
b^{(k)}_j = & -\,\frac{1}{h_0}\cdot\frac{z_0}{z_0-z_j(h_0)}\cdot\frac{z_j(h_0)^{k-1}}{\frac{\partial I}{\partial h}\bigl(h_0^{-1},z_j(h_0)\bigr)}\cr
= & \frac{1}{h_0}\cdot\frac{z_0}{z_0-z_j(h_0)}\cdot\frac{z_j(h_0)^{k-1}}{\frac{\partial I}{\partial z}\bigl(h_0^{-1},z_j(h_0)\bigr)}.\cr}\eqno(5.26)$$
Therefore,
$$\eqalign{
b^{(k)}
&=\sum_{j=1}^p \frac{1}{h_0}\cdot\frac{z_0}{z_0-z_j(h_0)}\cdot\frac{z_j(h_0)^{k-1}}{\frac{\partial I}{\partial z}\bigl(h_0^{-1},z_j(h_0)\bigr)}\cr
&=\frac{z_0}{h_0}\left[\lim_{R\to\infty}\int_{|z|=R} \frac{z^{k-1}}{(z_0-z)\cdot I(h_0^{-1},z)}\,\frac{dz}{2\pi i}+\frac{z_0^{k-1}}{I(h_0^{-1},z_0)}\right]\cr
&=\frac{z_0^k}{h_0 I(h_0^{-1},z_0)} = \frac{z_0^k}{\hbox{det}\,(z_0 I -\Cal E(h_0))}\cr}\eqno(5.27)$$
and so finally we conclude that
$$\{\, H_{h_0,z_0}, \phi_k\,\}(\Cal E) = - 2 i b^{(k)} = \frac{-2iz_0^k}{\hbox{det}\,(z_0 I -\Cal E(h_0))}.\eqno(5.28)$$
But on the other hand, it follows from $\{\, J_{h_0,z_0}, \phi_k\,\}(\Cal E)=0$ and (5.1) that
$$\eqalign{&
\{\,H_{h_0,z_0}, \phi_k\,\} (\Cal E) \cr= & \{\,H_{h_0,z_0} + i J_{h_0,z_0}, \phi_k\,\}(\Cal E)\cr
= & \frac{1}{\hbox{det}\,(z_0 I -\Cal E(h_0))}\left( \sum_{j = -p/2}^{p/2} \{I_j, \phi_k\,\}(\Cal E) z_0^{j+ p/2}
-z_0^{p/2} (h_0 + h_0^{-1})\{P,\phi_k\}(\Cal E)\right).\cr}\eqno(5.29)$$
By equating (5.28) and (5.29), we conclude that
$$-2iz_0^k = \sum_{j =p/2}^{p/2} \{I_j, \phi_k\,\}(\Cal E) z_0^{j+ p/2}
-z_0^{p/2} (h_0 + h_0^{-1})\{P,\phi_k\}(\Cal E), \,\,1\leq k\leq p-1.\eqno(5.30)$$
We now divide into three cases.
\smallskip
\noindent Case 1: $k\geq p/2 +1$ 
\smallskip
In this case, all  brackets are zero except for $\{\,I_{k-p/2}, \phi_k\,\}(\Cal E) = -2i,$ i.e.,
$$\eqalign{&
\{\, I_j, \phi_k\,\}(\Cal E) = -2i \delta_{j, k-p/2}, \,\,\{\,\bar I_j, \phi_k\,\}(\Cal E) =0, 1\leq j\leq p/2-1,\cr
& \{\, P, \phi_k\,\}(\Cal E) =0,\,\, \{\,I_0,\phi_k\,\}(\Cal E) =0.\cr}\eqno(5.31)$$
Hence
$$\eqalign{&
\{\,\hbox{Re}\, I_j, \phi_k\,\}(\Cal E) = -i\delta_{j,k-p/2}, \,\, \{\,\hbox{Im}\, I_j, \phi_k\,\}(\Cal E) = -\delta_{j,k-p/2},
\,1\leq j\leq p/2-1,\cr
&\{\,P, \phi_k\,\}(\Cal E) =0,\,\,\{I_0, \phi_k\,\}(\Cal E)=0.\cr}\eqno(5.32)$$
\smallskip
\noindent Case 2: $k\leq p/2-1$
\smallskip
In this case, we have
$$\eqalign{&
\{\, \bar I_j, \phi_k\,\}(\Cal E) = -2i\delta_{j, p/2-k}, \,\, \{\, I_j, \phi_k\,\}(\Cal E) = 0,\,1\leq j\leq p/2-1,\cr
& \{\,P, \phi_k\,\}(\Cal E)=0,\,\,\{\,I_0, \phi_k\,\}(\Cal E)=0,\cr}\eqno(5.33)$$
which implies
$$\eqalign{&
\{\,\hbox{Re}\, I_j, \phi_k\,\}(\Cal E) = -i\delta_{j,p/2-k},\,\, \{\,\hbox{Im}\,I_j, \phi_k\,\}(\Cal E) = \delta_{j, p/2-k},
\, 1\leq j\leq p/2-1,\cr
& \{\,P,\phi_k\,\}(\Cal E) =0,\,\, \{\,I_0,\phi_k\,\}(\Cal E)=0.\cr}\eqno(5.34)$$
\noindent Case 3: $k=p/2$
\smallskip
In this case, all brackets are zero except for 
$$-2i = \{\, I_0, \phi_{p/2}\,\}(\Cal E) - (h_0 + h_0^{-1})\{\, P, \phi_{p/2}\,\}(\Cal E).\eqno(5.35)$$
Therefore,
$$\eqalign{&
\{\,\hbox{Re}\,I_j, \phi_{p/2}\,\}(\Cal E) =0, \,\, \{\,\hbox{Im}\, I_j, \phi_{p/2}\,\}(\Cal E) =0,\,1\leq j\leq p/2-1,\cr
& \{\,P, \phi_{p/2}\,\}(\Cal E) = 0, \,\, \{\,I_0, \phi_{p/2}\,\}(\Cal E) = -2i.\cr}\eqno(5.36)$$
\pf
\enddemo

\proclaim
{Proposition 5.5} With $\nu$ defined as in relation (5.7), we have the following Poisson bracket relation:
$$
\left\{P,\text{Im}\left(\frac{\nu}{4}\right)\right\}(\Cal E)=1\,.
$$
\endproclaim

\demo
{Proof}  It suffices to compute the Poisson brackets 
$\{\, H_{h_0,z_0}, \nu\,\}(\Cal E)$  and $\{\, J_{h_0,z_0}, \nu\,\}(\Cal E)$  on an
open dense set consisting of Floquet CMV matrices $\Cal E(h)$ which satisfy
conditions (a)-(d) in Proposition 5.4.   Indeed, by following the same method of calculation, we find
$$\{\, H_{h_0,z_0}, \nu\,\}(\Cal E)= i \sum_{j=1}^{p-1} \hbox{Res}_{P_j} F + 
i \sum_{j=1}^{p-1} \hbox{Res}_{P_j}  G,\eqno(5.37)$$
and
$$\{\,J_{h_0,z_0}, \nu\,\}(\Cal E) = \sum_{j=1}^{p-1} \hbox{Res}_{P_j} F -
\sum_{j=1}^{p-1} \hbox{Res}_{P_j} G =0,\eqno(5.38)$$
where 
$$\eqalign{&
F = \Big(e_{p-1}, \big(z_0 I -\Cal E(h_0)\big)^{-1}\Cal E(h_0)\widehat v(z,0)\Big) \xi^{m},\cr
& G = \Big(e_{p-1}, z_0\big(z_0 I -\Cal E(h_0^{-1})\big)^{-1} \widehat v(z,0)\Big)\frac{\xi^m}{1-h_0h}.\cr}\eqno(5.39)$$
Thus 
$$\{\, H_{h_0,z_0}, \nu\,\}(\Cal E)= 2 i \sum_{j=1}^{p-1} \hbox{Res}_{P_j}  G.\eqno(5.40)$$
Consider the meromorphic $1$-form $G.$   As $z\to 0,$  it follows from the asymptotics of $h^{\pm}$
(see Section 4) and $(\frac{\partial I}{\partial h})^{\pm}$ that
$$(\xi^{m})^{\pm} \sim \mp \frac{1}{P z}\, dz.\eqno(5.41)$$
Since the $f_j$'s are analytic at $Q_{-},$ it follows from (5.23) and (5.41) that $G$ has a simple
pole at $Q_{-}$  Indeed, it follows from Propositon 4.11 that 
$$\hbox{Res}_{Q_{-}} G =  \frac{z_0 \Big(\big(z_0 I -\Cal E(h_0^{-1})\big)^{-1}\Big)_{p-1,p-1}}{P}.\eqno(5.42)$$
In a similar way,  it follows from (5.23), (5.41) and Proposition 4.12 that $G$ also has a simple
pole at $Q_{+}$ and
$$\hbox{Res}_{Q_{+}} G =  \frac{z_0 \Big(\big(z_0 I -\Cal E(h_0^{-1})\big)^{-1}\Big)_{p-1,0}}{P}\cdot\frac{\rho_{p-1}}{h_0\alpha_{p-1}}.
\eqno(5.43)$$
Now consider the two points $P_{\pm}$ at infinity and let $u = \frac{1}{z}$ be the local coordinate.
Then from (5.23), (5.41) and Proposition 4.10, we have
$$f^{+}_{2j}(z,0) \frac{(\xi^{m})^+}{1-h_0h^+}\sim \frac{\bar \alpha_{2j}}{h_0 \prod_{i=2j}^{p-2} \rho_i} u^{j}\,du
\quad \hbox{as}\,\,\, u\to 0,\eqno(5.44)$$
while
$$f^{+}_{2j+1}(z,0) \frac{(\xi^{m})^+}{1-h_0h^+}\sim - \frac{1}{h_0 \prod_{i=2j+1}^{p-2} \rho_i} u^{j}\,du
\quad \hbox{as}\,\,\, u\to 0.\eqno(5.45)$$
From (5.44) and (5.45), we conclude that $G$ is analytic at $P_{+}.$    Similarly, by making use
of Proposition 4.7 and (5.23), (5.41), we find that $G$ has a simple pole at $P_{-}$ with
$$\eqalign{
\hbox{Res}_{P_{-}}\, G = &\frac{z_0 \Big(\big(z_0 I -\Cal E(h_0^{-1})\big)^{-1}\Big)_{p-1,p-2}}{P}\cdot \frac{\rho_{p-2}}{\alpha_{p-2}}\cr
 & - \frac{z_0 \Big(\big(z_0 I -\Cal E(h_0^{-1})\big)^{-1}\Big)_{p-1,p-1}}{P}\,.\cr}\eqno(5.46)$$
 Since $G$ obviously has poles at the points of $D$ and at the points $Q_j,$ $j=1,\cdots, p,$ it follows
 by the residue theorem, (5.40), (5.42), (5.43) and (5.46) that
 $$\eqalign{
 \{\, H_{h_0,z_0}, \nu\,\}(\Cal E)= & -2 i \sum_{j=1}^{p} \hbox{Res}_{Q_j}  G -{2i z_0\over P}\Big[
 \Big(\big(z_0I -\Cal E(h_0^{-1})\big)^{-1}\Big)_{p-1,0}\cdot \frac{\rho_{p-1}}{h_0\alpha_{p-1}} \cr
 & + \Big(\big(z_0I-\Cal E (h_0^{-1})\big)^{-1}\Big)_{p-1,p-2}\cdot \frac{\rho_{p-2}}{\alpha_{p-2}}\Big].\cr}\eqno(5.47)$$
 
To compute the second term on the right-hand side of (5.47), we introduce the following ad-hoc notation: if $A$
is a $p\times p$ matrix and $0\leq j_1\leq \cdots\leq j_l\leq p-1$, $0\leq k_1\leq\cdots\leq k_l\leq p-1$ are two sets of indices, with $l\geq1$, then we denote
by $A[j_1,\ldots,j_l;k_1,\ldots,k_l]$ the submatrix obtained from $A$ by deleting rows $j_1,...,j_l$ and columns $k_1,\ldots,k_l$. For simplicity of notation, let
$$
M=z_0 I-\Cal E(h_0^{-1})\,.
$$
Then the terms appearing on the right-hand side of (5.47) can be identified as
$$
\big(M^{-1}\big)_{p-1,0}=\frac{\text{det}(M[0,p-1])}{\text{det} (M)}\quad\text{and}\quad \big(M^{-1}\big)_{p-1,p-2}=\frac{\text{det}(M[p-2,p-1])}{\text{det} (M)}\,.
$$
To proceed, we expand both $\text{det}(M[0,p-1])$ and $\text{det}(M[p-2,p-1])$ along their 0th columns, which makes all the minors appearing the calculation $h_0$-independent, and so allows us to separate the $h_0$-dependent terms from the $h_0$-independent ones. This leads to
$$
\frac{\rho_{p-1}}{h_0\alpha_{p-1}}\cdot \big(M^{-1}\big)_{p-1,0}+\frac{\rho_{p-2}}{\alpha_{p-2}}\cdot \big(M^{-1}\big)_{p-1,p-2}
=m_1+m_2\,,\eqno(5.48)
$$
where the $h_0$-dependent term is
$$\eqalign{
m_1=\frac{1}{\text{det}(M)}\bigg(&\frac{1}{h_0}\cdot\rho_0\rho_{p-1}\text{det}\big(M[0,1;0,p-1]\big)\cr
&\quad+h_0\cdot \rho_{p-2}\rho_{p-1}\text{det}\big(M[p-2,p-1;0,p-1]\big)\bigg)\cr}\eqno(5.49)
$$
and $m_2$ is $h_0$-independent. While the minors appearing in $m_2$ have a structure which cannot be easily simplified, the minors appearing
in $m_1$ can be computed explicitely.

Indeed, for each $0\leq j\leq \frac{p}{2}-1$, consider the $2\times2$ blocks
$$
A_j=
\pmatrix
-\rho_{2j-1}\bar\alpha_{2j} & z_0+\alpha_{2j-1}\bar\alpha_{2j}\\
-\rho_{2j-1}\rho_{2j} & \alpha_{2j-1}\rho_{2j}
\endpmatrix\quad\text{and}\quad
\tilde A_j=
\pmatrix
-\rho_{2j}\bar\alpha_{2j+1} & -\rho_{2j}\rho_{2j+1}\\
z_0+\alpha_{2j}\bar\alpha_{2j+1} & \alpha_{2j}\rho_{2j+1}
\endpmatrix\,.
$$ 
Note $A_j$ and $\tilde A_j$ are, respectively, the left and right ``halves'' of the $2\times4$ blocks appearing in the extended matrix $z_0 I-\Cal E$.
In particular, direct investigation shows that $M[0,1;0,p-1]$ is a block bi-diagonal matrix, having the blocks $A_1,\ldots,A_{\frac{p}{2}-1}$ on the diagonal and
$\tilde A_1,\ldots,\tilde A_{\frac{p}{2}-2}$ on the upper diagonal, while $M[p-2,p-1;0,p-1]$ has $\tilde A_0,\ldots,\tilde A_{\frac{p}{2}-2}$ on the diagonal and $A_1,\ldots,A_{\frac{p}{2}-2}$ on the lower diagonal. This implies that
$$
\text{det}\big(M[0,1;0,p-1]\big)=\prod_{j=1}^{\frac{p}{2}-1} \text{det}\big(A_j\big)=\prod_{j=1}^{\frac{p}{2}-1} z_0\rho_{2j-1}\rho_{2j}
=z_0^{\frac{p}{2}-1}\prod_{j=1}^{p-2}
\rho_j\,,
$$ 
and
$$
\text{det}\big(M[p-2,p-1;0,p-1]\big)=\prod_{j=0}^{\frac{p}{2}-2} \text{det}\big(\tilde A_j\big)=\prod_{j=0}^{\frac{p}{2}-2} z_0\rho_{2j}\rho_{2j+1} 
=z_0^{\frac{p}{2}-1}\prod_{j=0}^{p-3}\rho_j\,.
$$
Plugging these two expressions in (5.49) leads to a very simple expression:
$$
m_1=\frac{z_0^{\frac{p}{2}-1}\cdot P}{\text{det}\big(z_0 I-\Cal E(h_0^{-1})\big)}\cdot\Big(h_0+\frac{1}{h_0}\Big)\,.\eqno(5.50)
$$

On the other hand, proceeding exactly as in the proof of Proposition 5.4, we find that
$$
\sum_{j=1}^{p} \hbox{Res}_{Q_j}  G= 
\frac{z_0^{p/2}\big(h_0+\frac{1}{h_0}\big)}{\text{det}\big(z_0I-\Cal E(h_0)\big)}\,.\eqno(5.51)
$$
By combining (5.48), (5.50), and (5.51) into (5.47), and using (5.38), we conclude that
$$
\{H_{h_0,z_0}+iJ_{h_0,z_0},\nu\}(\Cal E)
=\frac{1}{\text{det}\big(z_0I-\Cal E(h_0)\big)}\Big[-4iz_0^{p/2}\big(h_0+\frac{1}{h_0}\big)
-2i\frac{z_0}{P}\cdot m_2\Big]\,.
$$
By the analogue for $\nu$ of relation (5.29), together with the fact that $m_2$ is $h_0$-independent, we conclude that
$$
\{P,\nu\}(\Cal E)=4i\,,
$$
which leads directly to our claim.

\pf
\enddemo

\remark
{Remark 5.6} (a) In the Proposition above, we were unable to obtain the Poisson brackets between the $I_j$'s 
and $\nu.$   Nevertheless, the functional independence of the integrals follows by combining Proposition 5.4
and Proposition 5.5.  
\newline
(b) Proposition 5.4 shows that the Hamiltonian flow generated by $P$ does not give rise to nontrivial
motion on the Jacobi variety of $C,$ this is the reason for the introduction of the algebro-geometric
variable $\nu.$ (See also Proposition 6.2 below in this connection.)  Note that in particular, 
this means that the periodic defocusing Ablowitz-Ladik equation is not an 
algebraically completely integrable system in the sense of Adler and
van Moerbeke \c{AMV}.   For other examples of integrable systems involving
spectral curves which are not algebraically integrable, we refer the reader
to \c{DLT} and \c{L2}.
\newline
\endremark

\bigskip
\bigskip

\subhead
6. \ Solving the equations via factorization problems
\endsubhead

\bigskip

In this section, we will solve the Hamiltonian equations of motion generated
by the commuting integrals in Section 5 via factorization problems.

We begin by writing down the Hamiltonian equations of motion by using
Proposition 3.9, Theorem 3.6 and (5.4).  To do so, for a map 
$F:gl(p,\Bbb C)\longrightarrow \Bbb C,$  we let $\nabla F(M) = (\partial F/\partial m_{ij})$
denote its gradient.   In what follows, we will regard $\hbox{Re}\,\,I_j,$ 
$\hbox{Im}\,\,I_j$ and $P$ as functions of $g^e$, $g^o,$  where
$\E(h) = g^eg^o(h),$ and $\tilde\E(h) =g^o(h) g^e.$  We will use the notation
of Proposition 3.9 (b).  As an example, if $H = \hbox{Im}\,\, I_{p/2-j},$  then the
corresponding central function $\varphi$ is given by 
$\varphi(X) = \hbox{Im}\,\,\oint_{|h|=1} E_{j}(X(h)) \frac{dh}{2\pi i h}.$

\proclaim
{Proposition 6.1} The Hamiltonian equations of motion generated by 
$H =\hbox{Im}\, I_{p/2-j},$ $\hbox{Re}\,I_{p/2-j},$ and $P$ are given by
$$\eqalign{
&\dot g^e = (\pibtw(D\varphi(\E(h))))g^e -g^e(\piktw(D\varphi(\tilde\E(h)))),\cr
&\dot g^o(h) = (\pibtw(D\varphi(\tilde\E(h))))g^o(h)-g^o(h)(\pibtw(D\varphi(\E(h)))),\cr}
\eqno(6.1)
$$
where
$$D\varphi(\E(h)) =\cases \E(h)\nabla^{T} E_{j}(\E(h)), &
                          \text{for $H=\hbox{Im}\, I_{p/2-j}$}\\
                          i\E(h)\nabla^{T} E_{j}(\E(h)), & 
                          \text{for $H=\hbox{Re}\, I_{p/2-j}$}\\
                          -i h\E(h)\nabla^{T} E_{p/2}(\E(h)), &
                          \text{for $H=P$}, \endcases
\eqno(6.2)
$$
and similarly for $D\varphi(\tilde\E(h)).$
\endproclaim

In (6.1), the parameter $h$ is on the unit circle, however, we will remove
this restriction later on.  (See (6.29) below.)  Before solving
these equations, let us spell out $\nabla^{T} E_{j}(x)$ more
explicitly.  To do so, observe that
$\nabla^{T}E_j(x)$  (see (5.3) above) obey the recursion
relations
$$\nabla^{T} E_{j+1}(x)= x\nabla^{T}E_j(x) -E_j(x)I, \quad
0\leq j\leq p,\eqno(6.3)$$
where by convention $\nabla^{T}E_{p+1}(x)\equiv 0.$   Since
$\nabla^{T}E_0(x)=0,$ by solving the recursion relations backwards,
we obtain
$$\nabla^{T}E_j(x) = -\sum_{i=0}^{j-1} E_{j-1-i}(x) x^i,\quad
j=1,\cdots, p.\eqno(6.4)$$

The equations of motion generated by $P$ are the simplest to solve.   
Although we could easily write down the solutions of these equations without recourse to 
Proposition 6.1 above, however,  we will do it by the proposition in order to
achieve uniformity in our treatment.

\proclaim
{Proposition 6.2}   The Hamiltonian equations of motion generated by $P$
simplify to
$$\eqalign{
& \dot g^e = g^e \,\Lambda_1 - \Lambda_2\, g^e\cr
& \dot g^o(h) = g^o(h) \,\Lambda_2 - \Lambda_1 \,g^o(h)\cr}
\eqno(6.5)
$$
where 
$$\Lambda_1 =  i\hbox{Re}\,\left(A_1\left(p\over 2\right)\right)_0
                   =  \hbox{diag}\,(0, iP, 0, iP,\cdots, 0, iP), \eqno(6.6)$$
and
$$\Lambda_2 =  i\hbox{Re}\,\left(A_{-1}\left(p\over 2\right)\right)_0
                   =  \hbox{diag}\,(iP, 0, iP, 0, \cdots, iP, 0).\eqno(6.7)$$
The solutions of (6.5) are therefore given by
$$g^e(t) = e^{-t\Lambda_2(0)} g^e(0) e^{t\Lambda_1(0)},
    g^o(h,t) = e^{-t\Lambda_1(0)} g^o(h,0) e^{t\Lambda_2(0)}.\eqno(6.8)$$
Hence
$$\alpha_j(t) = \alpha_j(0) e^{itP(0)}, \quad \rho_j(t) = \rho_j(0),\,\,\, j=0,\cdots,p-1.\eqno(6.9)$$
\endproclaim

\demo
{Proof}  According to (6.2) and (6.3),
$$D\varphi(\E(h)) = ih\sum_{j=0}^{{p\over 2} -1} E_j(\E(h)) \E(h)^{{p\over 2} -j}.\eqno(6.10)$$
For  $j=1, \cdots, {p\over 2} -1,$ it follows from (2.11) that
$$\Pi_{\btw} \left(ihE_j(\E(h))\E(h)^{{p\over 2}-j}\right) = ihE_j(\E(h))\E(h)^{{p\over 2}-j}.\eqno(6.11)
$$
Similarly, by using the fact that $A_{-1}\left(p\over 2\right)$  is lower triangular,  we find
$$\Pi_{\btw}\left(ihE_{0}(\E(h))\E(h)^{p\over 2}\right) = ih\E(h)^{p\over 2} -i\hbox{Im}\,\left(iA_{-1}
\left(p\over 2\right)\right)_0\eqno(6.12)$$
as $E_0 \equiv 1.$  Therefore, on using (6.11) and (6.12), we obtain
$$\Pi_{\btw} D\varphi(\E(h)) =  D\varphi(\E(h)) - i\hbox{Re}\,\left(A_{-1}\left(p\over 2\right)\right)_0.
\eqno(6,13)$$
By a similar calculation, we find
$$\Pi_{\btw} D\varphi(\widetilde\E(h)) = D\varphi(\widetilde\E(h)) - i\hbox{Re}\,
\left(A_1\left(p\over 2\right)\right)_0.\eqno(6.14)$$
Hence, on substituting (6.13) and (6.14) into (6.1), and using the obvious facts that
$\E(h)^{{p\over 2}-j} g^e = g^e \widetilde\E(h)^{{p\over 2}-j},$
$g^{o}(h) \E(h)^{{p\over 2}-j} = \widetilde\E(h)^{{p\over 2}-j} g^{o}(h),$
we obtain the equations in (6.5).  The formulas in (6.6) and (6.7) for 
$\Lambda_1$ and $\Lambda_2$ in terms of $P$  then follow
from Lemma 2.2 and (2.19).  As $P$ is a conserved quantity,
it is easy to verify that the expressions in (6.8) give solutions to the equations
in (6.5).   Finally the solution formulas in (6.9) are obtained from (6.8) by
multiplying out.
\pf
\enddemo

To solve the Hamiltonian equations generated by
$H = \hbox{Im}\, I_{p/2-j},\hbox{Re}\,I_{p/2-j},$ 
 we will make use of Proposition 3.9 (c),
which means we have to solve explicitly for each $t>0$
the following factorization problems
$$e^{tD\varphi(\E(h,0))} = k_1(h,t)b_{1}(h,t)^{-1},\quad
e^{t D\varphi(\tilde\E(h,0))}= k_2(h,t) b_{2}(h,t)^{-1}\eqno(6.15)$$
for $k_i(\cdot,t)\in \Ktw,$ $b_i(\cdot,t)\in \Btw, i=1,2.$
However, from the definition of $\Ktw,$ it is easy to show that it
is enough to solve
$$\aligned
& e^{-tD\varphi(\E(h,0))}e^{-tD\varphi(\E(h,0))^*} = b_1(h,t)b_1(h,t)^*,\\
& e^{-tD\varphi(\tilde\E(h,0))}e^{-tD\varphi(\tilde\E(h,0))^*} = b_2(h,t)b_2(h,t)^{*}\\
\endaligned
\eqno(6.16)$$
for $b_1(\cdot,t), b_2(\cdot, h)\in\Btw.$  Note that for $i=1,2,$
$b_i(\cdot,t)$ (resp.~$b_i(\cdot,t)^*$) can be extended analytically 
in the interior (resp.~exterior) of the unit circle $|h|=1.$   So (6.16) is a
Riemann-Hilbert problem.   In order to solve this problem explicitly,
we will first transform the product on the left hand side of (6.16)
into a form which makes the problem more tractable.  To this end,
note that it follows from (6.4) that
$$\nabla^{T}E_j(\E(h,0)) = -\sum_{i=0}^{j-1} E_{j-1-i}(\E(h,0))\E(h,0)^i,\quad
j=1,\cdots, p.\eqno(6.17)$$
As $\E(h,0)$ commutes with $\E(h,0)^*$ for $h\in \partial \Bbb D,$ it follows
from (6.17) and (6.2) that $D\varphi(\E(h,0))$ commutes with $D\varphi(\E(h,0))^*$
for $h\in \partial \Bbb D.$    In a similar way, we see that
$D\varphi(\tilde\E(h,0))$ commutes with $D\varphi(\tilde\E(h,0))^*.$
Consequently, we can rewrite (6.16) as 
$$\aligned
& e^{-t(D\varphi(\E(h,0)) +D\varphi(\E(h,0))^*)} = b_1(h,t)b_1(h,t)^*,\\
& e^{-t(D\varphi(\tilde\E(h,0)) + D\varphi(\tilde\E(h,0))^*)} = b_2(h,t)b_2(h,t)^{*},\,\,
h\in\partial\Bbb D .\\
\endaligned
\eqno(6.18)$$
Now if we compute $D\varphi(\E(h,0))^*$ (resp.~$D\varphi(\tilde\E(h,0))$) 
more carefully, we find
$$\aligned
&D\varphi(\E(h,0))^* = \pm D\varphi(\E(h,0)^{-1}),\\
&D\varphi(\tilde\E(h,0))^* = \pm D\varphi(\tilde\E(h,0)^{-1}),\quad h\in\partial \Bbb D ,\\
\endaligned
\eqno(6.19)$$
where we pick the $+$ sign for $H = \hbox{Im}\, I_{p/2-j}$ and the $-$ sign
for $H = \hbox{Re}\, I_{p/2-j}.$  Hence (6.18) becomes
$$\aligned
& e^{-t(D\varphi(\E(h,0))\pm D\varphi(\E(h,0)^{-1}))} = b_1(h,t)b_1(h,t)^*,\\
& e^{-t(D\varphi(\tilde\E(h,0)) \pm D\varphi(\tilde\E(h,0)^{-1}))} = b_2(h,t)b_2(h,t)^{*},\,\,
h\in\partial\Bbb D ,\\
\endaligned
\eqno(6.20)$$
where the choice of sign is described in the previous sentence.

As the explicit solution of the factorization problem in (6.20) will involve constructing
$b_i(h,t), i=1,2,$ for values of $h$ not on unit circle, it is necessary to introduce some Lie 
algebras and projection
operators which complements those in Section 3.  For this purpose, let ${\Cal A}$ be
the ring of Laurent polynomials in the variable $h$ and let $gl_{p}(\Cal{A})$ be
the Lie algebra of $p\times p$ matrix functions with entries in ${\Cal A}$ equipped
with the pointwise Lie bracket.   We will consider the following Lie subalgebras
of $gl_{p}({\Cal A})$:
$$\eqalign{
& \widetilde \fb = \left\{\sum_{j=0}^{n} X_jh^j\in gl_{p}(\Bbb C[h])\mid X_0\in \frak b \right\},\cr
& \widetilde \fk = \left\{ X\in gl_{p}({\Cal A})\mid X(h) + X(\bar h^{-1})^* =0 \right\}.\cr}
\eqno(6.21)$$
Then analogous to (3.8), we have the splitting
$$gl_{p}({\Cal A}) = \widetilde \fk \oplus \widetilde \fb.\eqno(6.22)$$
For $X\in gl_{p}({\Cal A}),$ define
$$(\Pi_{+} X)(h) = \sum_{j>0} X_jh^j,\quad (\Pi_{-}X)(h) =\sum_{j<0} X_jh^j,\eqno(6.23)$$
then the projection operator onto $\widetilde \fb$ associated with the splitting in (6.22)
is given by
$$(\Pi_{\widetilde \fb} X)(h) = (\Pi_{+} X)(h) + \Pi_{\fb} X_0 + ((\Pi_{-}X)(\bar h^{-1}))^{*}.
\eqno(6.24)$$

\smallskip

\proclaim
{Theorem 6.3}  For $H = \hbox{Re}\, I_{p/2-j},$$\hbox{Im}\, I_{p/2-j}$,
there exist unique holomorphic matrix-valued functions
$$\aligned
&b_1(\cdot,t):\IC\IP^{1}\setminus \{\infty\}\longrightarrow GL(p,\IR),\\
&b_2(\cdot,t):\IC\IP^{1}\setminus \{\infty\}\longrightarrow GL(p,\IR)\\
\endaligned
\eqno(6.25)
$$
which are smooth in $t$, solve the factorization problems
$$\aligned
& e^{-t(D\varphi(\E(h,0)) \pm D\varphi(\E(h,0)^{-1}))} = b_1(h,t)b_1(\bar h^{-1},t)^{*},\\
& e^{-t(D\varphi(\tilde\E(h,0)) \pm D\varphi(\tilde\E(h,0)^{-1}))}= b_2(h,t)b_2(\bar 
h^{-1},t)^{*},
\\
& h\in \IC\IP^{1}\setminus \{0,\infty\}\\
\endaligned
\eqno(6.26)$$
(where the $+$ sign corresponds to $H = \hbox{Im}\, I_{p/2-j}$ and
the $-$ sign corresponds to $H=\hbox{Re} \,I_{p/2-j}$)
and satisfy
$$b_i(0,t)\in B,\, b_i(h,t)^{-1}\dot b_i(h,t)\in \hbox{Im}\,\Pi_{\widetilde \fb}\,\,
 \,\, i=1,2.\eqno(6.27)$$
Moreover, for $h\in \IC\IP^{1}\setminus \{0,\infty\},$ the formulas
$$\aligned
g^e(t) & = b_1(h,t)^{-1} g^e(0)\, b_2(h,t)= b_1(\bar h^{-1},t)^{*} g^e(0)(b_2(\bar h^{-1},t)^{*})^{-1},\\
g^o(h,t) & = b_2(h,t)^{-1} g^o(h,0)\, b_1(h,t)= b_2(\bar h^{-1},t)^{*}g^o(h,0) (b_1(\bar h^{-1},t)^{*})^{-1}\\
\endaligned
\eqno(6.28)$$
give solutions of the equations
$$\eqalign{
&\dot g^e = (\Pi_{\widetilde\fb} D\varphi(\E(\cdot)))(h) g^e
                   - g^e(\Pi_{\widetilde\fb} D\varphi(\widetilde\E(\cdot)))(h),\cr
&\dot g^o(h) = (\Pi_{\widetilde\fb} D\varphi(\widetilde\E(\cdot)))(h) g^o(h)
                   - g^o(h)(\Pi_{\widetilde\fb}  D\varphi(\E(\cdot)))(h).\cr}
\eqno(6.29)$$
Finally, for generic initial data $g^e(0)$ and
$g^{o}(h,0)$,  $b_1(h,t)$ and $b_2(h,t)$
can be constructed by means of theta functions associated with the Riemann
surface of the spectral curve $\Cal{C} =\{ (h,z)\mid \hbox{det}\,(zI - g^e(0)g^{o}(h,0)) =0\}$
for values of $t >0$ for which $\alpha_j(t)\neq 0,$ $j=0,\cdots, p-1.$
\endproclaim

\demo
{Proof}  We will prove the result for $H = \hbox{Im}\,I_{{p/2}-j}.$  The argument
for the other case is similar.  We start
with uniqueness of the factors $b_i(h,t), i=1,2.$   To prove this, suppose
$b_i^0(h,t), i=1,2$ is a second pair of solutions of the factorization problem.
Then from $b_i(h,t)b_i(\bar h^{-1},t)^* = b_i^0(h,t) b_i^0(\bar h^{-1},t)^*,$
we have 
$$g_i(h,t)\equiv b_i^{0}(h,t)^{-1} b_i(h,t) = b_i^{0}(\bar h^{-1},t)^{*}(b_i(\bar h^{-1},t)^{*})^{-1},
\,\,\,h\in \IC\IP^{1}\setminus \{0,\infty\}.\eqno(6.30)$$
Clearly the function $g_i(\cdot, t)$ defined in (6.30) above can be extended to an analytic function everywhere, hence by Liouville's theorem, $g_i(h,t) = c_i(t).$    To determine $c_i(t),$ note that
$g_i(0,t) = b_i^{0}(0,t)^{-1} b_i(0,t)\in \fb.$   On the other hand,
$g_i(\infty, t)  = (g_i(0,t)^{-1})^*$ 
is upper triangular with positive diagonal entries on the diagonal.   Hence $c_i(t)\equiv I.$ 
  
To establish the existence of $b_i(h,t), i=1,2,$  note that $\E(h,t)$  (resp.~$\widetilde\E(h,t)$)
exists for
$h\in \IC\IP\setminus \{0,\infty\}$ since it exists for $h\in \partial \Bbb D.$
Hence we can obtain $b_i(h,t), i=1,2$ as solutions of the equations
$$\dot b_1(h,t) = -b_1(h,t) (\Pi_{\fb} D\varphi(\E(\cdot,t)))(h), \eqno(6.31)$$
and
$$\dot b_2(h,t) = -b_2(h,t) (\Pi_{\fb} D\varphi(\widetilde \E(h,t)))(h).\eqno(6.32)$$
Clearly, the analyticity properties and (6.27) are satisfied by definition.
We next consider the product $b_1(h,t)b_1(\bar h^{-1},t)^{*}.$   By differentiating,
we have
$$\eqalign{
& {d\over dt}(b_1(h,t) b_1(\bar h^{-1},t)^*)\cr
=\,\, & -b_1(h,t) \left((\Pi_{\widetilde\fb} D\varphi(\E(\cdot,t)))(h)
        + ((\Pi_{\widetilde\fb} D\varphi(\E(\cdot,t)))(\bar h^{-1}))^{*}\right) b_1(\bar h^{-1},t)^{*}.\cr}
\eqno(6.33)
$$
Now, from the definition of $\Pi_{\widetilde\fb},$ it is straightforward to check that
$$
 (\Pi_{\widetilde\fb} X)(h) + ((\Pi_{\widetilde\fb} X)(\bar h^{-1}))^{*}
=\,\, X(h) + (X(\bar h^{-1}))^{*}.
\eqno(6.34)
$$
Consequently,  we obtain
$$\eqalign{&
 (\Pi_{\widetilde\fb} D\varphi(\E(\cdot,t)))(h) + ((\Pi_{\widetilde\fb} D\varphi(\E(\cdot,t)))(\bar h^{-1}))^{*}\cr
=\,\,& D\varphi(\E(h,t)) + (D\varphi(\E(\bar h^{-1},t)))^{*}\cr
=\,\, & D\varphi(\E(h,t)) + D\varphi(\E(h,t)^{-1}).\cr}
\eqno(6.35)
$$
Substitution of (6.35) into (6.33) therefore gives the relation
$$\eqalign{
& {d\over dt} (b_1(h,t)b_1(\bar h^{-1},t)^*)\cr
= \,\, & -b_1(h,t)(D\varphi(\E(h,t)) + D\varphi(\E(h,t)^{-1})) b_1(\bar h^{-1},t)^*\cr
= \,\, & -b_1(h,t)b_1(\bar h^{-1},t)^{*}\left( D\varphi({\Cal F}(h,t))
           + D\varphi({\Cal F}(h,t)^{-1})\right)\cr}
\eqno(6.36)$$
where we have used the fact that $\varphi$ is a central function, and where
${\Cal F}(h,t) = (b_1(\bar h^{-1},t)^{*})^{-1} \E(h,t) b_1(\bar h^{-1},t)^{*}.$
Now, by direct differentiation, using (6.31), the equation 
$$\dot \E(h,t) = (\Pi_{\widetilde\fb} D\varphi(\E(\cdot,t)))(h) \E(h,t) -
\E(h,t)(\Pi_{\widetilde\fb} D\varphi(\E(\cdot,t)))(h),
\eqno(6.37)$$
and (6.35), we find that
$$\eqalign{
{d\over dt} {\Cal F}(h,t) = & (b_1(\bar h^{-1},t)^{*})^{-1} (D\varphi(\E(h,t)) + D\varphi(\E(h,t)^{-1}))
\E(h,t)b_1(\bar h^{-1},t)^{*}\cr
& -(b_1(\bar h^{-1},t)^{*})^{-1}\E(h,t) (D\varphi(\E(h,t)) + D\varphi(\E(h,t)^{-1})) b_1(\bar h^{-1},t)^{*}\cr
= & 0.\cr}\eqno(6.38)
$$
Therefore, ${\Cal F}(h,t) = \E(h,0)$ and so (6.36) becomes
$$\eqalign{
& {d\over dt} (b_1(h,t)b_1(\bar h^{-1},t)^*)\cr
= \,\, & -b_1(h,t)b_1(\bar h^{-1},t)^{*}\left( D\varphi(\E(h,0))
           + D\varphi(\E(h,0^{-1}))\right).\cr}
\eqno(6.39)$$
This shows $b_1(h,t)$ satisfies the first relation in (6.26).   In a similar fashion, we 
can show that $b_2(h,t)$ satisfies the second relation in (6.26).

Finally, we will show that $g^e(t)$ and $g^o(h,t)$ as defined in (6.28) satisfy (6.29).
First, note that by using the relation $\widetilde\E(h,0) = g^e(0)^{-1} \E(h,0) g^e(0)$
and (6.26), 
we have
$$\aligned
& b_1(h,t)^{-1} g^e(0)\, b_2(h,t)= b_1(\bar h^{-1},t)^{*} g^e(0)(b_2(\bar h^{-1},t)^{*})^{-1},\\
& b_2(h,t)^{-1} g^o(h,0)\, b_1(h,t)= b_2(\bar h^{-1},t)^{*}g^o(h,0) (b_1(\bar h^{-1},t)^{*})^{-1}.\\
\endaligned
\eqno(6.40)$$
 Differentiate $g^e(t) = b_1(h,t)^{-1} g^e(o)\,b_2(h,t)$ and
$g^o(h,t) = b_2(h,t)^{-1} g^o(h,0)\,b_1(h,t)$ with respect to $t,$ we find
$$\eqalign{
& \dot g^e(t) = -b_1(h,t)^{-1}\dot b_1(h,t)\, g^e(t) + g^e(t)\, b_2(h,t)^{-1} \dot b_2(h,t),\cr
& \dot g^o(h,t) = -b_2(h,t)^{-1}\dot b_2(h,t)\, g^o(h,t) + g^o(h,t) \,b_1(h,t)^{-1}\dot b_1(h,t).\cr}
\eqno(6.41)$$
On the other hand, by differentiating the first relation in (6.26) with respect to $t,$
and multiply the resulting expression on the left by $b_1(h,t)^{-1}$ and on the
right by $(b_1(\bar h^{-1},t)^{*})^{-1},$  we obtain
$$\eqalign{
& -b_1(\bar h^{-1},t)^{*}(D\varphi(\E(h,0)) + D\varphi(\E(h,0)^{-1}))(b_1(\bar h^{-1},t)^{*})^{-1}\cr
=& b_1(h,t)^{-1}\dot b_1(h,t)  + \dot b_1(\bar h^{-1},t)^{*} (b_1(\bar h^{-1},t)^{*})^{-1}.\cr}
\eqno(6.42)
$$
But from the fact that $\varphi$ is a central function and (6.40), we see that
$$\eqalign{
& b_1(\bar h^{-1},t)^{*}(D\varphi(\E(h,0)) + D\varphi(\E(h,0)^{-1}))(b_1(\bar h^{-1},t)^{*})^{-1}\cr
=\,\, &  D\varphi(\E(h,t)) + D\varphi(\E(h,t)^{-1})\cr}
\eqno(6.43)
$$
where $\E(h,t) = g^e(t) g^o(h,t).$   Thus (6.42) becomes
$$-(D\varphi(\E(h,t)) +D\varphi(\E(h,t)^{-1}))
=\, b_1(h,t)^{-1} \dot b_1(h,t) + (b_1(\bar h^{-1},t)^{-1}\dot b_1(\bar h^{-1},t))^{*}.
\eqno(6.44)
$$
Now let 
$$\aligned
&X(h,t) = b_1(h,t)^{-1} \dot b_1(h,t) -(b_1(\bar h^{-1},t)^{-1}\dot b_1(\bar h^{-1},t))^{*},\\
&Y(h,t) = D\varphi(\E(h,t)) -D\varphi(\E(\bar h^{-1},t)^{-1}).
\endaligned
\eqno(6.45)
$$   Clearly, $X(\cdot, t)\in \widetilde k,$ thus $\Pi_{\widetilde\fb} X(\cdot, t) =0.$
On the other hand, as $D\varphi(\E(h,t)^{-1}) = D\varphi(\E(\bar h^{-1}, t))^{*},$
we also have $Y(\cdot, t)\in \widetilde k$ and $\Pi_{\widetilde\fb} Y(\cdot, t) =0.$ 
Consequently, when we apply $\Pi_{\widetilde\fb}$ to both sides of (6.44),
we obtain
$$b_1(h, t)^{-1} \dot b_1(h,t) = -(\Pi_{\widetilde \fb} D\varphi(\E(\cdot,t)))(h).\eqno(6.46)$$
Similarly, from the second relation in (6.26), we can show that
$$b_2(h,t)^{-1} \dot b_2(h,t) = -(\Pi_{\widetilde\fb} D\varphi(\widetilde\E(\cdot,t)))(h).\eqno(6.47)$$
Finally, substituting  (6.46) and (6.47) into (6.41) gives (6.29).  This completes the
proof of the theorem modulo the assertion on the construction of $b_1(h,t)$
and $b_2(h,t)$ via Riemann theta functions.
\pf
\enddemo

We now turn to the construction of $b_1(h,t)$ and $b_2(h,t)$ via theta functions.
Again, we will give details for $H= \hbox{Im}\, I_{{p/2} -j},$ leaving the other
case to the interested reader.  The following proposition shows we can construct
$b_2(h,t)$ from $b_1(h,t)$ and the solution of a finite dimensional factorization
problem.

\proclaim
{Proposition 6.4}  Let $l(t)\in B,$ $u(t)\in K$ be the solution of the factorization
problem
$$b_1(0,t)^{-1} g^e(0) = u(t) l(t)^{-1}.\eqno(6.48)$$
Then 
$$b_2(h,t) = g^e(0)^{*} b_1(h,t) u(t).\eqno(6.49)$$
\endproclaim

\demo
{Proof}  Since $\widetilde\E(h,0) = g^e(0)^{-1}\E(h,0)g^e(0),$ the factorization
problem for $b_2(h,t)$  in (6.26) can be rewritten as
$$e^{-t(D\varphi(\E(h,0)) + D\varphi(\E(h,0)^{-1}))} = g^e(0)b_2(h,t)b_2(\bar h^{-1},t)^{*}
     (g^e(0))^{-1}.\eqno(6.50)$$
Therefore, when we compare this with the first relation in (6.26), we obtain
$$b_2(h,t) b_2(\bar h^{-1},t)^{*} = g^e(0)^{*}b_1(h,t)(g^e(0)^{*}b_1(\bar h^{-1},t))^{*}.
\eqno(6.51)$$
Now let $l(t)\in B,$ $u(t)\in K$ be the solution of the factorization problem in (6.48).
Then
$$g^e(0)^{*}b_1(h,t) = (l(t) + O(h))u(t)^{*}\eqno(6.52)$$
and
$$g^e(0)^{*}b_1(\bar h^{-1},t) = (l(t) + O(\bar h^{-1}))u(t)^{*}.\eqno(6.53)$$
Substitute (6.52) and (6.53) into (6.51), we obtain
$$b_2(h,t)b_2(\bar h^{-1},t) = (l(t) + O(h))(l(t)^{*} + O(h^{-1})).\eqno(6.54)$$
Therefore, from the uniqueness of solution of the factorization problem
(Theorem 6.3), we conclude that
$$b_2(h,t) = l(t) + O(h) = g^e(0)^{*}b_1(h,t)u(t).$$
\pf
\enddemo

To construct $b_1(h,t),$ we invoke the formula in (6.2) and (6.17), according to
which we have
$$(D\varphi(\E(h(P),0)) + D\varphi(\E(h(P),0)^{-1})) \widehat v(P)
=\mu_j(P) \widehat v(P)\eqno(6.55)$$
where $\mu_j$ is meromorphic on the hyperelliptic Riemann surface $C$.
From the first relation in (6.26), 
$$\eqalign{
& e^{-t\mu_j(P)} b_1(h(P),t)^{-1} \widehat v(P)\cr
= & b_1(h(P),t)^{-1}e^{-t(D\varphi(\E(h(P),0)) + D\varphi(\E(h(P),0)^{-1}))}\widehat v(P)\cr
= & b_1(\overline{h(P)}^{\, -1},t)^{*} \widehat v(P)\cr}
\eqno(6.56)
$$
for $h(P)\in \IC\IP^{1}\setminus\{0,\infty\}.$
Since $\E(h,t) = b_1(h,t)^{-1}\E(h,0)b_1(h,t),$ we have
$$\E(h(P),t)b_1(h(P),t)^{-1}\widehat v(P) = z(P)b_1(h(P),t)^{-1}\widehat v(P).\eqno(6.57)$$
On the other hand, as
$$\aligned
\E(h(P), t) = & (\E(\overline{h(P)}^{\,-1},t)^{*})^{-1}\\
                =  & b_1(\overline{h(P},t)^{*}\E(h(P),0)(b_1(\overline{h(P)}^{\,-1},t)^{*})^{-1},
                \endaligned
                \eqno(6.58)
$$
we also have
$$\E(h(P),t)b_1(\overline{h(P)}^{\,-1},t)^{*} \widehat v(P) = z(P) b_1(\overline{h(P)}^{\,-1},t)^{*}
\widehat v(P).\eqno(6.59)$$
Thus if we let
$$\aligned
& v^{t}_{+}(P) = b_1(h(P),t)^{-1} \widehat v(P),\\
& v^{t}_{-}(P) = b_1(\overline{h(P)}^{\,-1},t)^{*} \widehat v(P),
\endaligned
\eqno(6.60)
$$
then (6.56), (6.57) and (6.59) give
$$\aligned
& e^{-t\mu_j(P)} v^{t}_{+}(P) = v^{t}_{-}(P), \quad h(P)\in \IC\IP^{1}\setminus\{0,\infty\}\\
& \E(h(P),t) v^{t}_{\pm}(P) = z(P) v^{t}_{\pm}(P).
\endaligned
\eqno(6.61)
$$
In this way, we are led to scalar factorization problems.  Note that because
$b_1(0,t)$ is lower triangular, it follows from (4.50) and (6.60) that
$$\aligned
&((v^{t}_{+})_{2j}) \geq -D(0) + \left({\frac{p}{2}} -j-1\right)P_{-} + \left(\frac{p}{2}-j\right)Q_{-}\\
&((v^{t}_{+})_{2j+1}) \geq -D(0) + \left({\frac{p}{2}}-j-1\right)P_{-} + \left(\frac{p}{2}-j-1\right)Q_{-}
\endaligned
\eqno(6.62)
$$
on $C \setminus \{h(P) =\infty\}.$
Similarly, because $b_1(0,t)^{*}$ is  upper triangular, we find that
$$\aligned
&((v^{t}_{-})_{2j}) \geq -D(0) - \left({\frac{p}{2}} -j-1\right)P_{+} -\left(\frac{p}{2}-j\right)Q_{+}\\
&((v^{t}_{-})_{2j+1}) \geq -D(0) - \left({\frac{p}{2}}-j-1\right)P_{+} -\left(\frac{p}{2}-j-1\right)Q_{+}
\endaligned
\eqno(6.63)
$$
on $C\setminus \{h(P) =0\}.$

We will first solve the following scalar factorization problem  (cf. \c{RSTS}, \c{DL})
$$\aligned
& e^{-t\mu_j(P)} \omega^{t}_{+}(P) = \omega^{t}_{-}(P), \,\,\, P\in C\setminus\{h(P) =0,\infty\},\\
& (\omega^{t}_{+}) \geq -D(0), \,\,\, \hbox{on}\,\,C\setminus \{h(P)=\infty\},\\
& (\omega^{t}_{-})\geq -D(0), \,\,\, \hbox{on}\,\, C\setminus \{h(P)=0\}.
\endaligned
\eqno(6.64)
$$
To do so, we fix a canonical homology basis $\{a_j,b_k\}_{1\leq j,k\leq g}$ of the
Riemann surface associated with the spectral curve, and let $\{\omega_i\}_{1\leq i\leq g}$
be a cohomology basis dual to $\{a_j, b_k\},$ i.e.,
$$\int_{a_j} \omega_i = \delta_{ij}, \quad \int_{b_j} \omega_i =\Omega_{ij},\eqno(6.65)$$
where $(\Omega_{ij})$ is the Riemann matrix.  Let
$$\theta(z) =\theta(z,\Omega) =\sum_{m\in \Bbb Z^g} \exp\{2\pi i(m,z) + \pi i(m,\Omega m)\}
\eqno(6.66)
$$
be the Riemann theta function associated with the matrix $\Omega.$
Let $\omega = (\omega_1,\cdots, \omega_g).$  Choose a nonsingular
$e\in \Bbb C^g$ in the theta divisor, i.e. $\theta (e) =0,$ the prime form
$E_{e}(x,y) \equiv \theta\left( e + \int_{x}^{y} \omega\right)\not\equiv 0$ 
with the additional property that $E_{e}(P_{\pm}, P)$ are not identically
zero in $P.$  (See Lemma 3.3 of \c{M}.)   Let $P_0$ be a fixed point on the
finite part of the Riemann surface, then by Corollary 3.6 of \c{M},
there exists an effective divisor $D_{g-1}$ of degree $g-1$ such that 
$e = \Delta -\int_{(g-1)P_0}^{D_{g-1}}\omega,$ where $\Delta$ is
the vector of Riemann constants.  Also note that 
$$\theta\left( e + \int_{D(0)}^{D_{g-1} + P} \omega\right)
= \theta\left(\Delta + \int_{D(0)}^{(g-1)P_0 + P}\omega\right) =0 \iff P\in D(0).
\eqno(6.67)$$
Now let $d\phi_{+}$ be the unique meromorphic differential of the second kind
with vanishing $a$-periods with poles only at $\{h(P)=\infty\}$ such that
$(d\phi_{+} -d\mu_j)(P)$ is regular in $C\setminus\{h(P)=0\}.$ 
Set
$$\omega^{t}_{+}(P) = \exp\left(t\phi_{+}(P)\right)\frac{\theta\left( e + \int_{D(0)}^{D_{g-1}+P}
\omega + t \Phi_{+}\right)}{\theta\left( e + \int_{D(0)}^{D_{g-1} + P} \omega\right)},
\eqno(6.68)$$
where $\phi_{+}(P) = \int_{P_0}^{P} d\phi_{+}$ and $\Phi_{+}$ is the vector
of $b$-periods of $d\phi_{+},$ i.e.,
$$(\Phi_{+})_{j} = \frac{1}{2\pi i} \int_{b_j} d\phi_{+},\,\, j=1,\cdots, g.\eqno(6.69)$$
By using (6.69), the basic property
$$\theta(z+ \gamma^{\prime} + \Omega\gamma) = \theta(z)\exp\left(2\pi i(-(\gamma, z)
    -{1\over 2} (\gamma, \Omega\gamma))\right),$$
and (6.67), it is straightforward to check that $\omega^{t}_{+}$ is single-valued and 
meromorphic in $C\setminus \{h(P) =\infty\}$ with $(\omega^{t}_{+})\geq -D(0)$ there.
Moreover, $e^{-t\mu_j(P)}\omega^{t}_{+}(P)$ is meromorphic in
$C\setminus \{h(P) =0\}$ with $(e^{-t\mu_j}\omega^{t}_{+})\geq -D(0)$
 because $d\phi_{+}(P) -d\mu_j(P)$ is regular in $C\setminus \{h(P) =0\}.$
 Thus $\omega^{t}_{+}$ and $\omega^{t}_{-}\equiv e^{-t\mu_j}\omega^{t}_{+}$
 solves the scalar factorization problem (6.64).
 
 Next, we compare (6.64) and (6.61), this gives
 $$\omega^{t}_{+}(P)^{-1} v^{t}_{+}(P) = \omega^{t}_{-}(P)^{-1} v^{t}_{-}(P), \,\,\,
 C\setminus \{h(P) =0,\infty\}.\eqno(6.70)$$
 Therefore,
 $$\widetilde{v}^{t}(P) \equiv \cases \omega^{t}_{+}(P)^{-1} v^{t}_{+}(P), & P\in C\setminus
 \{h(P) =\infty\}\\
 \omega^{t}_{-}(P)^{-1} v^{t}_{-}(P), & P\in C\setminus \{h(P) =0\}\endcases
 \eqno(6.71)$$
 is meromorphic on $C.$   Moreover, it follows from (6.62), (6.63) and 
 the expressions for $\omega^{t}_{\pm}$ that
 $$\aligned
  (\widetilde{v}^{t}_{2j}) \geq -\widetilde{D}(t) & - \left({p\over 2}-j-1\right)P_{+}
     +\left({p\over 2} -j-1\right)P_{-}\\
     & -\left(\frac{p}{2}-j\right)Q_{+} + \left(\frac{p}{2}-j\right)Q_{-},\\
  (\widetilde{v}^{t}_{2j+1}) \geq -\widetilde{D}(t) & - \left({p\over 2} -j-1\right)P_{+}
      +\left({p\over 2} -j-1\right) P_{-}\\
      &-\left(\frac{p}{2}-j-1\right)Q_{+} + \left(\frac{p}{2} -j-1\right)Q_{-},\\
\endaligned
\eqno(6.72)$$
where $\widetilde{D}(t)$ is the divisor of zeros of the function
$$\theta\left( e + \int_{D(0)}^{D_{g-1} + P} \omega + t \Phi_{+}\right).\eqno(6.73)$$
Since $\E(h(P),t)v^{t}_{\pm}(P) =z(P) v^{t}_{\pm}(P),$ it follows from the definition
of $\widetilde{v}^{t}(P)$ that
$$\E(h(P),t) \widetilde{v}^{t}(P) = z(P) \widetilde{v}^{t}(P).\eqno(6.74)$$
But on the other hand, 
$$\E(h(P),t) \widehat{v}(t,P) = z(P) \widehat{v}(t,P),\eqno(6.75)$$
where the last component of $\widehat{v}(t,P)$ is equal to $1$ and
$$\aligned
 (\widehat{v}_{2j}(t,\cdot)) \geq -D(t) & -\left({p\over 2}-j-1\right)P_{+} +
   \left({p\over 2} -j-1\right) P_{-}\\
   & -\left(\frac{p}{2}-j\right)Q_{+} + \left(\frac{p}{2}-j\right)Q_{-},\\
 (\widehat{v}_{2j+1}(t, \cdot))\geq -D(t) & - \left({p\over 2} -j-1\right) P_{+} +
    \left({p\over 2} -j-1\right) P_{-}\\
    &-\left(\frac{p}{2}-j-1\right)Q_{+} + \left(\frac{p}{2} -j-1\right)Q_{-}.
\endaligned
\eqno(6.76)$$
Clearly, $\widetilde{v}^{t}(P) = \widetilde{v}^{t}_{p-1}(P) \widehat{v}(t,P).$
Let $\widetilde{D}_{0}(t)$ be the divisor of zeros of $\widetilde{v}^{t}_{p-1}(P)$
so that
$$(\widetilde{v}^{t}_{p-1}) = \widetilde{D}_{0}(t) -\widetilde{D}(t).\eqno(6.77)$$
Then it follows from the relation connecting $\widetilde{v}^{t}(P)$ and $\widehat{v}(t,P)$
above, (6.72) and (6.77) that
$$\aligned
(\widehat{v}_{2j}(t,\cdot))  = & (\widetilde{v}^{t}_{2j}) - (\widetilde{v}^{t}_{p-1})\\
                                         \geq &  -\widetilde{D}_{0}(t) - \left({p\over 2}-j-1\right) P_{+} + 
                                         \left( {p\over 2} -j-1\right) P_{-}\\
                                         &\quad -\left(\frac{p}{2}-j\right)Q_{+} + \left(\frac{p}{2} -j\right)Q_{-}.
                                         \endaligned
                                         \eqno(6.78)$$
Similarly, 
$$\eqalign{
(\widehat{v}_{2j+1}(t,\cdot)) \geq -\widetilde{D}_{0}(t) & -  \left({p\over 2}-j-1\right) P_{+} 
    + \left({p\over 2}-j-1\right) P_{-}\cr
    & -\left(\frac{p}{2}-j-1\right)Q_{+} + \left(\frac{p}{2}-j-1\right)Q_{-}.\cr}
    \eqno(6.79)$$
Clearly, we must have $\widetilde{D}_{0}(t)\geq D(t).$   Since 
$\hbox{deg}\,\, D(t) = \hbox{deg}\,\, \widetilde{D}_{0}(t) =g,$  we must have
$\widetilde{D}_{0}(t) = D(t).$  But $D(t)$ is a general divisor by Proposition 4.12, so from
$(\widetilde{v}^{t}_{p-1}) = \widetilde{D}_{0}(t) -\widetilde{D}(t) = D(t) -\widetilde{D}(t),$
we must have $\widetilde{v}^{t}_{p-1} =\hbox{constant}$ and this implies
$\widetilde{D}(t) = D(t).$  Thus we can solve for $\widetilde{v}^{t}(P)$ and
hence $v^{t}_{\pm}(P)$  up to multiples because of Proposition 4.12.
Indeed, by making use of the prime form,
we can write down the explicit expression
$$\eqalign{
\widetilde{v}^{t}_{2j}(P) = & c_{2j}(t)\left(\frac{E_{e}(P_{-},P)}{E_{e}(P_{+},P)}\right)^{{p\over 2}-j-1}\cdot
\left(\frac{E_{e}(Q_{-},P)}{E_{e}(Q_{+},P)}\right)^{{p\over 2}-j}\cr
&\times \frac{\theta\left( e + \int_{D(t) + \left({p\over 2} -j-1\right)P_{+}+ \left({p\over 2}-j\right)Q_{+}}^{D_{g-1} + P + 
\left({p\over 2} -j-1\right) P_{-} + \left({p\over 2}-j\right)Q_{-}} \omega \right)}
{\theta\left(e + \int_{D(t)}^{D_{g-1} + P} \omega \right)},\cr}
\eqno(6.80)
$$
where $c_{2j}(t)$ has to be determined.   Similarly, we have
$$\eqalign{
\widetilde{v}^{t}_{2j+1}(P) = & c_{2j+1}(t)\left(\frac{E_{e}(P_{-},P)}{E_{e}(P_{+},P)}\right)^{{p\over 2}-j-1}\cdot
\left(\frac{E_{e}(Q_{-},P)}{E_{e}(Q_{+},P)}\right)^{{p\over 2}-j-1}\cr
&\times \frac{\theta\left( e + \int_{D(t) + \left({p\over 2} -j-1\right)P_{+}+ \left({p\over 2}-j-1\right)Q_{+}}^{D_{g-1} + P + 
\left({p\over 2} -j-1\right) P_{-} + \left({p\over 2}-j-1\right)Q_{-}} \omega \right)}
{\theta\left(e + \int_{D(t)}^{D_{g-1} + P} \omega \right)},\cr}
\eqno(6.81)
$$
where $c_{2j+1}(t)$ is also as yet undetermined.

We are now ready to construct $b_1(h,t)$ and in the process, we will also determine
$c_j(t).$   For given $h\in \IC\IP^{1}$ which is not a
branch point of the coordinate function $h(P),$ there exist $p$ points
$P_0(h), \cdots, P_{p-1}(h)$ of the Riemann surface $C$ lying over $h.$
Therefore we can define the matrices
$$\aligned
& V_{\pm}(h,t) = (v^{t}_{\pm}(P_0(h)) \cdots v^{t}_{\pm}(P_{p-1}(h)))\\
& \widehat{V}^{\theta}(h) = (\widehat{v} (P_0(h)) \cdots \widehat{v}(P_{p-1}(h)))
\endaligned
\eqno(6.82)
$$
where $\widehat{v}(P)$ can be obtained from the formula for
$\widetilde{v}^{t}(P)$ by setting $t=0$ (since $\omega^{t=0}_{\pm}(P)\equiv 1$
and $b_1(h, t=0) = I$) and so can be computed in terms of theta functions.
Of course, $v^{t}_{\pm}(P)$ are also in terms of theta functions.  With these
matrices, it follows that
$$b_1(h,t) = \widehat{V}^{\theta}(h) V_{+}(h,t)^{-1}.\eqno(6.83)$$
Of course, $v^{t}_{\pm}(P) = \omega^{t}_{\pm}(P) \widetilde{v}^{t}(P)$
are determined only up to the $c_j(t)$'s.   Write
$$v^{t}_{\pm}(P) = c(t) v^{\theta}_{\pm}(t,P)\eqno(6.84)$$
where $v^{\theta}_{\pm}(t,P)$ are known and $c(t) =\hbox{diag}\,\,(c_0(t),\cdots, c_{p-1}(t))$
is to be determined.  Then
$$V_{\pm}(h, t) = c(t) V^{\theta}_{\pm}(h,t)\eqno(6.85)$$
where  $V^{\theta}_{\pm}(h,t) = ( v^{\theta}_{\pm}(t, P_0(h))\cdots v^{\theta}_{\pm}(t, P_{p-1}(h))).$
With these definitions,
$$b_1(h,t) = \widehat{V}^{\theta}(h) V^{\theta}_{+}(h,t)^{-1} c(t)^{-1}.\eqno(6.86)$$
As $b_1(h,0) =I,$ the above relation determines $c(0)$ via the formula
$$c(0) = \widehat{V}^{\theta}(h) V^{\theta}_{+}(h,0)^{-1}.\eqno(6.87)$$
To determine $c(t)$ for $t>0,$ we bring in
$$\aligned
k_1(h,t) = & e^{t D\varphi(\E(h,0))} b_1(h,t)\\
             = &  e^{t D\varphi(\E(h,0))} \widehat{V}^{\theta}(h) V^{\theta}_{+}(h,t)^{-1} c(t)^{-1}
             \endaligned
             \eqno(6.88)$$
which is required to be unitary for $h\in \partial \Bbb D.$
Therefore, if we equate the expression for $k_1(1,t)$ from (6.88) with the corresponding one
for $(k_1(1,t)^*)^{-1},$ we obtain $|c(t)|^2 = c(t)^{*}c(t).$   Explicitly,
$$ |c(t)|^2 = (\widehat{V}^{\theta}(1)V^{\theta}_{+}(1,t)^{-1})^{*}
e^{t(D\varphi(\E(1,0)) + D\varphi(\E(1,0)^{-1}))}\widehat{V}^{\theta}V^{\theta}_{+}(1,t)^{-1}.
\eqno(6.89)
$$
Write $c(t) =|c(t)|e^{i\eta(t)},$  where 
$e^{i\eta(t)} =\hbox{diag}\,\,(e^{i\eta_0(t)},\cdots,e^{i\eta_{p-1}(t)}).$
It remains to determine $e^{i\eta(t)}.$
However, this is fixed by the condition that
$$b_1(0,t) = \widehat{V}^{\theta}(0)V^{\theta}_{+}(0,t)^{-1}|c(t)|^{-1} e^{-i\eta(t)}\in B
\eqno(6.90)$$
as the diagonal entries of the elements in $B$ are positive.
\bigskip
\noindent{Acknowledgments.} The first author would like to thank
Jiang-Hua Lu for reminding him of the Bruhat-Poisson structure during
a visit to the University of Hong Kong.  He is also grateful to
MSRI and the organizers of the Program on Random matrices, Interacting
Particle Systems and Integrable Systems for the hospitality during his stay
in Berkeley in the Fall of 2010 where part of this work was being done.  The
second author would like to acknowledge the support of NSF grant
DMS-0701026.

\enddocument